# First M87 Event Horizon Telescope Results. VIII. Magnetic Field Structure near The Event Horizon

The Event Horizon Telescope Collaboration
(See the end matter for the full list of authors.)

## Abstract

Event Horizon Telescope (EHT) observations at 230 GHz have now imaged polarized emission around the supermassive black hole in M87 on event-horizon scales. This polarized synchrotron radiation probes the structure of magnetic fields and the plasma properties near the black hole. Here we compare the resolved polarization structure observed by the EHT, along with simultaneous unresolved observations with the Atacama Large Millimeter/submillimeter Array, to expectations from theoretical models. The low fractional linear polarization in the resolved image suggests that the polarization is scrambled on scales smaller than the EHT beam, which we attribute to Faraday rotation internal to the emission region. We estimate the average density $n_e \sim 10^{4\text{–}7}$ cm$^{-3}$, magnetic field strength $B \sim$ 1–30 G, and electron temperature $T_e \sim (1\text{–}12) \times 10^{10}$ K of the radiating plasma in a simple one-zone emission model. We show that the net azimuthal linear polarization pattern may result from organized, poloidal magnetic fields in the emission region. In a quantitative comparison with a large library of simulated polarimetric images from general relativistic magnetohydrodynamic (GRMHD) simulations, we identify a subset of physical models that can explain critical features of the polarimetric EHT observations while producing a relativistic jet of sufficient power. The consistent GRMHD models are all of magnetically arrested accretion disks, where near-horizon magnetic fields are dynamically important. We use the models to infer a mass accretion rate onto the black hole in M87 of $(3\text{–}20) \times 10^{-4}\, M_\odot\, \text{yr}^{-1}$.

## 1. Introduction

The Event Horizon Telescope (EHT) Collaboration has recently published total intensity images of event-horizon-scale emission around the supermassive black hole in the core of the M87 galaxy (M87*; Event Horizon Telescope Collaboration et al. 2019a, 2019b, 2019c, 2019d, hereafter EHTC I, EHTC II, EHTC III, EHTC IV). The data reveal a $42 \pm 3\,\mu$as diameter ring-like structure that is broadly consistent with the shadow of a black hole as predicted by Einstein's Theory of General Relativity (Event Horizon Telescope Collaboration et al. 2019e, 2019f; hereafter EHTC V, EHTC VI). The brightness temperature of the ring at 230 GHz ($\gtrsim 10^{10}$ K) is naturally explained by synchrotron emission from relativistic electrons gyrating around magnetic field lines. The ring brightness asymmetry results from light bending and Doppler beaming due to relativistic rotation of the matter around the black hole.

M87* is best known for launching a kpc-scale FR-I type relativistic jet, whose kinetic power is estimated to be $\sim 10^{42\text{–}44}$ erg s$^{-1}$ (e.g., Stawarz et al. 2006; de Gasperin et al. 2012). The structure of the relativistic jet has been resolved and studied at radio to X-ray wavelengths (e.g., Di Matteo et al. 2003; Harris et al. 2009; Kim et al. 2018; Walker et al. 2018).

The published EHT image of M87* together with multi-wavelength observations are consistent with the picture that the supermassive black hole in M87 is surrounded by a relativistically hot, magnetized plasma (Rees et al. 1982; Narayan & Yi 1995; Narayan et al. 1995; Yuan & Narayan 2014; Reynolds et al. 1996; Yuan et al. 2002; Di Matteo et al. 2003). However, it is not clear whether the compact ring emission is produced by plasma that is inflowing (in a thick accretion flow), outflowing (at the jet base or in a wind), or both. Furthermore, the total intensity EHT observations also could not constrain the structure of magnetic fields in the observed emission region. In order to find out which physical scenario is realized in M87*, additional information is necessary.

Event Horizon Telescope Collaboration et al. (2021, hereafter EHTC VII) reports new results from the polarimetric EHT 2017 observations of M87*. The polarimetric images of M87* are reproduced in Figure 1. These images reveal that a significant fraction of the ring emission is linearly polarized, as expected for synchrotron radiation. The EHT polarimetric measurements are consistent with unresolved observations of the radio core at the same frequency with the Submillimeter Array (SMA; Kuo et al. 2014) and the Atacama Large Millimeter/submillimeter Array (ALMA; Goddi et al. 2021). They also provide a detailed view of the polarized emission region on event-horizon scales near the black hole. Polarized synchrotron radiation traces the underlying magnetic field configuration and magnetized plasma properties along the line of sight (Bromley et al. 2001; Broderick & Loeb 2009; Mościbrodzka et al. 2017). These polarimetric measurements allow us to carry out new *quantitative tests* of horizon-scale scenarios for accretion and jet launching around the M87* black hole. In this Letter we present our interpretation of the EHTC VII resolved polarimetric images of the ring in M87*.

[126] NASA Hubble Fellowship Program, Einstein Fellow.
[127] EACOA Fellow.
[128] UKRI Stephen Hawking Fellow.

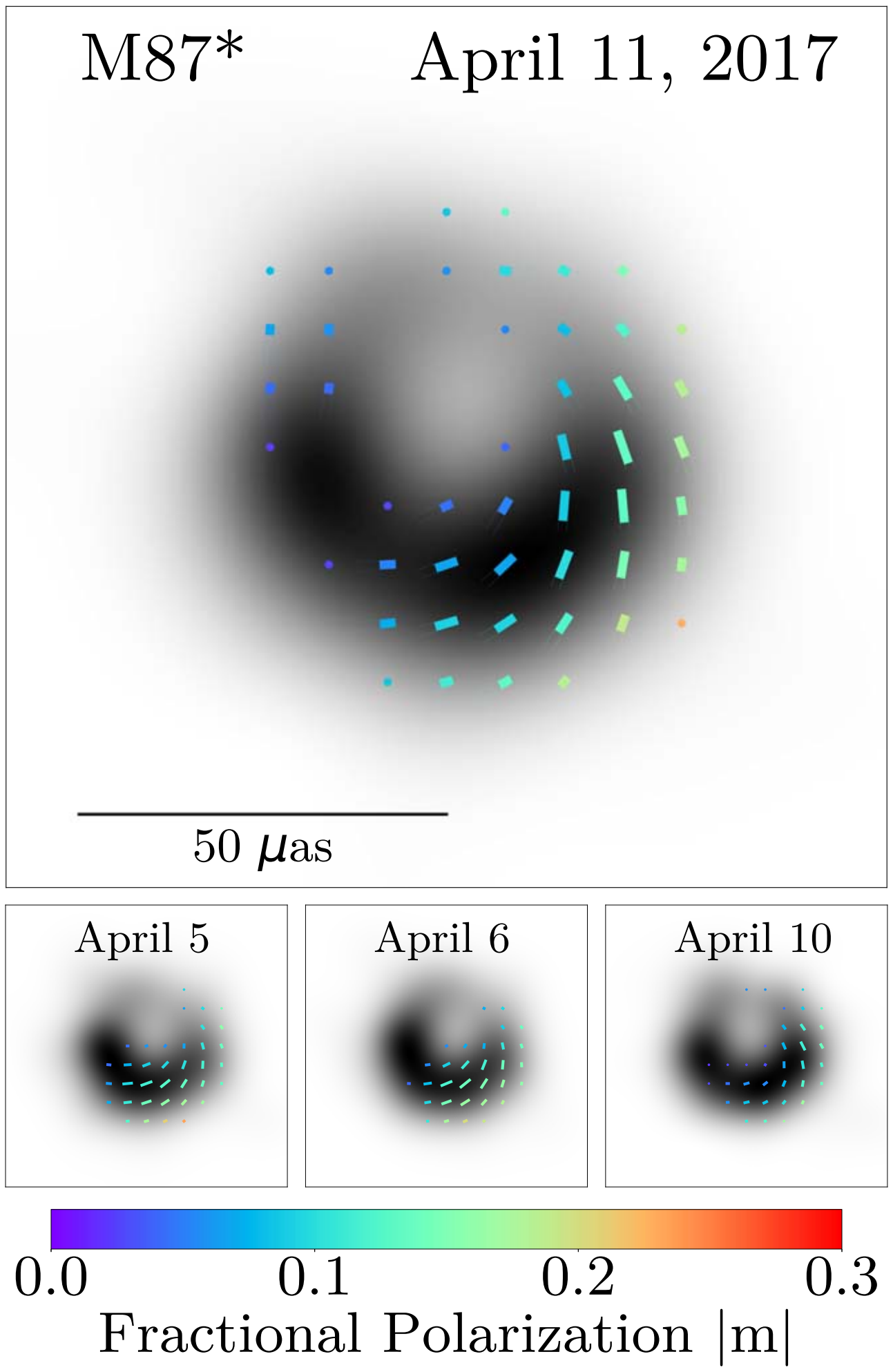

**Figure 1.** Top panel: 2017 April 11 fiducial polarimetric image of M87* from EHTC VII. The gray scale encodes the total intensity, and ticks illustrate the degree and direction of linear polarization. The tick color indicates the amplitude of the fractional linear polarization, the tick length is proportional to $|\mathcal{P}| \equiv \sqrt{\mathcal{Q}^2 + \mathcal{U}^2}$, and the tick direction indicates the electric-vector position angle (EVPA). Polarization ticks are displayed only in regions where $\mathcal{I} > 10\% \, \mathcal{I}_{\max}$ and $|\mathcal{P}| > 20\% |\mathcal{P}|_{\max}$. Bottom row: polarimetric images of M87* taken on different days.

Our analysis is presented as follows. In Section 2 we report polarimetric constraints from M87* EHT 2017 and supplemental observations, and argue that they can be used for scientific interpretation, focusing on several key diagnostics of the degree of order and spatial pattern of the polarization map. In Section 3 we present one-zone estimates of the properties of the synchrotron-emitting plasma. In the transrelativistic parameter regime relevant for the M87 core, a full calculation of polarized radiative transfer using a realistic description of the plasma properties is essential. To that end, in Section 4 we describe a set of numerical simulations of magnetized accretion flows that we use to compare with our set of observables. In Section 5 we show that only a small subset of the simulation parameter space is consistent with the observables. The favored simulations feature dynamically important magnetic fields. We discuss limitations of our models in Section 6, and discuss how future EHT observations can further constrain the magnetic field structure and plasma properties near the supermassive black hole's event horizon in Section 7.

## 2. Polarimetric Observations and Their Interpretation

### 2.1. Conventions in Observations and Models

Throughout this Letter we use the following definitions and conventions for polarimetric quantities, following the International Astronomical Union (IAU) definitions of the Stokes parameters ($\mathcal{I}$, $\mathcal{Q}$, $\mathcal{U}$, $\mathcal{V}$) (Hamaker & Bregman 1996; Smirnov 2011). The complex linear polarization field $\mathcal{P}$ is

$$\mathcal{P} = \mathcal{Q} + i\mathcal{U}. \tag{1}$$

Then, the electric-vector position angle (EVPA) is defined as

$$\mathrm{EVPA} \equiv \frac{1}{2}\arg(\mathcal{P}). \tag{2}$$

The EVPA is measured east of north on the sky. Therefore, positive $\mathcal{Q}$ is aligned with the north–south direction and negative $\mathcal{Q}$ with the east–west direction. Positive $\mathcal{U}$ is at a $+45$ deg angle with respect to the positive $\mathcal{Q}$ axis (in the northeast–southwest direction). Positive Stokes $\mathcal{V}$ indicates right-handed circular polarization, meaning in our convention that the electric field vector of the incoming electromagnetic wave is rotating counter-clockwise as seen by the observer. In the synchrotron radiation models that we consider, a positive value of emitted Stokes $\mathcal{V}$ is associated with an angle $\theta_B$ between the wavevector $k^\mu$ and magnetic field $b^\mu$ as measured in the frame of the emitting plasma in the range $\theta_B \in [0, 0.5\pi]$. Negative $\mathcal{V}$ corresponds to $\theta_B \in [0.5\pi, \pi]$.

The linear and circular polarization fractions at a point in the image are defined as

$$|m| \equiv \frac{|\mathcal{P}|}{\mathcal{I}}, \tag{3}$$

$$|v| \equiv \frac{|\mathcal{V}|}{\mathcal{I}}. \tag{4}$$

We also define the rotation measure between two wavelengths $\lambda_1$ and $\lambda_2$

$$\mathrm{RM} \equiv \frac{\mathrm{EVPA}(\lambda_1) - \mathrm{EVPA}(\lambda_2)}{\lambda_1^2 - \lambda_2^2}. \tag{5}$$

Unresolved observations measure the *net* (image-integrated) polarization fractions

$$|m|_{\mathrm{net}} = \frac{\sqrt{\left(\sum_i \mathcal{Q}_i\right)^2 + \left(\sum_i \mathcal{U}_i\right)^2}}{\sum_i \mathcal{I}_i}, \tag{6}$$

$$v_{\mathrm{net}} = \frac{\sum_i \mathcal{V}_i}{\sum_i \mathcal{I}_i}, \tag{7}$$

**Table 1**
ALMA-only Measurements of M87*'s Unresolved Polarization Properties at $\nu = 221$ GHz (Goddi et al. 2021)

| Day | $F$ (Jy) | $\|m\|_{\rm net}$ (%) | $\|v\|_{\rm net}$ (%) | RM ($10^5$ rad m$^{-2}$) |
|---|---|---|---|---|
| April 5 | $1.28 \pm 0.13$ | $2.42 \pm .03$ | $\leqslant 0.2$ | $(0.6 \pm 0.3)$ |
| April 6 | $1.31 \pm 0.13$ | $2.16 \pm .03$ | $\leqslant 0.3$ | $(1.5 \pm 0.3)$ |
| April 10 | $1.33 \pm 0.13$ | $2.73 \pm .03$ | $\leqslant 0.3$ | $(-0.2 \pm 0.2)$ |
| April 11 | $1.34 \pm 0.13$ | $2.71 \pm .03$ | $\leqslant 0.4$ | $(-0.4 \pm 0.2)$ |

where the sums are over all pixels $i$ in the resolved image. In addition to the signed circular polarization fraction $v_{\rm net}$, we also frequently consider the absolute value $|v_{\rm net}|$, as circular polarization measurements of the M87* core at 230 GHz do not constrain its sign (Goddi et al. 2021).

In describing the *resolved* linear polarization in EHT images, we define the image-average linear polarization fraction, weighted by the total intensity of each image pixel, as

$$\langle |m| \rangle = \frac{\sum_i \sqrt{\mathcal{Q}_i^2 + \mathcal{U}_i^2}}{\sum_i \mathcal{I}_i}. \tag{8}$$

Note that $\langle |m| \rangle$ depends on the imaging resolution (beam size), while $|m|_{\rm net}$ is the usual unresolved linear polarization fraction and does not depend on resolution.

### 2.2. Unresolved Polarization and Rotation Measure Measurements toward M87's Core from ALMA

As part of the EHT 2017 observation campaign, we obtained ALMA array measurements of the unresolved, net, near 230 GHz, polarimetric properties of M87's core and jet on 2017 April 5, 6, 10, and 11 (hereafter these observations are referred to as ALMA-only observations). ALMA-only measurements are given at $\nu = 221$ GHz, a central frequency of ALMA Band 6, which has four spectral windows, each centered at 213, 215, 227, and 229 GHz. These results, along with details on the observations and data calibration, are presented in Goddi et al. (2021); we summarize them here in Table 1. From the ALMA-only data, the net linear polarization fraction (Equation (6)) of the core is $|m|_{\rm net} \simeq 2.7\%$. The data also provide an upper limit on the net circular polarization fraction (Equation (7)) of the core of $|v|_{\rm net} \lesssim 0.3\%$, with a magnitude and sign that vary over the four observed epochs. Goddi et al. (2021) also measured an EVPA that varies with wavelength across the ALMA band; the slope of EVPA with wavelength is consistent with EVPA $\propto \lambda^2$, as expected for Faraday rotation. The inferred rotation measure (Equation (5), for frequencies/wavelengths in ALMA Band 6) is also time variable and changes sign between 2017 April 5 and 11, with a maximum magnitude $|\mathrm{RM}| \simeq 1.5 \times 10^5$ rad m$^{-2}$.

The ALMA-only measurements include extended ~arcsecond scale structures that are entirely resolved out of the EHT maps of M87's core region. As a result, the total 221 GHz flux density of M87* measured by ALMA alone is a factor of $\simeq$2 larger than that captured by the resolved EHT images (see also EHTC IV). For that reason, we adopt a more conservative upper limit of $|v|_{\rm net} < 0.8\%$, which would account for the case where the large-scale emission is not circularly polarized.

### 2.3. Spatially Resolved Linear Polarization of M87's Core in EHT 2017 Data

The resolved polarimetric images of the M87 core reported in EHTC VII display robust features between different image-reconstruction algorithms and across four days of observations (2017 April 5, 6, 10, and 11). At 20 $\mu$as resolution, those images consistently show a region of the highest linear polarized intensity in the southwest portion of the ring, with a fractional linear polarization $|m| \lesssim 30\%$ at its maximum. The image-average linear polarization fraction takes on values $5.7\% \leqslant \langle |m| \rangle \leqslant 10.7\%$ across the different observation days and image-reconstruction techniques. The range of the image-integrated net polarization fraction is $1.0\% \leqslant |m|_{\rm net} \leqslant 3.7\%$ (see EHTC VII, their Table 2). Because polarized emission outside the EHT field of view but inside the ALMA-only core is unconstrained, we adopt the EHT $|m|_{\rm net}$ range when evaluating models.

The EHT images on all four observing days reveal a characteristic azimuthal pattern of the EVPA angle around the emission ring. To quantify this pattern across the image, we use a decomposition of the complex linear polarization $\mathcal{P} = \mathcal{Q} + i\mathcal{U}$ into azimuthal modes with complex coefficients $\beta_m$ (Palumbo et al. 2020, hereafter PWP). For a polarization field in the image plane given in polar coordinates ($\rho$, $\varphi$), the $\beta_m$ coefficients are

$$\beta_m = \frac{1}{I_{\rm ann}} \int_{\rho_{\rm min}}^{\rho_{\rm max}} \int_0^{2\pi} \mathcal{P}(\rho, \varphi)\, e^{-im\varphi}\, \rho d\varphi d\rho, \tag{9}$$

where $I_{\rm ann}$ is the Stokes $\mathcal{I}$ flux density contained inside the annulus set by the limiting radii $\rho_{\rm min}$ and $\rho_{\rm max}$. We take $\rho_{\rm min} = 0$ and $\rho_{\rm max}$ to be large enough to include the entire EHT image.

Within the library of polarized images from general relativistic magnetohydrodynamic (GRMHD) simulations presented in EHTC V, PWP found that the $m = 2$ coefficient, $\beta_2$, was the most discriminating in identifying the underlying magnetized accretion model. The phase of $\beta_2$ maps well onto the qualitative behavior expected of polarization maps with idealized magnetic field configurations. In our convention, radial EVPA patterns have positive real $\beta_2$ ($\angle\beta_2 = 0$ deg), azimuthal EVPA patterns have negative real $\beta_2$ ($\angle\beta_2 = 180$ deg), and left- (right-) handed spiral patterns have positive (negative) pure imaginary $\beta_2$ ($\angle\beta_2 = 90$ deg and $-90$ deg, respectively). These idealized EVPA pattern configurations and their corresponding $\beta_2$ coefficients are summarized in Appendix A and Figure 1 of PWP. The measured range of the complex $\beta_2$ coefficient across the different image-reconstruction methods and observing days reported in EHTC VII, their Table 2, is $0.04 \leqslant |\beta_2| \leqslant 0.07$ for the amplitude and $-163$ deg $\leqslant \arg[\beta_2] \leqslant -129$ deg for the phase.

Appendix A demonstrates that the information in the $\beta_2$ coefficient can be obtained in the visibility domain using measurements of the linear polarization $E-$(gradient) and $B-$(curl) modes of the polarization field (e.g., Kamionkowski & Kovetz 2016). Trends in $\beta_2$ metric computed across the GRMHD image library (Section 4) can be obtained in the visibility domain using only $E$- and $B$-mode measurements taken on EHT 2017 baselines, as long as the visibilities are accurately phase calibrated. Because accurate phase calibration of EHT data is non-trivial and requires fully modeling the polarized source structure, we use image-domain comparisons

**Table 2**
Parameter Ranges for the Quantities Used in Scoring Theoretical Models in this Letter

| Parameter | Min | Max |
|---|---|---|
| $\|m\|_{\rm net}$ | 1.0% | 3.7% |
| $\|v\|_{\rm net}$ | 0 | 0.8% |
| $\langle\|m\|\rangle$ | 5.7% | 10.7% |
| $\|\beta_2\|$ | 0.04 | 0.07 |
| $\angle\beta_2$ | −163 deg | −129 deg |

**Note.** The ranges for $|m|_{\rm net}$, $\langle|m|\rangle$, and $\beta_2$ were taken from EHTC VII Table 2. These ranges represent the minimum $-1\sigma$ error bound and maximum $+1\sigma$ error bound across five different image-reconstruction methods, and incorporate both statistical uncertainty in the polarimetric image reconstruction and systematic uncertainty in the assumptions made by different reconstruction algorithms. The upper limit on $|v|_{\rm net}$ was taken as $\simeq 2\times$ the maximum value found by Goddi et al. (2021).

to the reconstructions presented in EHTC VII for the constraints in this Letter.

As in total intensity, both the unresolved and resolved polarimetric properties of the 230 GHz M87* image changed over the week between 2017 April 5 and April 11. Notably, the integrated EVPA in the EHT image changes by ≈30–40 deg (while the ALMA-only EVPA changes by $\lesssim$10 deg). We will not interpret this variability further in this work; however, Section 6 discusses expectations for time variability from viable simulation models. The observational ranges of the key parameters that we use in comparing theoretical models to data in Section 5—namely $|m|_{\rm net}$, $|v|_{\rm net}$, $\langle|m|\rangle$, and $\beta_2$ amplitude and phase—are summarized in Table 2.

### 2.4. External and Internal Faraday Rotation

Faraday rotation in a uniform plasma with rotation measure (RM) rotates the EVPA away from its intrinsic value $\mathrm{EVPA}_0$ according to Equation (5). The change in EVPA from its intrinsic value at 230 GHz ($\lambda \simeq 1.3$ mm) is

$$\Delta\mathrm{EVPA} \simeq 9.7\left(\frac{\mathrm{RM}}{10^5\ \mathrm{rad\ m^{-2}}}\right)\ \mathrm{deg}. \tag{10}$$

Polarized light rays passing through a uniform medium are subject to the same RM. The net source polarization angle is then coherently rotated away from its intrinsic value without any depolarization. This scenario of "external" Faraday rotation has been used to infer the mass accretion rate for sources where an RM is measured or constrained (e.g., Bower et al. 2003; Marrone et al. 2006, 2007; Kuo et al. 2014), by assuming that the observed radiation passes through the bulk of the accretion flow. Because relativistic electrons suppress the Faraday rotation coefficient as $\propto 1/T_e^2$ (e.g., Jones & Hardee 1979), these models assume that the RM is produced outside the emission region at the radius where $\Theta_e = kT_e/m_e c^2 = 1$, usually $r \sim 100\, r_g$ (where $r_g = GM/c^2$ is the gravitational radius).

However, in accreting systems like M87*, it is unclear whether this external Faraday rotation model is a good approximation. As we estimate below, one-zone emission models of M87* predict substantial RM within the emission region itself at radii $r \lesssim 5\, r_g$. At its low viewing inclination, the observed polarized radiation emitted near the horizon may not pass through the bulk of the high-density, infalling gas. Therefore, "internal" Faraday rotation, which can depolarize the emission as well as rotate the EVPA (Burn 1966), is also an important effect to consider.

The observed ≃10% linear polarization of the ring at the EHT scale of ∼20 μas is much lower than the intrinsic synchrotron polarization fraction $\gtrsim$70% expected locally. This could result from synchrotron self-absorption of the emitted radiation, but one-zone estimates and theoretical models (e.g., EHTC V, and references therein) suggest that the 230 GHz emission is mostly optically thin. It is more likely that the observed depolarization of the resolved emission could be the result of polarization structure that is scrambled at resolutions finer than the EHT beam. Turbulent magnetic fields and Faraday rotation internal to the emission region could produce this scrambling. In Section 4.3 we show that turbulence in GRMHD models alone is insufficient to produce this level of depolarization. Significant internal Faraday rotation of polarization vectors on different rays by different amounts can produce a sufficiently scrambled image that is depolarized when spatially averaged over a telescope resolution element (beam,e.g., Burn 1966; Agol 2000; Quataert & Gruzinov 2000; Beckert & Falcke 2002; Ruszkowski & Begelman 2002; Ballantyne et al. 2007).

From the simultaneous ALMA-only M87* observations, the RM implied by changes in the EVPA across the ALMA band is $|\mathrm{RM}| \lesssim 1.5 \times 10^5$ rad m$^{-2}$. These values are consistent with, but much more constraining than, the range determined from past SMA observations ($-3.4$–$7 \times 10^5$ rad m$^{-2}$, Kuo et al. 2014). The ALMA-only EVPA difference varies by order unity in magnitude and sign over the observing campaign, and includes a large flux contribution from extended emission not captured by EHT 2017 imaging (EHTC IV). Using a two-component model, Goddi et al. (2021) show that the RM toward the core emission in the EHT field of view could exceed the RM computed from the ALMA-only data, with allowed values as large as $|\mathrm{RM}| \lesssim 10^6$ rad m$^{-2}$. Because of this uncertainty, we do not use the observed RM as an observational constraint in our analysis. We account for uncertainty related to the observed time variability by using reconstructed polarized EHT images from both 2017 April 5 and 11 to define the acceptable ranges (see Table 2) of the observational parameters used to score theoretical models in Section 5.

## 3. Estimates and Phenomenological Models

In this Section, we take a first look at the importance of internal Faraday rotation and magnetic field structure in determining the characteristics of the 230 GHz EHT image. In Section 3.1 we obtain order-of-magnitude estimates of the plasma properties in M87* by interpreting the observed depolarization as entirely due to the effect of internal Faraday rotation on small scales. In Section 3.2 we explore the effects of different idealized magnetic field configurations on the observed polarization pattern from plasma orbiting a black hole in the absence of Faraday effects.

### 3.1. Parameter Estimates from One-zone Models

Based on a one-zone isothermal sphere model, EHTC V derived order-of-magnitude estimates of the plasma number density $n_e$ and magnetic field strength $B$ in the emitting region around M87* as constrained by the Stokes $\mathcal{I}$ image's brightness, size, and total flux density:

$$n_e \simeq 2.9 \times 10^4\ \mathrm{cm}^{-3}, \tag{11}$$

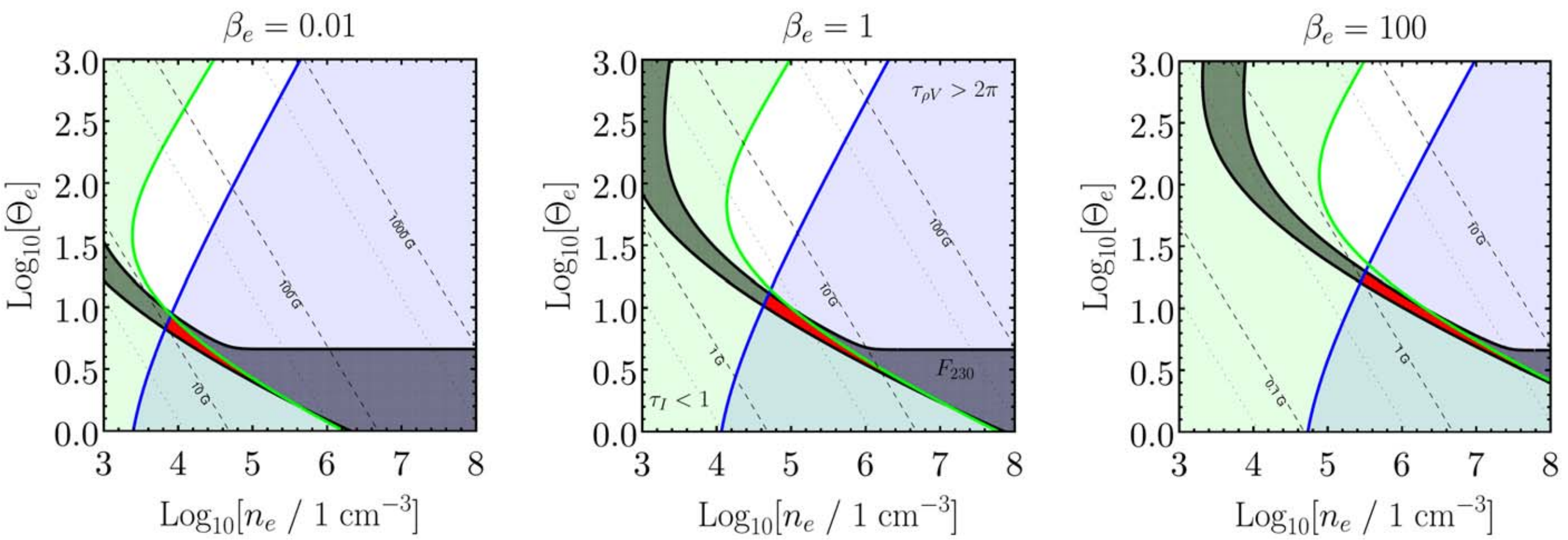

**Figure 2.** Allowed parameter space in number density and dimensionless electron temperature ($n_e$, $\Theta_e$) (red region) for the one-zone model described in Section 3.1 for three constant values of $\beta_e = 8\pi n_e m_e c^2 \Theta_e / B^2$. We require that the optical depth $\tau_I < 1$ (green region), the Faraday optical depth $\tau_{\rho_V} > 2\pi$ (blue region), and the total flux density $0.2 < F_\nu < 1.2$ Jy (black region). Contours of constant magnetic field strength are denoted by labeled dashed lines.

$$B \simeq 4.9\ \mathrm{G}. \tag{12}$$

In this model, the emission radius was assumed to be $r \simeq 5r_g$, and the electron temperature was assumed to be $T_{\mathrm{e}} = 6.25 \times 10^{10}$ K, based on the observed brightness temperature of the EHT image. This temperature corresponds to $\Theta_e = kT_e / m_e c^2 = 10.5$, so the emitting electrons have moderately relativistic mean Lorentz factors $\bar{\gamma} \approx 3\Theta_e \approx 30$. The angle between the magnetic field and line of sight is set at $\theta = \pi/3$. This model ignores several physical effects that are included in more sophisticated models and simulations and which are necessary to fully describe the emission from M87*. The plasma is considered to be at rest and so there is no Doppler (de)boosting of the emitted intensity from relativistic flow velocities. Redshift from the gravitational potential of the black hole is also not included.

Given $n_{\mathrm{e}}$, $B$, and $T_{\mathrm{e}}$, we can estimate the strength of the Faraday rotation effect at 230 GHz quantified by the optical depth to Faraday rotation $\tau_{\rho_V}$:

$$\tau_{\rho_V} \approx r \times \rho_V \simeq 5.2 \left(\frac{r}{5r_g}\right), \tag{13}$$

where $\rho_V$ is the Faraday rotation coefficient (e.g., Jones & Hardee 1979). For emission entirely behind an external Faraday screen, $\tau_{\rho_V}$ is related to the rotation measure RM via $\tau_{\rho_V} = 2\mathrm{RM}\lambda^2$, which follows from the radiative transfer equations for spherical Stokes parameters in the absence of other effects (see e.g., Appendix A of Mościbrodzka et al. 2017) and the fact that $\rho_V \propto \lambda^2$.

Our estimated $\tau_{\rho_V}$ indicates that Faraday rotation internal to the emission region is an important effect and could thus explain the depolarization observed in M87*. Faraday effects are even more important for the case of polarized light emitted by relativistic electrons that travel through a dense, colder accretion flow (e.g., Mościbrodzka et al. 2017; Ricarte et al. 2020). In addition, for the same parameters, Faraday conversion of linear to circular polarization may also be important ($\tau_{\rho_Q} \simeq 0.5$), while self-absorption is weak ($\tau_I \simeq 0.05$). Requiring an internal Faraday optical depth $\tau_{\rho_V} > 2\pi$ (large enough to produce significant depolarization) provides an additional constraint on one-zone models independent of those used in EHTC V, which fixed the electron temperature at an assumed value. Assuming $\tau_{\rho_V} > 2\pi$ allows us to break the degeneracy between magnetic field strength, electron temperature, and plasma number density.

Hence, we consider the same model as in EHTC V at several different values of $\beta_e = 8\pi n_e k T_e / B^2$, constrained by the requirement that the Faraday optical depth $\tau_{\rho_V} > 2\pi$. To be consistent with the 230 GHz EHT data, we also require that the observed image have a total flux $F_\nu$ between 0.2 and 1.2 Jy, and that the model has a maximum total intensity optical depth $\tau_I < 1$. Figure 2 shows what constraints these requirements put on the electron number density $n_e$ and the dimensionless electron temperature $\Theta_e$ at three different values of $\beta_e$. For values of $0.01 < \beta_e < 100$, in this simple model the electron temperature is constrained to lie in a mildly relativistic regime $2 \lesssim \Theta_e \lesssim 20$ ($10^{10} < T_e < 1.2 \times 10^{11}$ K), and the magnetic field strength is $1 \lesssim B \lesssim 30$ G. The number density of the emitting electrons depends more sensitively on the assumed value of $\beta_e$, taking on values between $10^4\ \mathrm{cm}^{-3}$ and $10^7\ \mathrm{cm}^{-3}$.

The one-zone model estimates suggest that both the total intensity and polarized emission can be produced in a mildly relativistic plasma in a magnetic field of relatively low strength $B \lesssim 30$ G. For higher values of $B$, the electron temperature would be too small to explain the observed maximum brightness temperature ($\simeq 10^{10}$ K) in the M87* EHT image (EHTC IV). Very high values of $B$ are independently disfavored by the small degree of circular polarization $|v|_{\mathrm{net}} \lesssim 1\%$ seen in M87*. For $B \simeq 100$ G, the ratio of the Stokes $\mathcal{V}$ emissivity to the Stokes $\mathcal{I}$ emissivity $j_V / j_I \simeq 1\%$. For $B \simeq 10^3$ G, $j_V / j_I \simeq 10\%$, for all temperatures $> 10^{10}$ K. We also note that magnetic fields of $B \gtrsim 5$ G are sufficient to produce jet powers of $P_{\mathrm{jet}} \gtrsim 10^{42}\ \mathrm{erg\ s}^{-1}$ (e.g., EHTC V) via the Blandford & Znajek (1977) process.

### 3.2. EVPA Pattern and Field Geometry

To demonstrate how the intrinsic magnetic field structure in the emission region influences the observed polarization pattern, in this section we present the polarization configurations from three idealized magnetic field geometries around a black hole—a purely toroidal field, a purely radial field, and a purely vertical field— as seen by a distant observer. In Figure 3 we show polarimetric images from these simple field configurations computed with two methods: a numerical model of an optically thin emission region around the black hole (top row of Figure 3), and an analytic treatment of the parallel transport of the polarization vector that is originally perpendicular to the magnetic field (R. Narayan et al. 2021, in preparation, middle row of Figure 3). We show the polarization maps from both methods for the three idealized magnetic field configurations viewed at an inclination angle of $i = 163$ deg. Both the analytical and numerical calculations assume a zero-spin black hole (Schwarzschild metric), though we have found that the main features of these polarization patterns are insensitive to spin.

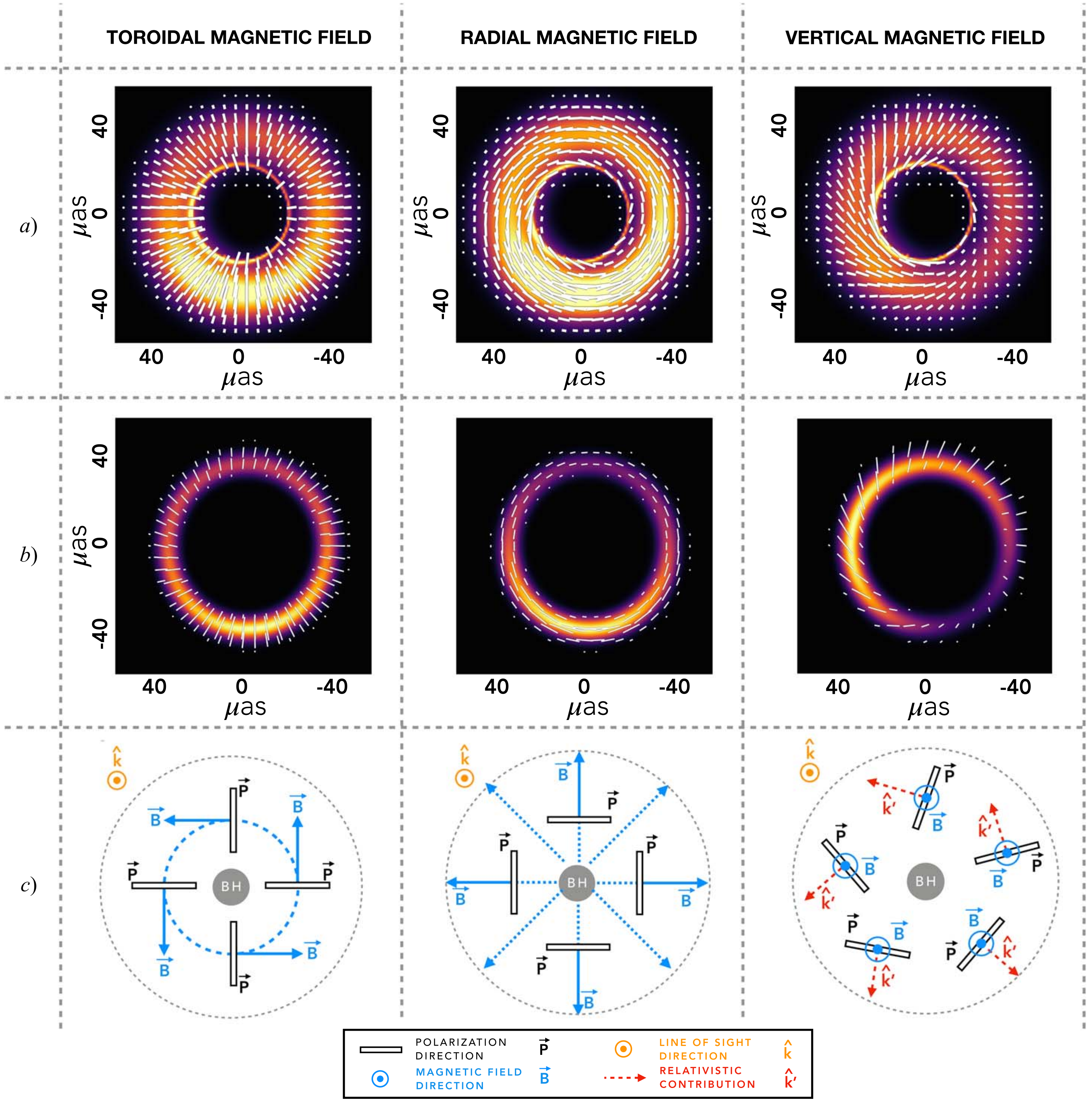

**Figure 3.** (a) Numerical calculations of the polarization configuration generated by an orbiting emission region in the shape of a torus at $8r_g$ in three imposed magnetic field geometries and viewed at $i = 163$ deg (with material orbiting clockwise on the sky). The orbital angular momentum vector is pointing away from the observer and to the east (to the left). Total intensity is shown in the background with higher brightness temperature regions shown as being lighter in color. In the foreground, the observed EVPA direction is shown with white ticks, with the tick length proportional to the polarized flux. (b) Analytic calculations of the polarization configuration from a thin ring of magnetized fluid at $8r_g$ inclined by 163 deg to the observer in the same magnetic field geometries as in (a). While the distribution of emitting material is different in the two models, both the sense of asymmetry in the brightness distributions and the polarization patterns match those from the numerical calculations. (c) Schematic cartoons showing the emitting frame wavevector $\hat{k}$, magnetic field direction $\hat{B}$, and polarization vector $\hat{P} = \hat{k} \times \hat{B}$ for each case. In the bottom-right panel, $\hat{k}'$ denotes the approximate light bending contribution to the wavevector.

In the top row of Figure 3 we show the result of numerical calculations performed with the general relativistic ray tracing code `grtrans` (Dexter & Agol 2009; Dexter 2016) of polarized emission from an optically and Faraday-thin compact emission region, or "hotspot", in Keplerian orbit around a black hole in the equatorial plane. The hotspot has a radial extent of $3r_g$ and moves in an imposed and idealized magnetic field geometry in a circular orbit at a radius of $8r_g$ (following Gravity Collaboration et al. 2018, 2020). We construct a phenomenological model of a torus of emitting, rotating plasma by studying the time-averaged polarized emission images from one revolution of this hotspot around the black hole. We have verified that a semi-analytic implementation (Broderick & Loeb 2006) of a hot accretion flow model (Yuan et al. 2003) produces consistent polarization patterns when using the same field geometry.

In the second row of Figure 3, we compare these numerical results to results from an analytic calculation of the observed polarization pattern generated by the emission of polarized light on a thin ring of radius $8r_g$ in the equatorial plane. In this model (R. Narayan et al. 2021, in preparation) the polarization vectors are emitted perpendicular to the imposed magnetic field geometry in the fluid rest frame; they are transformed on their way to the observer using an approximate, analytic treatment of the effects of light bending, parallel transport, and Doppler beaming. This calculation includes radial inflow as well as rotation in the velocity field; the models shown use purely toroidal motion (clockwise on the sky) with the same idealized magnetic field geometries as in the numerical case. The models match the asymmetric brightness distributions and polarization patterns of the numerical calculations. In particular, both models produce consistent helical EVPA pattern in the case of a vertical magnetic field.

The linear polarization direction $\bar{P}$ of synchrotron radiation in the emitted frame is perpendicular to the wavevector $\hat{k}$ and the magnetic field vector $\bar{B}$. We define the toroidal magnetic field as consisting only of magnetic field components in the azimuthal direction, while the poloidal magnetic field consists of the remainder, including both radial and vertical components. In a purely toroidal field case, the EVPA shows a radial pattern (left column in Figure 3). Purely radial magnetic fields (middle column) give a complementary result; the polarization has a toroidal configuration, similar to a 90 deg rotation of the linear polarization ticks from the toroidal case.

In a vertical magnetic field (right column in Figure 3), we might expect that $\bar{P}$ should be vertical (north–south) everywhere since a vertical $\bar{B}$ is tilted east–west for this viewing geometry. We might also expect that $\bar{P} \simeq 0$ when the black hole is viewed face on, because $\hat{k}||\bar{B}$. Instead, the linearly polarized emission from a purely vertical field shows a twisting pattern that wraps around the black hole. The twist results from a combination of light bending and relativistic aberration. Light bending in the emitting region near the black hole contributes a radial component $\hat{k}'$ to the emitted wavevector $\hat{k}$ that initially points away from the black hole (see the schematic cartoon in the bottom-right panel of Figure 3). As a result, close to the black hole, the total wavevector $\hat{k}_{\rm emit} = \hat{k} + \hat{k}'$ and the magnetic field $\bar{B}$ are no longer parallel, the polarization is non-zero, and the resulting EVPA pattern is north–south symmetric. Relativistic motion of the emitting material (aberration) breaks the symmetry and gives the twisting pattern a handedness corresponding to the orbital direction. For the pure vertical field considered here, the handedness depends on the rotation direction and the observed pattern is consistent with clockwise rotation. The dependence on direction of motion and magnetic field configuration are discussed in more detail in a forthcoming paper (R. Narayan et al. 2021, in preparation). The EVPA patterns in these images do not show a strong dependence on the black hole spin.

In a rotating flow, weak magnetic fields are sheared into a predominantly toroidal configuration (e.g., Hirose et al. 2004). In the absence of other effects (e.g., external Faraday rotation), the observed azimuthal EVPA pattern suggests the presence of dynamically important magnetic fields in the emission region, which can retain a significant poloidal component in the presence of rotation. In the next sections, we compare numerical simulations of the accretion flow and jet-launching region in M87* with different field configurations to the EHT2017 data to better constrain the magnetic field structure.

## 4. M87* Model Images from GRMHD Simulations

The low resolved fractional linear polarization observed by the EHT contradicts the results from an idealized magnetic field structure with no disorder. For typical parameters of the 230 GHz emission region, Faraday rotation and conversion are expected to be important. Magnetic field structure, plasma dynamics and turbulence, and radiative transfer effects including Faraday rotation can be realized in images from three-dimensional general relativistic magnetohydrodynamic (3D GRMHD) simulations of magnetized accretion flows. We use 3D GRMHD simulations (described in Section 4.1) in combination with polarized general relativistic radiative transfer (GRRT) models (described in Section 4.2) to model polarized images of M87*. In Section 4.3, we describe trends of the key observables ($|m|_{\rm net}$, $|v|_{\rm net}$, $\langle |m| \rangle$, and $\beta_2$) in our GRMHD polarimetric image library.

### *4.1. GRMHD Model Description*

The simulation library generated for the analysis of the EHT 2017 total intensity data in EHTC V consists of a set of 3D GRMHD simulations that were postprocessed to generate simulated black hole images via GRRT. For simulations using black holes with non-zero angular momentum, we only considered accretion flows in which the angular momentum of the flow and the hole were aligned (parallel or anti-parallel). Because the equations of non-radiating[129] GRMHD are scale invariant, each fluid simulation was thus fully parameterized by two values describing the angular momentum of the black hole and the relative importance of the magnetic flux near the horizon of the accretion system. A comparison of several contemporary GRMHD codes, including those used to generate the simulation library, can be found in Porth et al. (2019) and in H. Olivares et al. (2021, in preparation).

The black hole angular momentum $J$ is expressed in terms of the dimensionless black hole spin parameter $a_* \equiv Jc/GM^2$. In this Letter, we consider simulations run with the `iharm` code (Gammie et al. 2003; Noble et al. 2006) with $a_* = -0.94$, $-0.5$, 0, 0.5, and 0.94, where positive (negative) spin implies alignment (anti-alignment) between the accretion disk and the black hole angular momentum. Several studies of "tilted" disks have been conducted (e.g., Fragile et al. 2007; McKinney et al. 2013; Morales Teixeira et al. 2014; Liska et al. 2018; White et al. 2019; Chatterjee et al. 2020). As there does not yet exist a full library of tilted disk simulations spanning a range of spins, we limit our analysis to the aligned and anti-aligned simulations considered in EHTC V.

[129] We assume that M87* can be described by models in which radiative cooling is negligible so that it does not affect the dynamics of the plasma and images can be generated in postprocessing.

The strength of the magnetic flux near the horizon qualitatively divides accretion flow solutions into two categories: the Magnetically Arrested Disk (MAD) state (e.g., Bisnovatyi-Kogan & Ruzmaikin 1974; Igumenshchev et al. 2003; Narayan et al. 2003) in which the magnetic flux near the horizon saturates and significantly affects the dynamics of the flow, and the contrasting Standard and Normal Evolution (SANE) state (e.g., De Villiers et al. 2003; Gammie et al. 2003; Narayan et al. 2012). The relative importance of magnetic flux in a simulation is quantitatively described by the dimensionless quantity

$$\phi \equiv \Phi_{\rm BH}(\dot{M} r_g^2 c)^{-1/2}, \tag{14}$$

where $\Phi_{\rm BH}$ is the magnitude of the magnetic flux crossing one hemisphere of the event horizon (see Tchekhovskoy et al. 2011; Porth et al. 2019) and $\dot{M}$ is the mass accretion rate through the event horizon. The flux saturates at values of $\phi \gtrsim 50$, and the flow becomes MAD. The SANE simulations that we consider have lower values of $\phi \approx 5$.[130] Accreted material supplied at large scales could, in principle, supply any value of net vertical flux. Here, we do not explore cases with small or zero net vertical flux $\phi \lesssim 1$. We also do not consider values in the relatively narrow intermediate range $5 \lesssim \phi \lesssim 50$.

The SANE simulations considered here used a grid resolution of $288 \times 128 \times 128$, a fluid adiabatic index $\gamma = 4/3$, and an outer simulation domain of $r_{\rm out} = 50\, r_g$. The MAD simulations used a grid resolution of $384 \times 192 \times 192$, an adiabatic index $\gamma = 13/9$, and an outer simulation domain of $r_{\rm out} = 10^3\, r_g$. The simulations were carried out in modified spherical polar Kerr–Schild coordinates, where grid resolution is concentrated toward the midplane to help resolve the magnetorotational instability. All models in the EHT library are evolved for at least $t = 10^4 r_g/c$ and their accretion flows reach steady state within $r = 10$–$20\, r_g$.

### 4.2. Ray-traced Polarimetric Images from GRMHD Simulations

Unlike the equations of GRMHD, the equations of radiative transfer are not scale invariant, and so we must introduce two scales to the simulation when we ray-trace images from the numerical fluid data. The length (and time) scale is set by the mass of the black hole, assumed to be $M_{\rm BH} = 6.2 \times 10^9 M_\odot$ in accordance with the value used to generate the EHTC V simulation library. For our models, we also adopt the $D = 16.9$ Mpc distance to M87* used in EHTC V. The density scale of the accreting plasma (equal to the scale of the magnetic pressure) is chosen so that on average the simulated images reproduce the observed 230 GHz compact flux density, $F_\nu \simeq 0.5$ Jy.

Images were generated from the set of simulations for several values of the polar inclination angle $i$ chosen to be broadly consistent with observational estimates of the inclination angle of the M87 jet (e.g., Walker et al. 2018). The position angle on the sky can be changed after image generation by rotating both the image and the Stokes $\mathcal{Q}$ and $\mathcal{U}$ components appropriately. Each image has a $320 \times 320$ pixel resolution over a 160 $\mu$as field of view, where each pixel contains full Stokes $\mathcal{I}$, $\mathcal{Q}$, $\mathcal{U}$, $\mathcal{V}$ intensities.

In GRMHD simulations, we make the approximation that the plasma is thermal, i.e., that the electrons and ions are described by a Maxwell-Jüttner distribution function (Jüttner 1911). However, the plasma around M87* and in other hot accretion flows is most likely collisionless, with electrons and protons that are unable to equilibrate their temperatures (e.g., Shapiro et al. 1976; Ichimaru 1977). We mimic collisionless plasma properties in producing images from the GRMHD simulations by allowing the electron temperature $T_e$ to deviate from the proton temperature $T_i$. The simulations used in this work only track the total internal energy density $u_{\rm gas}$, not the distinct electron and ion temperatures. We set $T_{\rm e}$ after running the simulation according to local plasma parameters following the parameterization introduced by Mościbrodzka et al. (2016; see also Mościbrodzka et al. 2017 and EHTC V). The ratio between the ion and electron temperatures $R$ is determined by the local plasma $\beta = p_{\rm gas}/p_{\rm mag}$, where $p_{\rm gas} = (\gamma - 1)u_{\rm gas}$, and $p_{\rm mag} = B^2/8\pi$. The temperature ratio is then taken to be

$$R = \frac{T_i}{T_e} = R_{\rm high}\frac{\beta^2}{1+\beta^2} + R_{\rm low}\frac{1}{1+\beta^2}, \tag{15}$$

where $R_{\rm high}$ ($R_{\rm low}$) are the free parameters of the model and give the approximately constant temperature ratio at high (low) $\beta$. This approach allows us to associate the electron heating with magnetic properties of the plasma.

In calculating the electron temperature, we further assume that the plasma is purely ionized hydrogen and that ions are nonrelativistic with an adiabatic index $\gamma_p = 5/3$ and electrons are relativistic with $\gamma_e = 4/3$. Then, given $u_{\rm gas}$ from the simulation and $R$ from Equation (15), (EHTC V):

$$T_e = \frac{2\, m_p u_{\rm gas}}{3\rho k(2 + R)}. \tag{16}$$

We note that this procedure is not entirely self-consistent, as the $\gamma$ of the combined electron-ion fluid will change depending on the relative pressure contributions of electrons and protons while we assume it is fixed throughout the simulation domain. See Sądowski et al. (2017) for an alternative, self-consistent approach.

In this Letter, we consider a library of 72,000 simulated images composed of sets of 200 realizations of the same accretion system described by a fixed set of heating and observation parameters. Each set of 200 images is drawn from output files spaced by 25–50 $r_g/c$ from the set of 10 GRMHD simulations spanning five spin values in both MAD and SANE field configurations. The inclination angle for each image is set to one of either $i = 12, 17, 22$ deg (retrograde models, $a_* < 0$) or $i = 158, 163, 168$ deg (prograde models, $a_* \geqslant 0$), according to which parity is required to orient the brightest portion of the ring in the southern part of the image while ensuring the orientation of the approaching jet is consistent with large-scale observations.

We use electron heating parameters $R_{\rm low} = 1$, 10 and $R_{\rm high} = 1$, 10, 20, 40, 80, or 160 in Equation (15). EHTC V only considered models with $R_{\rm low} = 1$. Larger values of $R_{\rm low}$ correspond to lower electron-to-proton temperature ratios in the low $\beta$ regions (e.g., the jet funnel). This choice is physically motivated for M87*, where radiative cooling of the electrons may keep $T_e < T_i$ even in magnetized regions where electron heating is efficient (e.g., Mościbrodzka et al. 2011; Ryan et al. 2018; Chael et al. 2019). Lower electron temperatures in $R_{\rm low} = 10$ models increase the Faraday rotation depth and can result in increased depolarization in parts of the image.

[130] Note that the MAD threshold $\phi \gtrsim 50$ is given in Gaussian units where $[\Phi] = {\rm G\, cm^2}$. If the field strength is given in the Lorentz–Heaviside units typically used in simulations ($B_{\rm LH} = B_{\rm G}/\sqrt{4\pi}$), the MAD threshold on the dimensionless flux is $\phi \simeq 15$.

GRMHD simulations produce a highly magnetized jet funnel above the black hole's poles, away from the accretion disk. In the funnel, where the plasma magnetization parameter $\sigma \equiv B^2/4\pi\rho c^2 \gg 1$, our numerical methods typically fail to accurately evolve the plasma internal energy. In the image library, we cut off all emission in regions where $\sigma > 1$ to ensure that we limit the emitting region to plasma whose internal energy is safely evolved without numerical artifacts (as in EHTC V). We tested the importance of a $\sigma > 1$ electron population by generating a supplementary set of images from all models with a cut at $\sigma = 10$ and found that it did not change the overall distribution of the derived metrics we use for model scoring in Section 5.

Each set of 200 model images with the same parameters in the image library requires a unique density scaling factor that is determined by matching the average flux density from the model to the observed compact flux density of M87* measured by the EHT. Hence, the mass accretion rates, radiative efficiencies, and jet powers will differ between two models even if they are derived from the same underlying simulation (e.g., if $R_{\rm high}$, $R_{\rm low}$, or $i$ are changed). The additional models discussed in Section 6, which explore the effects of different $\sigma$ cutoff values and the inclusion nonthermal electrons, also require unique mass-scaling factors.

All of the polarimetric images from GRMHD simulations that we analyze in this Letter were generated using the `ipole` code (Mościbrodzka & Gammie 2018), which has been tested against analytic solutions and numerical ones produced by other numerical GRRT codes (Dexter 2016; Mościbrodzka 2020). A more comprehensive comparison of various GRRT codes that perform total intensity transport and fully polarized GRRT can be found in Gold et al. (2020) and B. Prather et al. (2021, in preparation), respectively. Preliminary results from B. Prather et al. (2021, in preparation) show that the tested codes are consistent at the fraction of 1% in all Stokes parameters. All calculated images in this work ignore light travel time delays through the emission region (the so-called "fast light" approach), and are calculated at a single frequency of $\nu = 230$ GHz, neglecting the finite observing bandwidth of the EHT. We confirm that neither of those effects are important for models of interest for M87*.

### 4.3. Sample GRMHD Model Images and Polarization Maps

In Figures 4 and 5 we show images and polarization maps for a subset of library models. In general, because the horizon-scale magnetic fields in MAD models are strong enough not to be advected with the accretion flow, they are more likely to have a significant poloidal component and produce azimuthal EVPA patterns (Figure 3). In contrast, SANE models tend to show more radial EVPA patterns. Some MAD $a_* = 0.94$ and SANE $a_* = 0$ images are notable exceptions to this trend. These trends are also apparent in the distributions of the $\beta_2$ phase across the full image library that we consider later in Figure 9.

The GRMHD models at their native resolution include notable disorder in the EVPA structure, resulting from both magnetic turbulence and Faraday rotation. Models with larger $R_{\rm high}$ have lower electron temperatures and higher Faraday rotation depths, resulting in the most disordered polarization maps. Many of the EVPA patterns seen in the images blurred with a 20 $\mu$as Gaussian kernel to simulate the limited EHT resolution resemble those from the idealized magnetic field models in Figure 3, indicating that the net EVPA pattern after blurring may trace the intrinsic magnetic field structure.

In Figure 6 we show a sample polarization map at full resolution compared to the same map blurred with circular Gaussian kernels of 10 $\mu$as and 20 $\mu$as FWHM. From tests with synthetic data, blurring (convolving with a circular Gaussian kernel) provides a reasonable approximation to image reconstruction from the EHT data at a comparable resolution (EHTC VII). The resolved average fractional polarization in the blurred images $\langle |m| \rangle$ traces the degree of order in the intrinsic polarization map. In the blurred images, disordered polarized structure on small scales produces beam depolarization. The degree of depolarization decreases with increasing spatial resolution (decreasing beam size).

The bottom row of Figure 6 shows the same unblurred and blurred polarization maps, but calculated without the effect of Faraday rotation ($\rho_V = 0$). Those images show more coherent EVPA structure, with much larger $|m|_{\rm net}$ and, particularly when blurred, much larger $\langle |m| \rangle$. Evidently, for this particular model, the depolarization visible in the corresponding top panels is due to Faraday rotation internal to the emission region. In addition, the net EVPA pattern shifts by a significant amount. The change in $\beta_2$ by $\simeq$80 deg would correspond to an apparent RM of $\simeq -4 \times 10^5$ rad m$^{-2}$. Our GRRT calculations include all Faraday rotation occurring inside the GRMHD simulation domain ($r_{\rm out} = 50$–$100\, r_g$), both external and internal to the 230 GHz emission region. The observables considered here, for the low viewing inclination of M87*, do not depend strongly on that outer radius, as long as it is at $r \gtrsim 40\, r_g$. We cannot rule out the presence of additional Faraday rotating material at larger radii $\gtrsim 100\, r_g$, and its effects are not included in our models. Appendix B discusses the origin of the RM in our models in more detail.

### 4.4. GRMHD Model Theory Metrics

We compute the polarimetric observables ($|m|_{\rm net}$, $|v|_{\rm net}$, $\langle |m| \rangle$, $\beta_2$) described in Section 2.3 from model images blurred with a circular Gaussian kernel with a FWHM of 20 $\mu$as in order to compare them to the ranges measured from EHT and ALMA-only data. Both $\langle |m| \rangle$ and $\beta_2$ depend on the resolution and hence the size of the Gaussian blurring kernel. The value of $\beta_2$ also depends on the choice of the image center. We do not shift the library images before computing $\beta_m$ coefficients for comparison with the range inferred from the EHT image reconstructions, which have been centered by aligning them to the centered, fiducial total intensity images in EHTC IV. As discussed in Palumbo et al. (2020), a centering offset $u$ expressed as a fraction of the diameter of a PWP $m = 2$ ring causes a quadratic falloff in $\beta_2$ power as $\delta\beta_2/|\beta_2| \approx 4u^2$. Effects on the $\beta_2$ phase enter at similar order. In the case of the EHT image, $u$ is likely less than one-fifth, meaning that centering errors in $\beta_2$ will be sub-dominant to other uncertainties, such as the choice of the blurring kernel or the variation across methods and days.

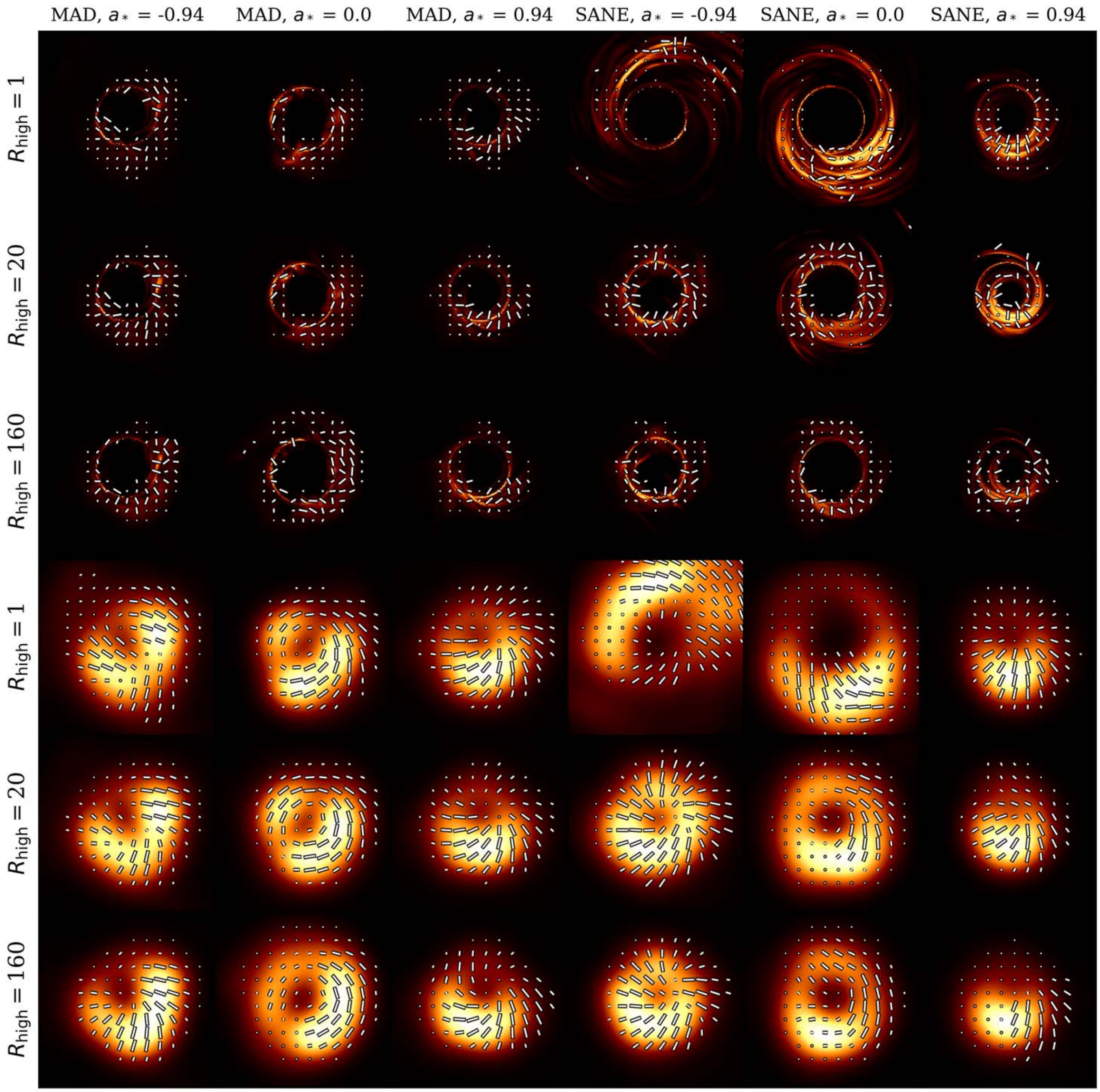

**Figure 4.** Sample snapshot false-color images and polarization maps for a subset of the models in the EHT M87* simulation image library at their native resolution (top three rows) and blurred with a 20 $\mu$as circular Gaussian beam (bottom three rows). The inclination angle for all images is either 17 deg (for negative $a_*$ models) or 163 deg (for positive $a_*$ model), with the black hole spin vector pointing to the left and away from the observer. The tick length is proportional to the polarized flux, saturated at 0.5 of the maximum value in each panel. Here models with $R_{\rm low} = 1$ are shown. In general, the EVPA pattern is predominantly azimuthal for MAD models (e.g., MAD $a_* = 0$ $R_{\rm high} = 1$) and radial for SANE models (e.g., SANE $a_* = 0.94$ $R_{\rm high} = 1$), although the SANE $a_* = 0$ models in particular are exceptions to this trend. All models show scrambling in the polarization structure on small scales from internal Faraday rotation, with more pronounced scrambling in models with cooler electrons (larger $R_{\rm high}$ parameter).

Figure 7 (right panel) shows the resolved average polarization fraction $\langle|m|\rangle$ as a function of their image-averaged Faraday rotation depth, $\langle\tau_{\rho_V}\rangle$. At small $\langle\tau_{\rho_V}\rangle$, the average polarization fraction is $\langle|m|\rangle \simeq 20\%$–$50\%$. Intrinsic disorder in the magnetic field structure due to turbulence is generally insufficient to produce the low observed image-average polarization fraction in EHT 2017 M87* data ($5.7\% \leqslant \langle|m|\rangle \leqslant 10.7\%$). This is especially evident for the SANE models with prograde black hole spin, which have the highest resolved polarization fractions. At large $\langle\tau_{\rho_V}\rangle$, strong scrambling from internal Faraday rotation typically results in small predicted polarization fractions of $<5\%$ at the scale of the EHT beam.

The clear exceptions to this trend are some SANE retrograde models ($a_* = -0.9375$ for large $R_{\rm high}$), which show $\langle|m|\rangle \simeq 10\%$–$20\%$ despite their large $\langle\tau_{\rho_V}\rangle \gtrsim 10^3$. In these models, most of the observed polarized flux originates in the forward jet, while most of the computed Faraday depth is accumulated near the midplane. Photons that travel from the forward jet to the observer do not encounter the large Faraday depth. For similar reasons, the inferred RM can be much lower than implied by their large values of integrated $\tau_{\rho_V}$.

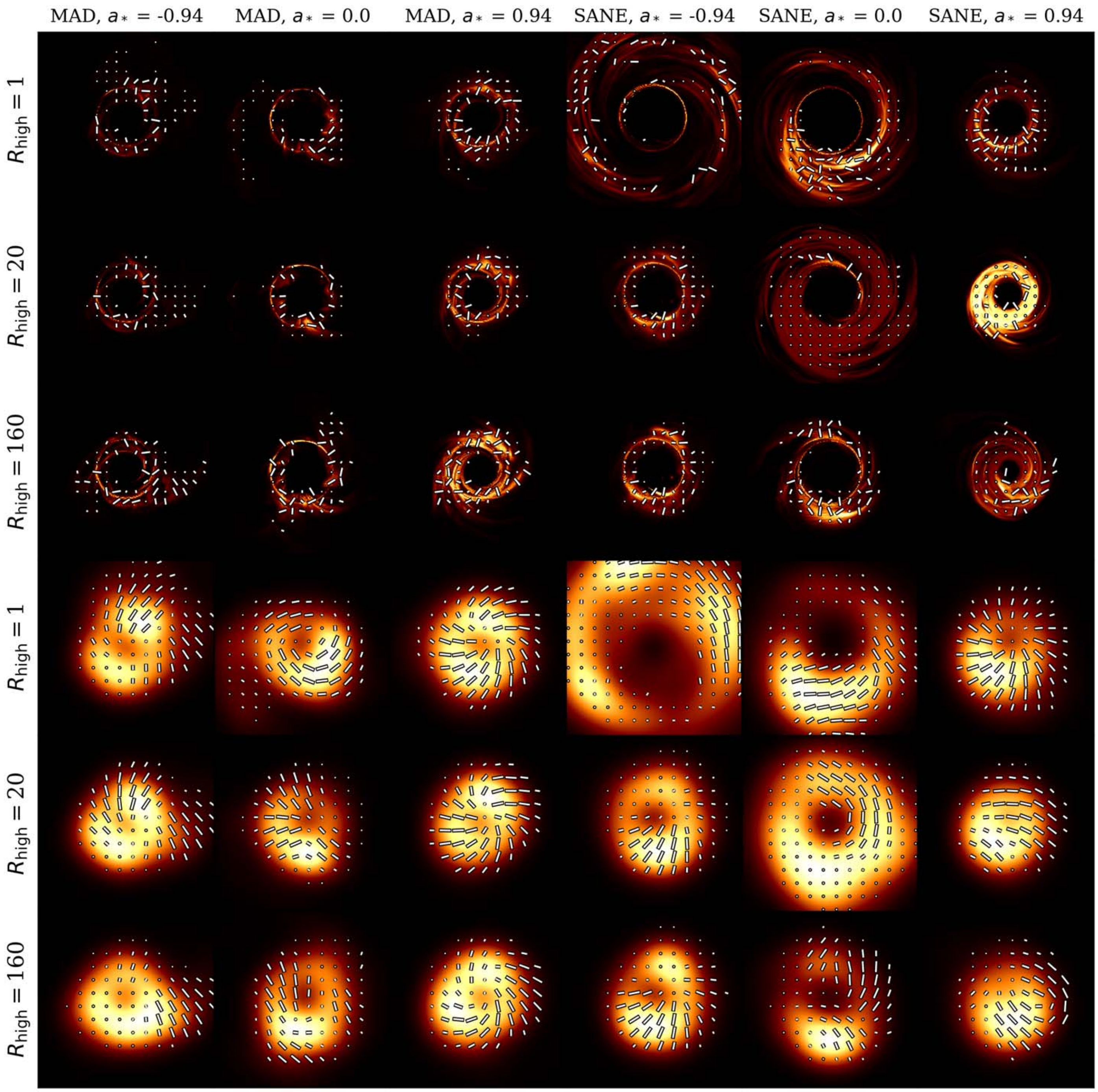

**Figure 5.** Same as in Figure 4 but for $R_{\rm low} = 10$. We find similar trends, but with more scrambling from larger Faraday depths due to lower electron temperatures.

Distributions of all observables are shown in Figure 7 ($\langle|m|\rangle$, left panel), Figure 8 ($|m|_{\rm net}$ and $|v|_{\rm net}$), and Figure 9 ($|\beta_2|$ and $\angle\beta_2$). SANE models tend to have a lower integrated polarization fraction and larger circular polarization fraction than M87* at 230 GHz. In many cases this is a result of very large Faraday rotation internal to the emission region. MAD models tend to have larger net linear polarization fraction than observed in M87*. The resolved average fractional polarization produces similar trends. Most SANE models with prograde spin are too scrambled and most MAD models are too ordered compared to the reconstructed polarization maps of M87*. Full distributions for all models, including their $R_{\rm high}$, $R_{\rm low}$, and $a_*$ dependence, are further discussed in Appendix C.

## 5. Model Evaluation

### 5.1. Model Constraints from Polarimetry

To evaluate whether a given GRMHD model is consistent with the EHT observations reported in EHTC VII, we require images from the model to satisfy constraints on the four parameters derived from the reconstructed EHT images and ALMA-only measurements presented in Table 2 and summarized again here.

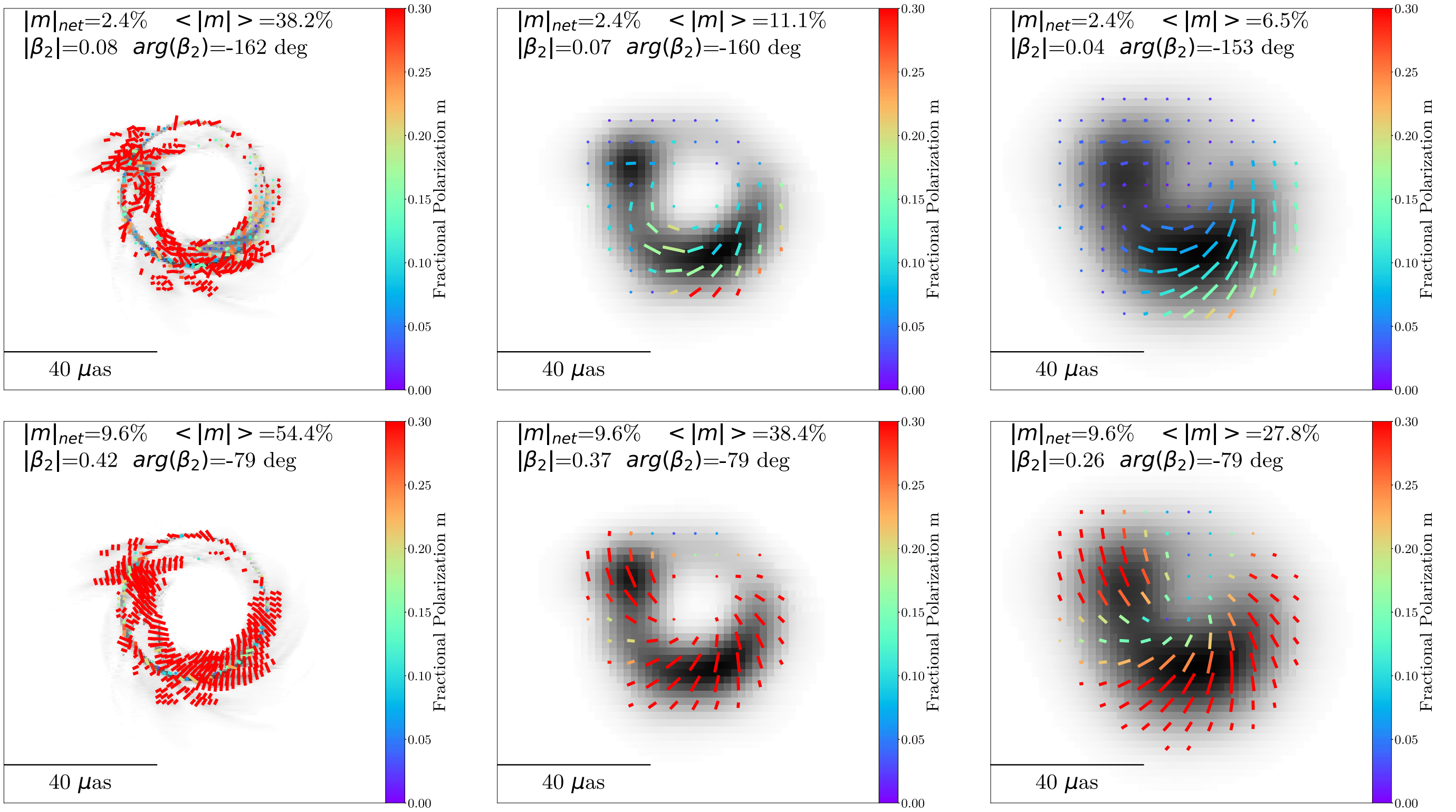

**Figure 6.** Top-left panel: a sample image library polarization map at original resolution, taken from the MAD $a_* = 0.5$ ($R_{\rm low} = 10$, $R_{\rm high} = 80$) model. Top-middle and top-right panels: the same map but convolved with a 10 $\mu$as and 20 $\mu$as FWHM circular Gaussian beam, respectively. The position angle (PA) of the black hole spin in all frames is PA = +90 deg and the inclination angle is $i = 158$ deg, meaning that the black hole spin points left and away from the observer. The bottom row shows the same model but calculated with $\rho_V = 0$ (no Faraday rotation). When Faraday rotation is excluded, the EVPA pattern is more coherent, resulting in much larger values of $|m|_{\rm net}$ and $\langle|m|\rangle$. There is also a net rotation of the EVPA pattern between the two cases, by $\simeq$80 deg in the phase of $\beta_2$.

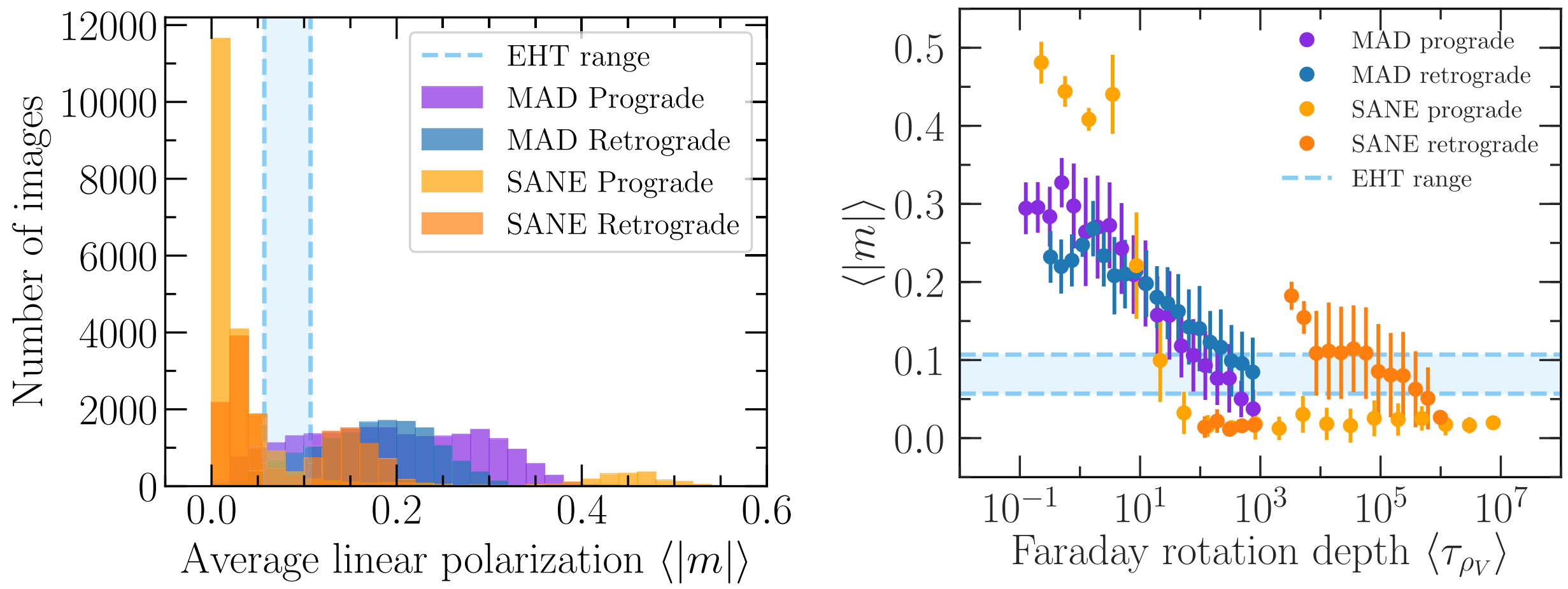

**Figure 7.** Left panel: distribution of image-averaged fractional polarization $\langle|m|\rangle$ over the M87* library images blurred with a 20 $\mu$as beam. The measured range from reconstructed polarimetric images of M87* is shown in dashed lines. Right panel: $\langle|m|\rangle$ as a function of the intensity-weighted Faraday depth across each image for library images blurred with the same 20 $\mu$as circular Gaussian beam. The Faraday depth is calculated as the intensity-weighted sum of $|\rho_V|$ integrated along each ray. For clarity, we show the median (points) and standard deviations (error bars) of the full distributions. The Faraday depth increases monotonically with increasing $R_{\rm high}$ for fixed values of the other parameters. A large Faraday depth corresponds to scrambling of the polarization map, which decreases the coherence length of the EVPA (Jiménez-Rosales & Dexter 2018). Increased scrambling results in stronger depolarization at the scale of the EHT beam and lower values of $\langle|m|\rangle$.

1. The image-integrated net linear polarization $|m|_{\rm net}$ is in the measured range from the EHT image reconstructions: $1\% \leqslant |m|_{\rm net} \leqslant 3.7\%$.
2. The image-integrated net circular polarization $|v|_{\rm net}$ satisfies an upper limit from ALMA-only measurements reported in Goddi et al. (2021): $|v|_{\rm net} \leqslant 0.8\%$.
3. The image-averaged linear polarization $\langle|m|\rangle$ is in the measured range from the EHT image reconstructions at 20 $\mu$as scale resolution: $5.7\% \leqslant |m|_{\rm net} \leqslant 10.7\%$.

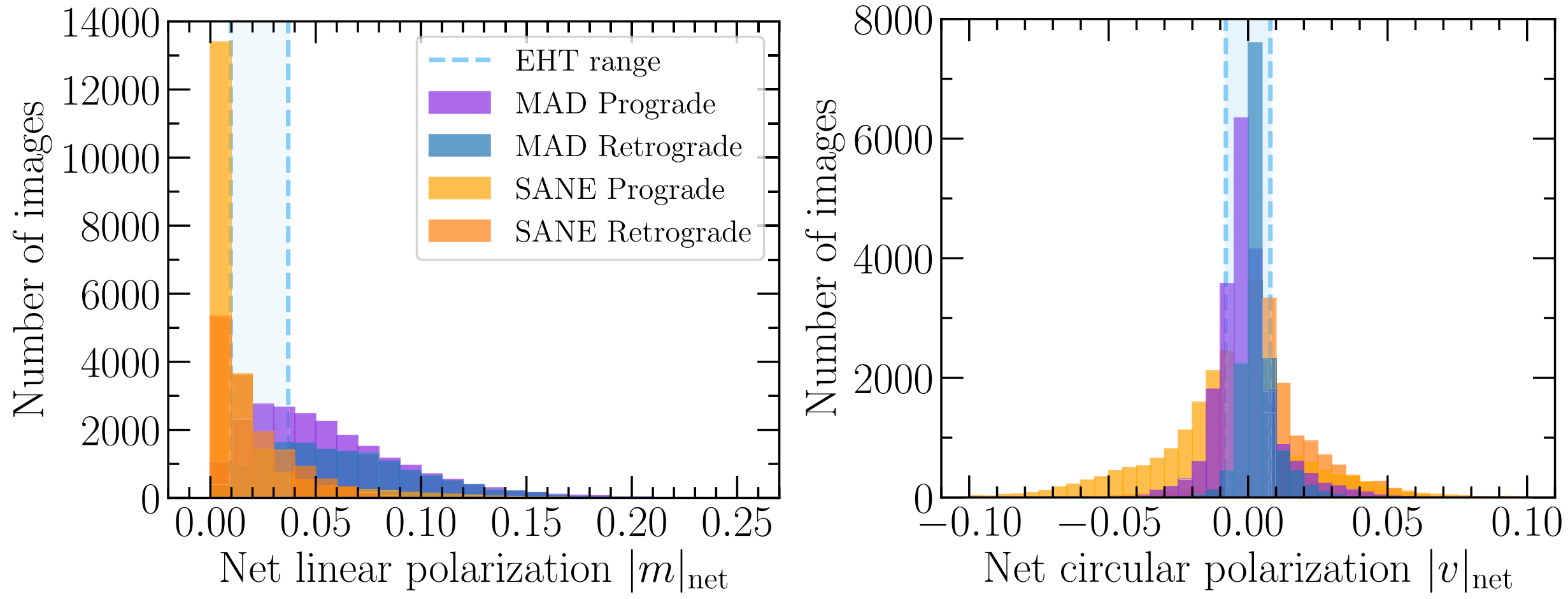

**Figure 8.** Distributions of image-integrated net linear (left panel) and circular (right panel) polarization fractions for all EHT M87* library images. The dashed lines show the allowed range inferred from EHT image reconstructions (for $|m|_{\rm net}$) and ALMA-only data (for $|v|_{\rm net}$).

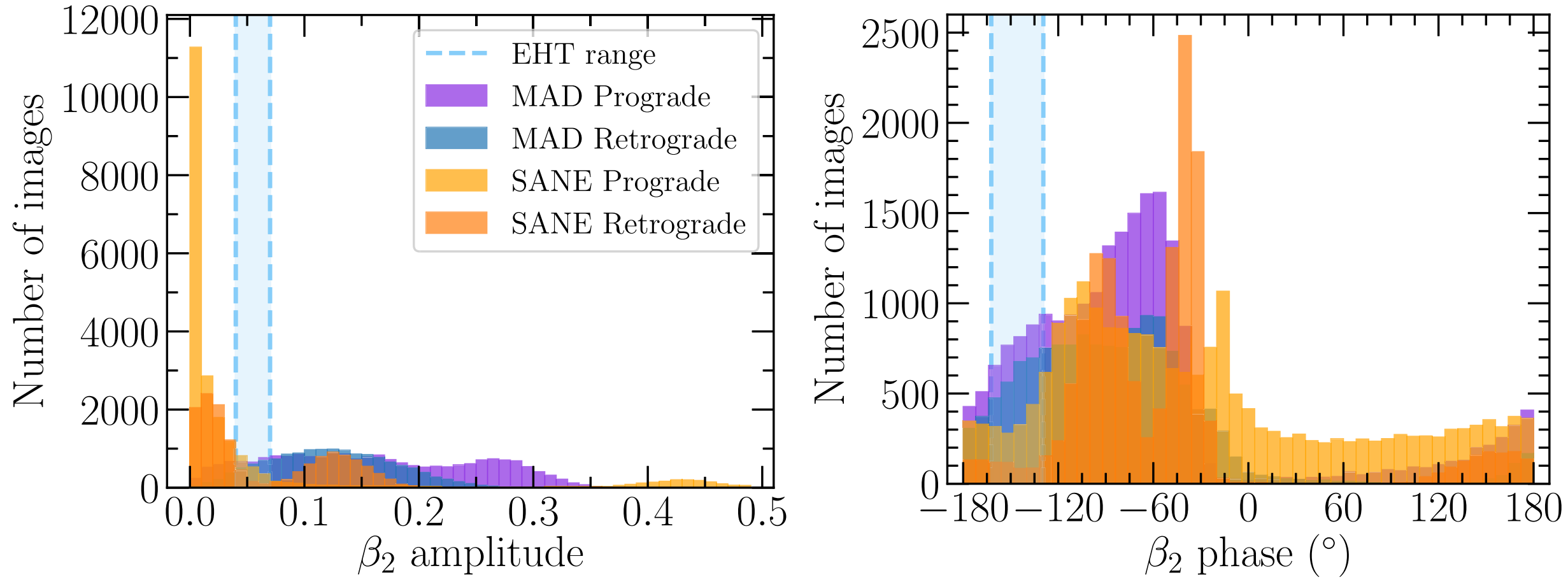

**Figure 9.** Distributions of $\beta_2$ amplitude (left panel) and phase (right panel) for EHT M87* library images blurred with a 20 $\mu$as beam. The measured ranges from reconstructed images of M87* are shown as dashed lines.

4. The amplitude and phase of the complex $\beta_2$ coefficient quantifying coherent azimuthal structure are in the measured range: $0.04 \leqslant |\beta_2| \leqslant 0.07$ and $-163$ deg $\leqslant \arg[\beta_2] \leqslant -129$ deg.

We use 72,000 library images (from Section 4) with 200 time snapshots per model at three inclination angles, six values of $R_{\rm high} = 1$, 10, 20, 40, 80, 160, two values of $R_{\rm low} = 1$, 10, five values of $a_* = -0.9375$, $-0.5$, 0, $+0.5$, $+0.9375$, and realized with both MAD and SANE magnetic field configurations.

In comparing models to observables, the $\beta_2$ metric is the most constraining. Only 790 snapshot images out of the 72,000 considered fall in the range of those reconstructed in both $\beta_2$ amplitude and phase, compared to 11,526 snapshots for both $|m|_{\rm net}$ and $|v|_{\rm net}$ and 7,727 for the resolved image-average linear polarization fraction $\langle |m| \rangle$.

Below we explore two quantitative methods for scoring models, either by requiring that at least one single snapshot image from a model simultaneously passes all constraints ("simultaneous scoring;" see Section 5.2) or that each observational constraint is satisfied by at least one snapshot image from a given model ("joint scoring;" see Section 5.3).

### *5.2. Simultaneous Snapshot Model Scoring*

In the simultaneous scoring procedure, we rule out models where none of the 600 snapshot images (200 time samples at three inclination angles) can simultaneously satisfy the constraints on all of the polarimetric observables. Only 73 out of 72,000 snapshot images across 15 of 120 models simultaneously pass all of the constraints. Of those, all but two viable snapshot images come from a MAD model. The only models with more than five passing images are MAD $a_* = 0$ $R_{\rm low} = 1$ $R_{\rm high} = 160$ and MAD $a_* = -0.5$ $R_{\rm low} = 1$ $R_{\rm high} =$ 80, 160.

Figure 10 shows three viable snapshot images from both SANE and MAD models as well as three snapshot images from models ruled out by simultaneous scoring (i.e., with no snapshots in the entire sample from the model simultaneously satisfying all constraints). These images are representative of the snapshots that simultaneously satisfy all constraints on the image-integrated metrics; they have not been selected based on detailed matching of the resolved polarization structure to the EHT images. Nonetheless, the top row of images show good qualitative agreement with the primary features of the EHT image in Figure 1. In contrast, the snapshots from the ruled-out models tend to be too polarized, too depolarized, or too radial in their EVPA pattern. Figure 11 shows the distributions of $|\beta_2|$ for all 600 snapshots from these three passing and three failing models. Although variability is present, the systematic differences over the five observables considered allow us to constrain the models. The left panel of Figure 12 shows the total number of images that pass simultaneous scoring as a function of model, summing over the six $R_{\rm high}$ values.

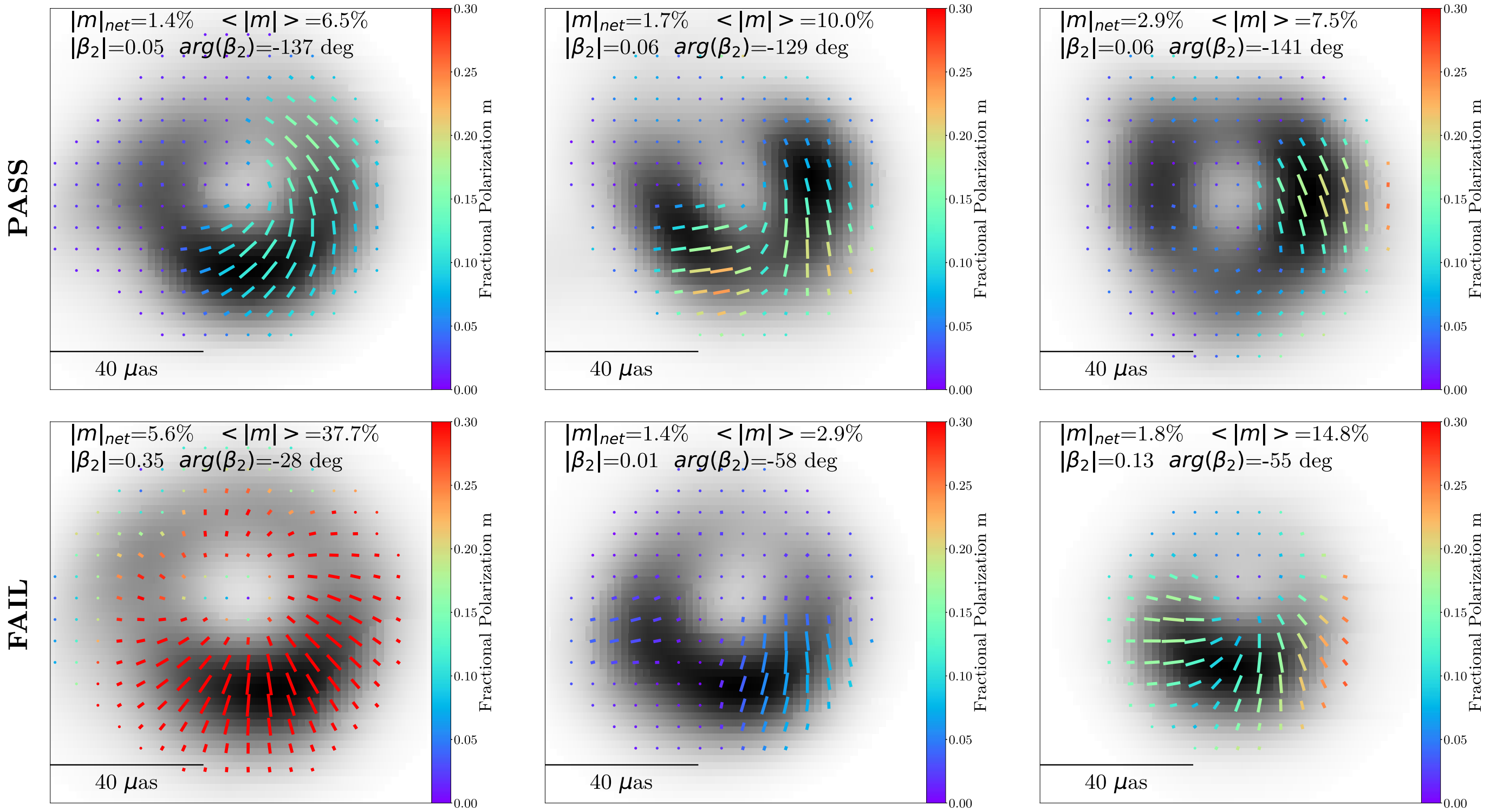

**Figure 10.** Sample 230 GHz image library polarization maps shown in the same style as the EHT image in Figure 1. All maps are blurred with a 20 $\mu$as circular Gaussian beam. In all images, the inclination angle is either 17 deg (for negative $a_*$ models) or 163 deg (for positive $a_*$ model), with the black hole spin vector pointing to the left and away from the observer. The top row displays SANE ($a_* = 0$, $R_{\rm high} = 80$) and two MAD snapshots (both $a_* = -0.5$ and $R_{\rm high} = 160$) from left to right. All of the top row images satisfy simultaneous constraints on image-integrated metrics ($|m|_{\rm net}$, $|v|_{\rm net}$, $\langle|m|\rangle$, $|\beta_2|$, $\angle\beta_2$) derived from the EHT2017 results. The bottom row displays snapshots from models that fail to produce any images that simultaneously satisfy the observational constraints. These snapshots are from two SANE ($a_* = 0.5$ and $R_{\rm high} = 1$ and 160) and one MAD ($a_* = 0.94$, $R_{\rm high} = 160$) model, from left to right. The failing images are either more polarized than the data (left and right panels) or too depolarized (middle panel). They also fail to match the observed EVPA pattern ($\beta_2$ phase).

### 5.3. Joint Distribution Model Scoring

In the joint scoring procedure, we use the measured distributions of the data metrics to ask whether the observed value of each metric for M87* is consistent with being drawn from the distribution seen in the GRMHD simulations. To do this, we measure $\chi^2$ values for the five metrics $x_j \in \{|m|_{\rm net}, |v|_{\rm net}, \langle|m|\rangle, |\beta_2|, \angle\beta_2\}$ for all snapshots $k$ from a given model as

$$\chi^2_{j,k} = \frac{(x_{j,k} - \bar{x}_j)^2}{\sigma_j^2}, \tag{17}$$

where $x_{j,k}$ are the values of a scoring metric $x_j$ for each of the 600 snapshots $k$ from a given model, $\bar{x}_j$ is the mean of those values for the model, and $\sigma_j$ is taken as one half of the *observed* data range from Table 2. Note that the scoring results of this method do not depend on the choice of $\sigma_j$. We then calculate an analogous $\chi^2_{j,{\rm data}}$ value for the midpoint of the measured range from Table 2. A likelihood value $\mathcal{L}_j$ of the data being drawn from the model distribution is defined as the fraction of images with $\chi^2_{j,k} > \chi^2_{j,{\rm data}}$. The joint likelihood of each model is the product $\mathcal{L} = \Pi_j \mathcal{L}_j$ of those for the five metrics $x_j$.

To produce a non-zero likelihood $\mathcal{L}$ in this method, at least one snapshot from a model must lie further from its mean than the data value does. That can be a different snapshot for each metric, which makes this method more lenient than the simultaneous scoring method. We also note that snapshots are allowed to have the *wrong sign* of the difference with their mean, due to the definition of $\chi^2$ and our use of the mean of the model snapshots themselves. In practice, this makes little difference in the results.

In this method, we consider models viable whose joint likelihood is >1% of the maximum found from any model. The right panel of Figure 12 shows the resulting joint likelihoods summed over $R_{\rm high}$.

### 5.4. Comparison of Scoring Results

The results of both scoring procedures are summarized in Figure 12, summed over $R_{\rm high}$. Both scoring methods prefer MAD models to SANE models, with most of the passing models coming from the MAD $a_* = 0$ and $a_* = \pm 0.5$ simulations.

The main difference between the two scoring procedures is that joint scoring prefers $R_{\rm low} = 10$ models, while $R_{\rm low} = 1$ is preferred by simultaneous scoring. SANE models with $a_* = 0.94$, $R_{\rm low} = 1$, 10, and $R_{\rm high} = 10$ are ruled out by simultaneous scoring, but score fairly well in joint scoring. For the favored MAD models, when $R_{\rm low} = 1$, there are more images that simultaneously satisfy all constraints, but when $R_{\rm low} = 10$, the distributions generally stay closer to the observed data ranges and are thus favored by the joint scoring method. Due to differences between simultaneous and joint scoring results, as well as other methods we have tried, we consider the inferred parameters of $R_{\rm low}$, $R_{\rm high}$, and $a_*$ from passing models to be less robust than the overall trend that MAD models are favored.

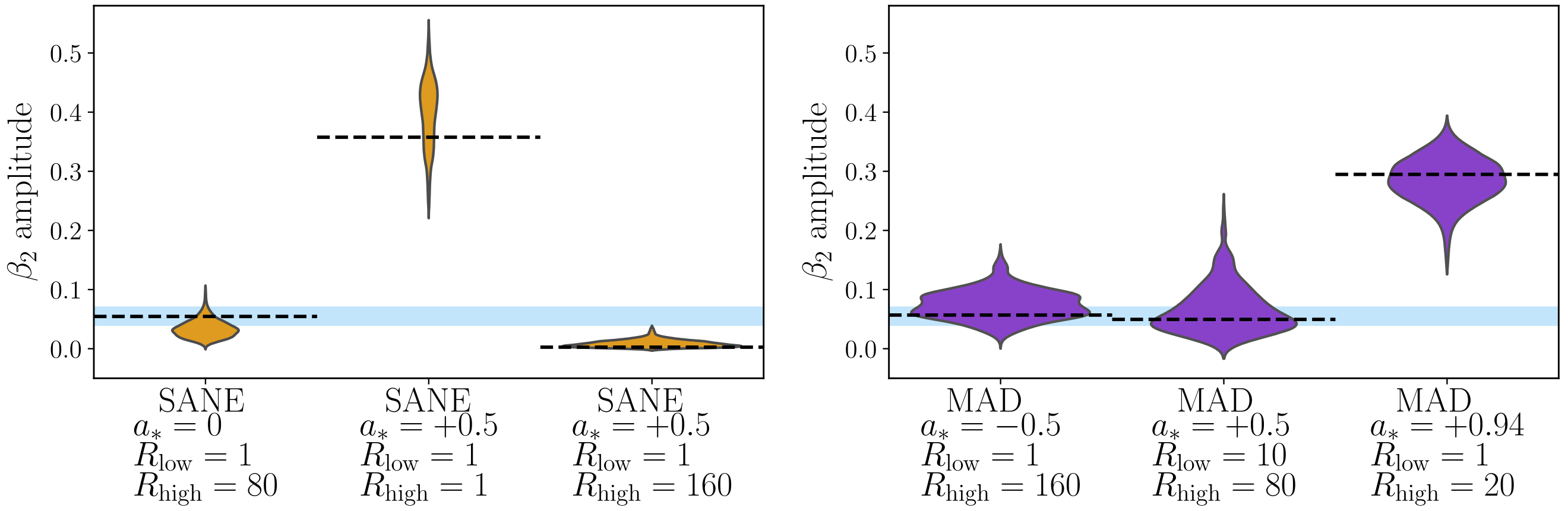

**Figure 11.** Distributions of $|\beta_2|$ for the sample passing and failing models in Figure 10. The dashed black lines mark the measured values for the snapshot images in Figure 10, and the blue bands show the range inferred from EHT M87* data. The models can be constrained using EHT observables even in the presence of significant scatter due to time variability.

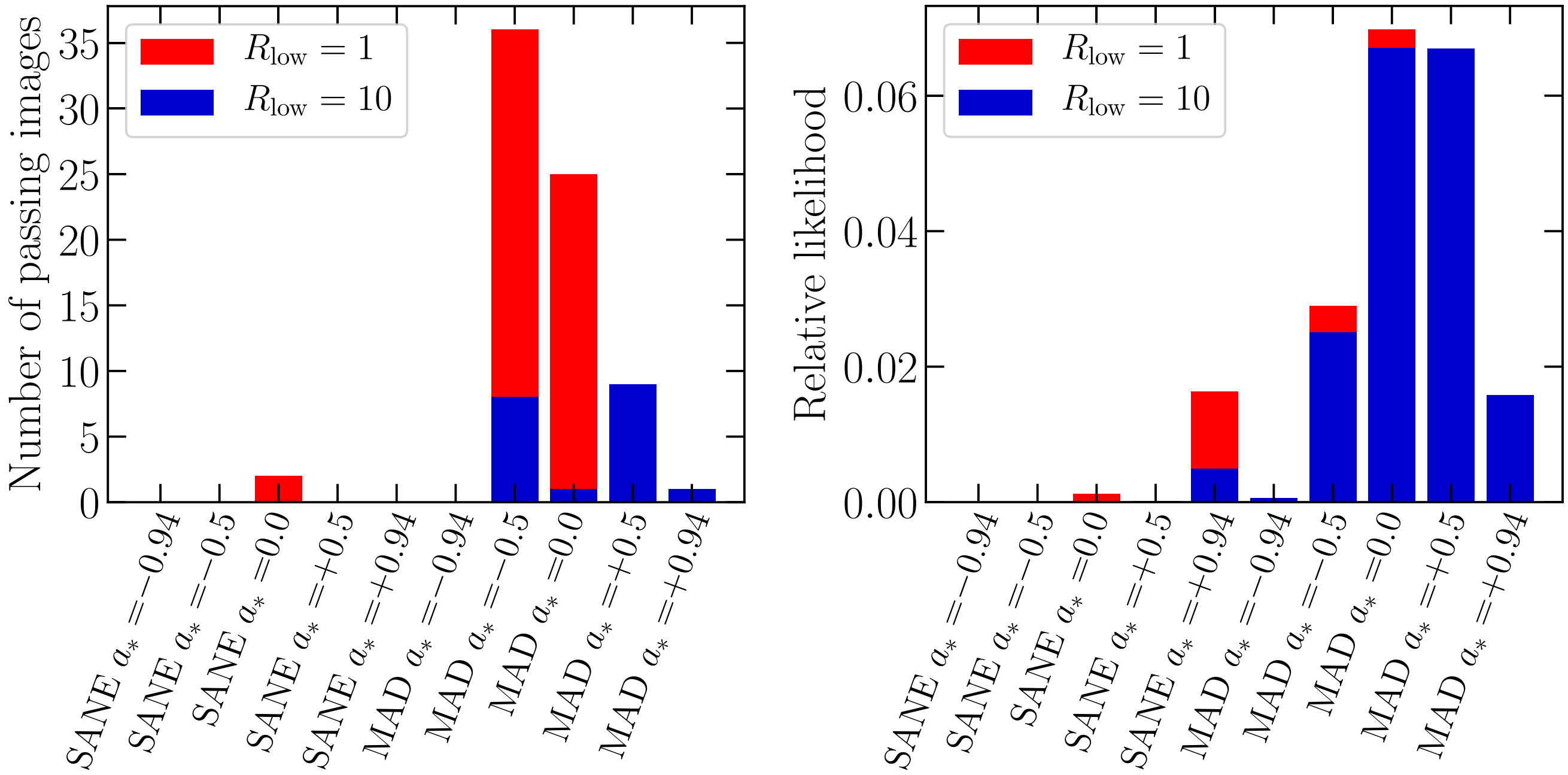

**Figure 12.** Results of the simultaneous (left panel) and joint (right panel) scoring methods for comparing GRMHD models to M87* observables. The simultaneous scoring method shows the total number of viable images for each image library model after summing over $R_{\rm high}$. Out of a total of 73 passing images, only two are from a SANE model. The right panel shows the joint likelihood of each library model after summing over $R_{\rm high}$. In this method, $R_{\rm low} = 10$ MAD models are preferred and SANE $a_* = +0.94$ $R_{\rm high} = 10$ models are also allowed.

The simultaneous scoring method has the advantage of conceptual simplicity, and of applying each constraint simultaneously per image (thus accounting for correlations between the scoring metrics). Simultaneous scoring is more strict and rules out more models than joint scoring, but it may be more limited by the finite number of images generated per model. The joint scoring procedure has the advantage of being more conservative in disfavoring models, but assumes the observational constraints are independent in calculating a joint likelihood. Instead, they are correlated (in particular $|m|_{\rm net}$, $\langle |m| \rangle$, and $|\beta_2|$).

The number of images in each model that pass each constraint individually (used in joint scoring) and that simultaneously pass all constraints (used in simultaneous scoring) are presented in Appendix D.

### 5.5. Combined EHTC V and Current Polarimetric Constraints

EHTC V presented constraints on the GRMHD simulation models based on fits to the EHT total intensity data, model self-consistency (requiring a radiative efficiency less than that of a thin accretion disk at the same black hole spin), and M87's measured jet power (requiring a simulation to produce a jet power consistent with a conservative lower limit of that from M87*, $>10^{42}$ erg s$^{-1}$). Those constraints ruled out MAD $a_* = -0.94$ models (from failing to satisfy the EHT image morphology), SANE models with $a_* = -0.5$, and all models with $a_* = 0$ (from failing to produce enough jet power). Here we retain only the jet power lower limit, which is the most constraining and straightforward to apply to the expanded image library considered in this work.

Relativistic jets launched in GRMHD simulations (defined here as in EHTC V, with a cutoff of $\beta\gamma > 1$) are fully consistent with being produced via the Blandford–Znajek process (e.g., McKinney & Gammie 2004; McKinney 2006). As a result, $a_* = 0$ models have small or zero jet power, $P_{\rm jet}$, and are rejected by this constraint. These models can still produce significant total outflow powers ($P_{\rm out}$ in EHTC V) in a mildly relativistic jet or wind. Many other models with low values of $R_{\rm high}$ or moderate black hole spin, including those of SANEs with $a_* = +0.94$, which are allowed by the joint scoring procedure, are also ruled out by the jet power constraint (see Table 3 in Appendix D and EHTC V). Combining the simultaneous scoring polarimetric constraints with the jet power constraint results in 15 remaining viable models: all MADs, and spanning the full range of non-zero $a_*$ explored. This conclusion does not depend on the choice of the simultaneous or joint model-scoring procedure.

## 6. Discussion

The resolved EHT 2017 linear polarization map of M87* shows a predominantly azimuthal linear polarization (EVPA) pattern, and relatively low fractional polarization of $\lesssim$20% on 20 $\mu$as scales. We interpret the low fractional polarization as the result of Faraday rotation internal to the emission region, which acts to rotate, scramble, and depolarize the resolved polarized emission. Adopting this constraint in a one-zone model, we estimate typical values of particle density $n_e$, magnetic field strength $B$, and electron temperature $T_{\rm e}$. In semi-analytic emission models with externally imposed, idealized magnetic field configurations, azimuthally dominated EVPA patterns are produced by poloidal (radial and/or vertical) magnetic field components. To fully capture the complicated combined effects from magnetic field structure, turbulence, relativity, and Faraday rotation on polarimetric images of M87*, we turn to radiative transfer calculations from GRMHD simulations.

We compared a large image library of emission models from GRMHD simulations with metrics designed to capture these salient features of the data. The combined constraints of a predominantly azimuthal EVPA pattern and a low but non-zero fractional polarization are inconsistent with most SANE GRMHD models with weaker horizon-scale magnetic fields. Some MAD models with relatively cold electrons, realized in our library by larger values of $R_{\rm high}$ and/or $R_{\rm low}$, remain consistent with the data. Here we discuss the implications of our results, and limitations in our set of theoretical models that may impact our interpretation.

### 6.1. Near-horizon Plasma and Magnetic Field Properties in Passing Models

Both our one-zone and GRMHD models find similar plasma conditions in the 230 GHz emission region, driven by the requirements of weak 230 GHz absorption and strong 230 GHz Faraday rotation. In viable GRMHD models, we find average, intensity-weighted plasma properties in the emission region of $n_e \sim 10^{4-5}$ cm$^{-3}$, $B \simeq 7$–30 G, and $\theta_e \sim 8$–60. These are in good agreement with our one-zone estimates (Section 3.1). We have also calculated the intensity-weighted values of the absorption and Faraday optical depth, $\tau_I$ and $\tau_{\rho_V}$, over snapshots that simultaneously satisfy all our observational constraints. The median values are $\tau_I \simeq 0.1$ and $\tau_{\rho_V} \simeq 50$. All of our viable images have $\tau_{\rho_V} > 2\pi$, while 2 out of 73 have $\tau_I \gtrsim 1$, consistent with our assumptions in Section 3.1 that the plasma Faraday depth is large while the Stokes $\mathcal{I}$ optical depth is small.

By quantitatively evaluating a large library of images based on GRMHD models (Section 5), we identify 25 out of 120 models that remain viable after applying constraints based only on EHT and ALMA-only polarimetric observations. Additionally applying a cut on jet power of $P_{\rm jet} > 10^{42}$ erg s$^{-1}$ (EHTC V) rules out the five viable SANE models and all $a_* = 0$ models. The precise number and identity of the viable models depends mildly on the chosen scoring procedure and on the Gaussian blurring kernel size applied to the EHT image reconstructions and library simulated images. The overall preference for MAD over SANE models is found from both the simultaneous and joint scoring procedures, as well as other variants. After applying the jet power constraint, no viable SANE models remain for any of the scoring methods that we explored.

MAD models are associated with dynamically important magnetic fields. The significant poloidal components of those fields can produce a predominantly azimuthal polarization pattern (Figure 4), similar to those seen in idealized models with prescribed poloidal magnetic fields (Figure 3). Strong Faraday effects complicate a direct interpretation of the observed EHT polarization map in terms of those idealized models. Still, our more detailed comparison favoring MADs suggests the presence of dynamically important magnetic fields in the emission region on event-horizon scales.

In Figure 13 we present mass accretion rate and jet power distributions both for the viable models identified in EHTC V and when adopting the new constraints from polarimetry.[131] Polarimetric constraints break degeneracies present in the single epoch total intensity data, allowing us to estimate a mass accretion rate onto the black hole of $\dot{M} \simeq (3-20) \times 10^{-4} M_\odot$ yr$^{-1}$. This corresponds to $\dot{m} = \dot{M}/\dot{M}_{\rm Edd} \simeq (2$–$15) \times 10^{-6}$, where $\dot{M}_{\rm Edd}$ is the Eddington accretion rate.[132] The measured radiative efficiency $\epsilon = L/\dot{M}c^2$ (where $L$ is the bolometric luminosity) of the passing models is relatively high for a hot accretion flow model: $\epsilon \lesssim 1$%. These models have jet powers of $P_{\rm jet} \simeq 10^{42-43}$ erg s$^{-1}$.

The mass accretion rate found here is much lower than the Bondi rate calculated from Chandra observations

[131] Note differences in some $\dot{M}$ and $P_{\rm jet}$ values compared to EHTC V. We have corrected minor tabulation errors from that work, and have used a slightly different time range for averaging the MAD $a_* = -0.94$ simulation.

[132] The Eddington rate is defined as $\dot{M}_{\rm Edd} = L_{\rm Edd}/\epsilon_{\rm Edd} c^2$, where $L_{\rm Edd} = 4\pi G M m_p c/\sigma_T$ is the Eddington luminosity and we adopt an efficiency factor $\epsilon_{\rm Edd} = 0.1$. Note that this assumed efficiency factor $\epsilon_{\rm Edd}$ is distinct from the reported radiative efficiency $\epsilon = L/\dot{M}c^2$ measured from the simulations.

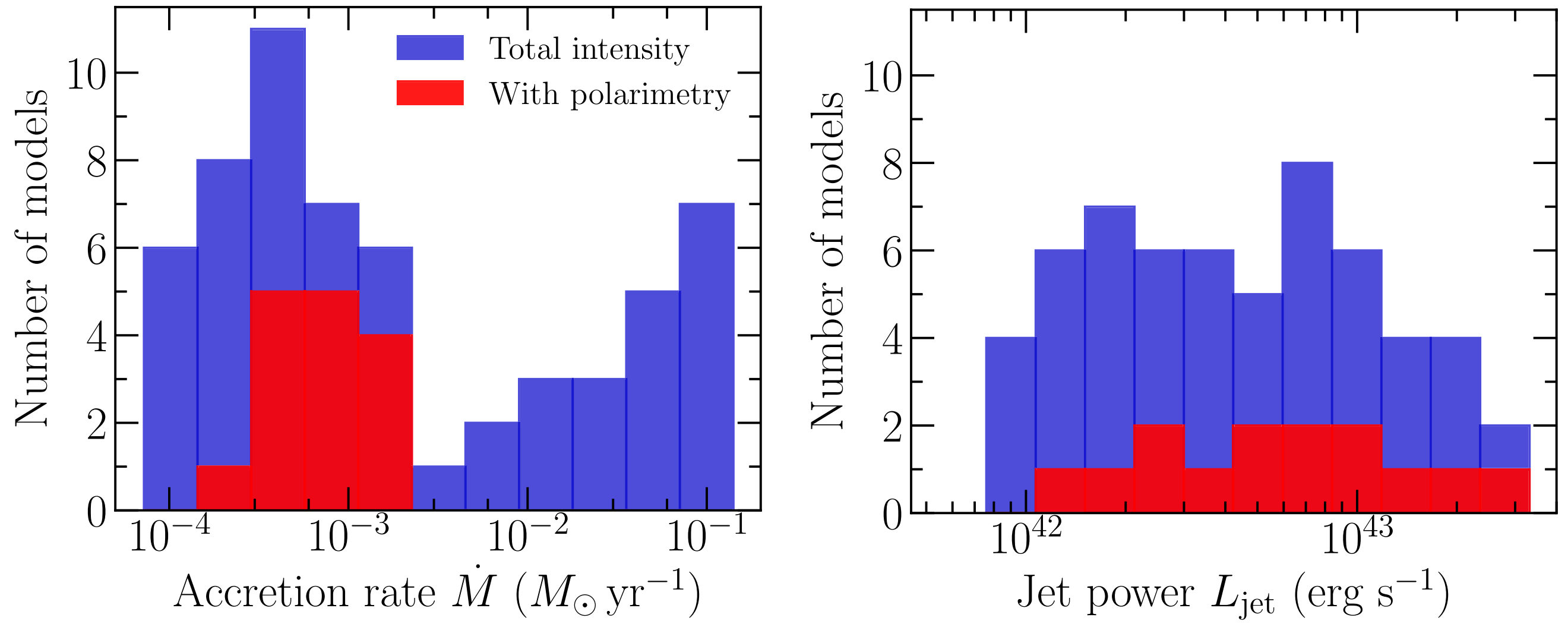

**Figure 13.** Average mass accretion rate (left panel) and jet power (right panel) for viable GRMHD models of M87* identified by selecting on total intensity data and jet power (blue, EHTC V), and when including polarimetric constraints (red). We estimate a mass accretion rate of $\dot{M} \simeq (3-20) \times 10^{-4} M_\odot$ yr$^{-1}$, resulting in a radiative efficiency $\epsilon \lesssim 1\%$ (see EHTC V). The jet powers produced by our models are $\sim 10^{42} - 10^{43}$ erg s$^{-1}$, and the jet efficiencies are $\simeq$5%–80%. The mass accretion rate is better constrained when including polarimetric information.

(Di Matteo et al. 2003; see also Russell et al. 2015), and higher than that found from hybrid disk+jet models of the M87* spectral energy distribution (SED; Prieto et al. 2016). Our inferred jet powers of $\lesssim 10^{43}$ erg s$^{-1}$ are toward the lower end of the observed range. In particular, the jet power measured at the location of Hubble Space Telescope (HST)-1 is $\sim 10^{43-44}$ erg s$^{-1}$ (Stawarz et al. 2006), and Low-Frequency Array (LOFAR) observations suggest that a jet power of $\sim 10^{44}$ erg s$^{-1}$ was necessary within the last ~million years to inflate the observed radio lobes on scales of ~80 kpc (de Gasperin et al. 2012).

Measurements of the accretion rate and the radiative efficiency can begin to constrain the microphysical plasma processes that heat electrons in M87*, for example by inferring the fraction of the dissipated energy in the system that heats electrons, $\delta_{\rm e}$. In axisymmetric, self-similar, hot accretion flow models, a system with $\dot{M} \sim 10^{-5} \dot{M}_{\rm Edd}$ and a radiative efficiency $\epsilon \lesssim 1\%$ has a value of $\delta_{\rm e}$ in the range 0.1–0.5 (see Figure 2 of Yuan & Narayan 2014). This range is consistent with that produced by simulations of turbulence and reconnection in the $\beta \sim 1$ regime (e.g., Rowan et al. 2017; Werner et al. 2018; Kawazura et al. 2019). Future studies using simulations with self-consistent electron heating and radiative cooling (Section 6.3) can better constrain $\delta_e$ and its dependence on local plasma parameters throughout the accretion flow and jet-launching region.

We have assumed that all effects responsible for the appearance of the EHT polarized image of M87* are captured within the relatively small GRMHD simulation spatial domain, $\lesssim 10^{2-3}\ r_g$. Goddi et al. (2021) developed a two-component model for the ALMA and image-integrated EHT data where each component is Faraday rotated by a different screen. The model demonstrates that the rotation measure of the compact component is unconstrained by the ALMA measurements alone, as the ALMA measurements are also sensitive to the Faraday rotation properties of the larger-scale component. In addition, the observed time variability in ALMA data (e.g., the RM sign change) can be explained by the observed EVPA variation of the compact core seen by the EHT. To produce the observed variability requires an RM of $\approx -6 \times 10^5$ rad m$^{-2}$. The ALMA data do not constrain the location or nature of this Faraday screen, except that it must be relatively close to the compact core, $r \lesssim 10^5\ r_g$.

For our favored plasma parameters for M87*, we expect substantial Faraday RM internal to the emission region itself, $\tau_{\rho_V} \gtrsim 2\pi$, consistent with that measured from viable GRMHD images. In a model of uniform, external Faraday rotation this Faraday depth at 230 GHz would correspond to an RM of $\lesssim 10^6$ rad m$^{-2}$. Figure 14 shows that the apparent RMs measured from our GRMHD images span a wide range, often comparable to or larger than that inferred from the Goddi et al. (2021) two-component model ($\lesssim 10^6$ rad m$^{-2}$). For the low inclination angle of M87*, the apparent RM measured from GRMHD images is not a good tracer of the mass accretion rate (Mościbrodzka et al. 2017), and originates close to the emission region and well within the simulation domain (Ricarte et al. 2020, and Appendix B). The RM inferred from low-inclination GRMHD models of M87* can also vary rapidly and change signs (Ricarte et al. 2020), as seen in the ALMA-only data. As a result, the RM inferred from the two-component model in Goddi et al. (2021) is apparently consistent with the intrinsic properties of the GRMHD models studied here, without invoking an additional, external Faraday screen. At the same time, we cannot rule out that such an external screen could be present. Future EHT observations with wider frequency spacing can directly measure the resolved RM of the core and address this uncertainty.

### 6.2. Electron Acceleration

Magnetic reconnection, magnetohydrodynamic (MHD) turbulence and collective plasma modes in collisionless hot accretion flows likely result in nonthermal particle acceleration. While pioneering attempts have been made (e.g., Ball et al. 2016; Chael et al. 2017; Davelaar et al. 2019), it is not yet known how to properly incorporate electron acceleration in global GRMHD simulations of hot accretion flows.

We adopt an empirical approach to investigate the impact of nonthermal (accelerated) electrons on 230 GHz polarimetric images of M87* to quantify whether and how neglecting particle acceleration in our models affects our conclusions. We use a specific, but widely explored (e.g., Özel et al. 2000; Markoff et al. 2001; Yuan et al. 2003), description for electron acceleration, namely that accelerated electrons add a power-law tail to the thermal distribution function. The power-law tail is described by the fraction of the thermal energy density in the power-law tail, $\eta$, the power-law slope, $p$, and the maximum Lorentz factor of the accelerated electrons, $\gamma_{\rm max}$. The minimum Lorentz factor, $\gamma_{\rm min}$, is calculated self-consistently by assuming a continuous transition between the thermal and power-law distribution functions (e.g., Yuan et al. 2003). In this model, the parameters $p$, $\eta$, and $\gamma_{\rm max}$ are constant across the accretion flow. We assume that $\gamma_{\rm max} \to \infty$ and we explore values of $p = 3.5$, 4.5 and $\eta = 0.01$, 0.1.

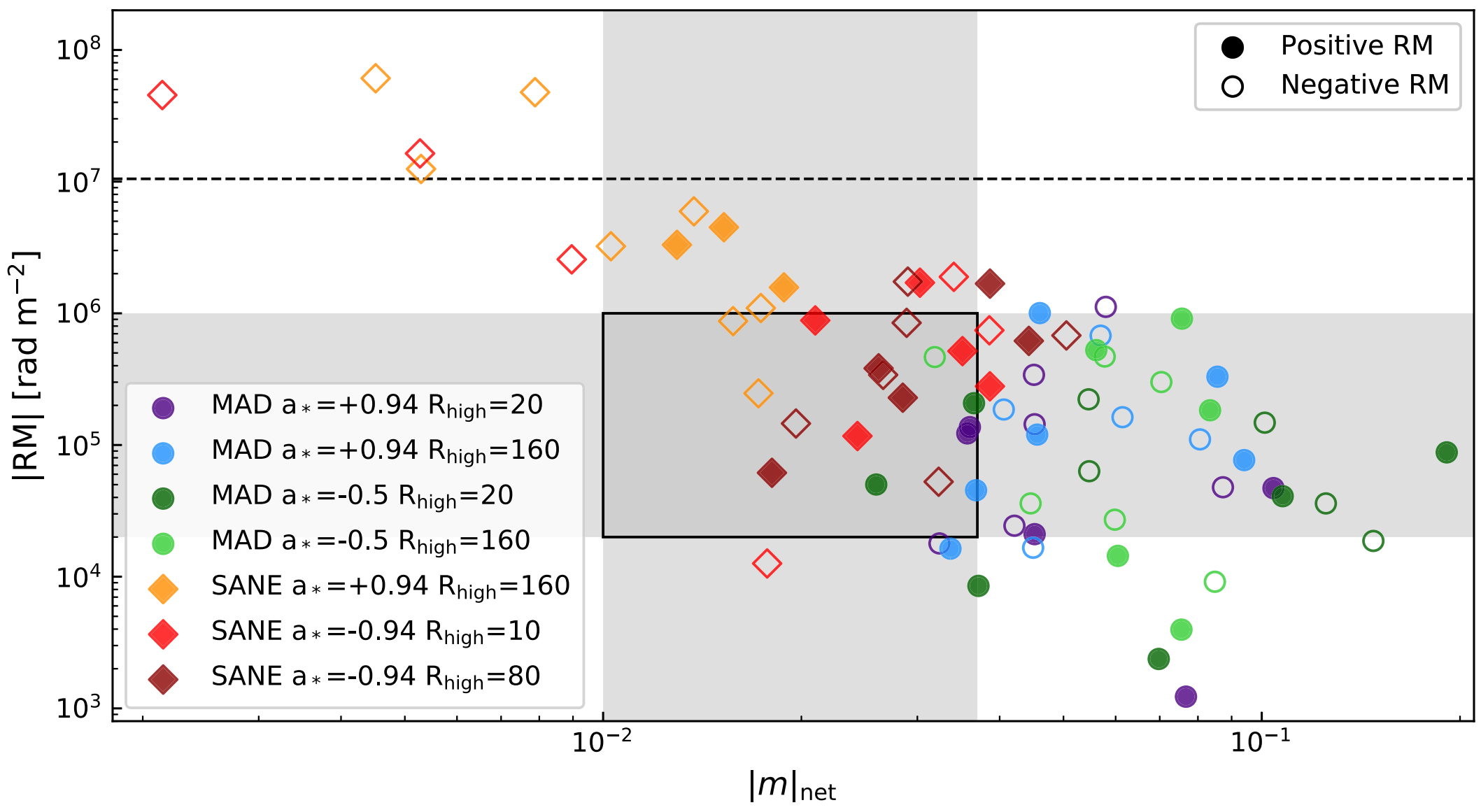

**Figure 14.** Absolute value of RM vs. net linear polarization $|m|_{\rm net}$ for a subset of our EHT GRMHD library models explored in more detail in Ricarte et al. (2020). Closed symbols represent positive RM while open symbols represent negative RM, revealing significant time variability across the 2500 $r_g/c$ spanned by these snapshots. In gray, we plot our allowed region of $|m|_{\rm net}$ and bracket the range of core RM inferred from contemporaneous ALMA-only observations, 2–100 × $10^4$ rad $m^{-2}$ (Goddi et al. 2021). The dashed horizontal line demarcates the RM at which an EVPA rotation by $\pi$ radians would have been observed between the 212 and 230 GHz frequency range used in the ALMA-only measurements, 1.05 × $10^7$ rad $m^{-2}$. Despite large Faraday depths, a large fraction of these snapshots exhibit RMs consistent with simultaneous ALMA-only constraints. RM and $|m|_{\rm net}$ are anti-correlated, as larger Faraday depths lead to greater scrambling of the intrinsic polarization.

In Figure 15 we present linear polarization maps from two MAD models and one SANE model comparing purely thermal and hybrid electron distributions. Using a hybrid distribution function does not affect the structure of the EVPA map ($\beta_2$ phase), but it changes the image-integrated and resolved linear polarization fractions. For example, in the MAD $a_* = -0.5$ (MAD $a_* = 0.5$) model, with the selected hybrid parameters, the $|m|_{\rm net}$, $v_{\rm net}$ and $\langle|m|\rangle$ ranges are 4.3%–4.6% (2.5%–3.8%), 0.25%–0.37% ((−0.5) to (−0.12)%), and 10.6%–11.5% (12%–14%), respectively. Slightly larger deviations from the thermal model are measured in the SANE $a_* = -0.94$ scenario, where the $|m|_{\rm net}$, $v_{\rm net}$ and $\langle|m|\rangle$ ranges are 2.2%–4.1%, −0.004% to 0.31%, and 14%–20%, respectively.

However, fixing the accretion rate to that used in the thermal model results in an increased total intensity flux density when we add high-energy nonthermal electrons to our models. If instead we compare the models at fixed flux density, we need to reduce the mass accretion rate of the hybrid model. Therefore, generalizing the distribution function introduces order unity uncertainties in the inferred mass accretion rate, radiative efficiency, and jet power. The changes in the polarimetric observables in a given snapshot are also larger at fixed flux density. For example, in the MAD $a_* = -0.5$ model, the $|m|_{\rm net}$ increases from 4.7% to 6% when adding nonthermal electrons. As a result, in principle, polarimetric observables constrain the distribution function as well.

Such constraints presumably depend on the details of the assumed particle acceleration scenario. Viable scenarios include hybrid electron distribution functions, or models where particle acceleration is a function of local gas conditions or magnetic energy density rather than fixed throughout the flow (e.g., Ball et al. 2016; Davelaar et al. 2019). More realistic particle acceleration scenarios could be considered using *resistive* GRMHD simulations (e.g., Ripperda et al. 2020).

### 6.3. Coherently Polarized Forward Jet Emission

As discussed above, some SANE retrograde model images in the library show coherently polarized features even when the Faraday depth through the entire emission region is large. The observed polarized flux in those cases originates on the near side of the midplane and is not scrambled from Faraday rotation along the line of sight. A similar effect might be possible if nonthermal electrons could be accelerated efficiently in the low-density, strongly magnetized funnel region in front of the black hole.

It is beyond the scope of this Letter to evaluate whether or how such a model might be realized physically, e.g., whether any process could fill the funnel with high-energy electrons efficiently enough to produce the observed 230 GHz luminosity from the funnel alone. Instead we carry out one sample calculation of polarized emission from the funnel of a prograde $a_* = 0.94$ SANE library snapshot. We assign a nonthermal energy density $u_{\rm nth} = \alpha u_{\rm mag}$ wherever the magnetization $\sigma > 1$, with $\alpha = 0.02$ the fraction of the magnetic energy density $u_{\rm mag}$ that is put into nonthermal particles. We calculate synchrotron radiation from a pure power-law distribution of electrons with $\gamma_{\rm min} = 100$ and $p = 3$.

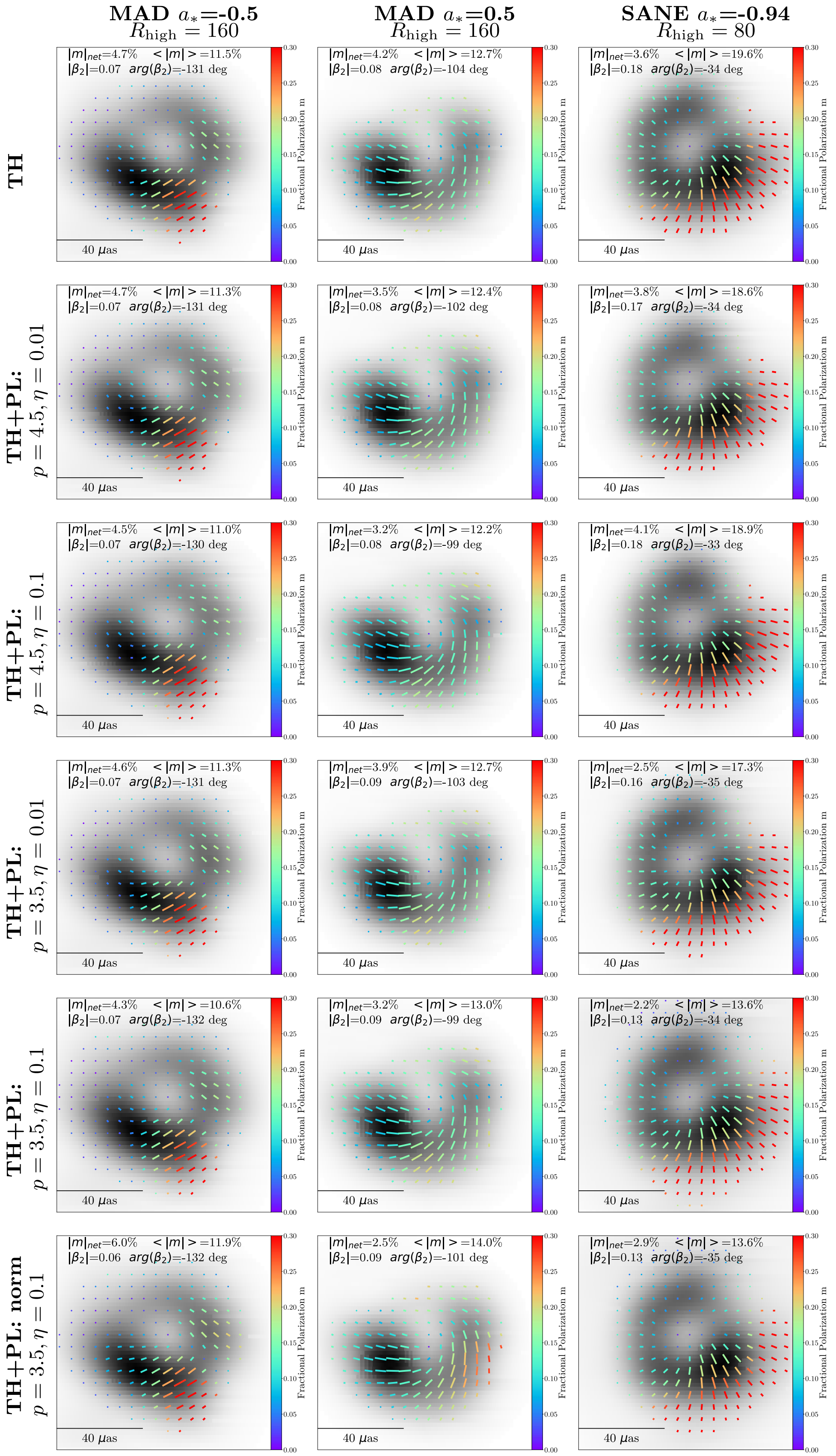

**Figure 15.** Sample polarization maps with varying electron distribution function. Columns display single snapshots from three selected models. Row 1 shows images with a thermal electron distribution function, as assumed in the standard EHT image library. Rows 2 through 5 are the same models but with emission from a hybrid distribution of electrons. Row 6 shows a hybrid model but the mass accretion rate of the model is adjusted to reproduce the same total intensity flux as the purely thermal snapshot. All maps are blurred with a 20 μas circular Gaussian. In all images, $i = 17$ deg (for negative $a_*$ models) or $i = 163$ deg (for positive $a_*$ model), with the black hole spin vector pointing to the left and away from the observer.

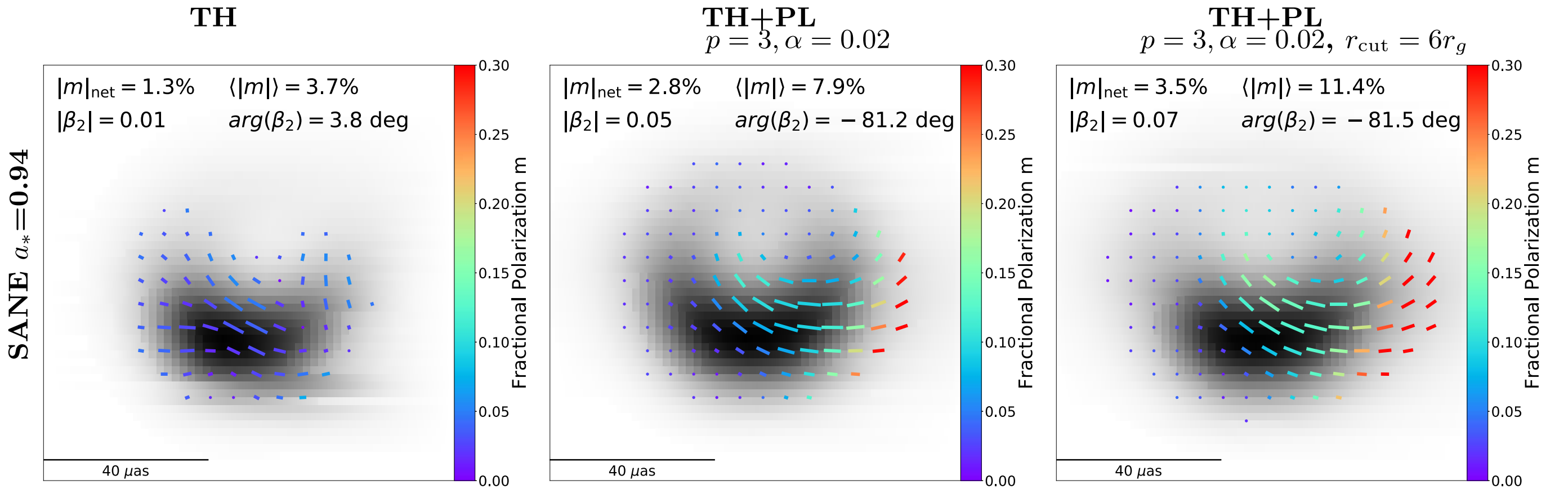

**Figure 16.** Sample library SANE $a_* = 0.94$ $R_{\rm high} = 160$ snapshot (left panel) and with power-law emission from nonthermal electrons added in $\sigma > 1$ regions (in the funnel near the pole, middle panel) and in the funnel only outside of a radius $r_{\rm cut} = 6\ r_g$ (right panel). The library model is heavily depolarized due to Faraday rotation. Nonthermal radiation from the forward jet is coherently polarized; these images look qualitatively similar to polarimetric images of the forward-jet dominated semi-analytic models of Broderick & Loeb (2009). Even a small total flux contribution can increase the net and image-averaged linear polarization fractions to lie within the observed range. However, in this example the EVPA patterns ($\beta_2$ phase) of the TH+PL images remain inconsistent with M87* data.

Figure 16 compares the original thermal snapshot with two realizations of this hybrid thermal+nonthermal funnel emission models. In the purely thermal case, Faraday rotation depolarizes the emission at the EHT beam scale, producing low fractional polarization across the image that is inconsistent with EHT observations of M87*. Adding power-law electrons in the funnel produces coherent linearly polarized emission. When we assume $u_{\rm nth} \propto u_{\rm mag}$ (middle panel), the power-law emission is concentrated close to the black hole and lensed into a ring (Dexter et al. 2012). The weak forward jet component is strongly polarized but lies inside the observed ring, and is thus potentially inconsistent with the EHT total intensity and polarimetric image. In the right panel, we exclude nonthermal emission from inside a radius $r_{\rm cut} = 6r_g$. Both nonthermal maps are consistent with our cuts on net and image-average linear polarization fraction, $|m|_{\rm net}$ and $\langle |m| \rangle$. However, both are inconsistent with the observed EVPA pattern of M87* (i.e., the $\beta_2$ phase).

For this example, we assume a plasma of protons and electrons rather than $e^+/e^-$ pairs. The latter are presumably more likely to form in the funnel (Mościbrodzka et al. 2011; Chen et al. 2018; Levinson & Cerutti 2018; Anantua et al. 2020; Crinquand et al. 2020; Wong et al. 2021), and have different circular polarization properties. Future observations that constrain the resolved circular polarization structure could potentially discriminate between pair and electron-ion plasmas in the emitting region. At longer wavelengths and larger scales, the limb-brightened jet structure of M87 (e.g., Walker et al. 2018) also suggests that the radiating electrons are not concentrated inside the funnel as modeled here.

### 6.4. Radiative Models

Our GRMHD images use the parameterization of Mościbrodzka et al. (2016) to model the electron and ion temperatures given the total gas temperature from a simulation. In this prescription, the electron-to-ion temperature ratio is a function entirely of the local plasma $\beta$. This functional form (Equation (15)) captures the general behavior seen in many simulations of electron heating in turbulent or reconnecting collisonless plasmas; namely, the electron heating is more efficient (and thus the temperature ratio is closer to unity) when $\beta < 1$ (e.g., Howes 2010; Rowan et al. 2017). However, the actual distribution of $T_e$ in a hot accretion flow reflects the balance of heating, cooling, and advection of hot electrons throughout the system. Furthermore, the GRMHD simulations in the library considered here do not include radiative cooling. Our passing models for M87* favor a radiative efficiency of $\epsilon \sim 1\%$ (Section 6.1), however, and we may begin to worry if cooling is dynamically important in M87*.

To assess these uncertainties, it will be useful to compare the results in this work with results from simulations performed with radiative GRMHD codes. These codes typically use either the M1 closure method (e.g., Sądowski et al. 2013; McKinney et al. 2014; Sądowski et al. 2017) or a Monte Carlo approach (e.g., Ryan et al. 2015) to track radiation and its interactions with the plasma near the black hole. In addition to the effects of cooling on the gas temperature, these codes can also evolve the separate electron and ion temperatures under the influence of cooling and different subgrid prescriptions for the electron heating efficiency (e.g Ressler et al. 2015, 2017; Chael et al. 2018, 2019; Ryan et al. 2018; Dexter et al. 2020).

In Figure 17, we show a comparison of the temperature ratio $T_i/T_e$ obtained directly from the two MAD radiative simulations of M87* in Chael et al. (2019; right column), with the temperature ratio obtained from the same simulations using Equation (15) with $R_{\rm low} = 1$, $R_{\rm high} = 20$ (left column). The two rows show simulations using different underlying models for electron heating: the top row (`H10`) uses the turbulent heating prescription of Howes (2010), and the bottom row (`R17`) uses the reconnection heating prescription of Rowan et al. (2017). The simulation data in both rows is averaged in time and azimuth.

Figure 17 shows that the temperature ratios obtained from the electron-ion evolution and from the postprocessing prescription both transition from smaller to larger values when moving from the jet/funnel region (low $\beta$) to the disk (high $\beta$). In simulation `H10` (top row), $T_i/T_e \approx 1$ in the funnel, which well matches the result from Equation (15) with $R_{\rm low} = 1$; in simulation `R17`, electrons are cooler in the funnel ($T_i/T_e \approx 5$), and even colder in the disk-jet interface directly outside the $\sigma = 1$ contour ($T_i/T_e \approx 15$). The cooler electrons in the funnel regions of `R17` reflect the decreased efficiency of low-$\beta$ electron heating in the Rowan et al. (2019) model than in Howes (2010). In Section 5.4, we note that the results of joint scoring (but not simultaneous scoring) favor $R_{\rm low} = 10$ models over $R_{\rm low} = 1$ models. Model `R17` shows that some radiative models can naturally produce $T_i/T_e > 1$ in low-$\beta$ regions. However, the funnel value of $T_i/T_e \approx 5$ in this simulation is in between the two cases $R_{\rm low} = 1$, $R_{\rm low} = 10$ that we considered in the image library.

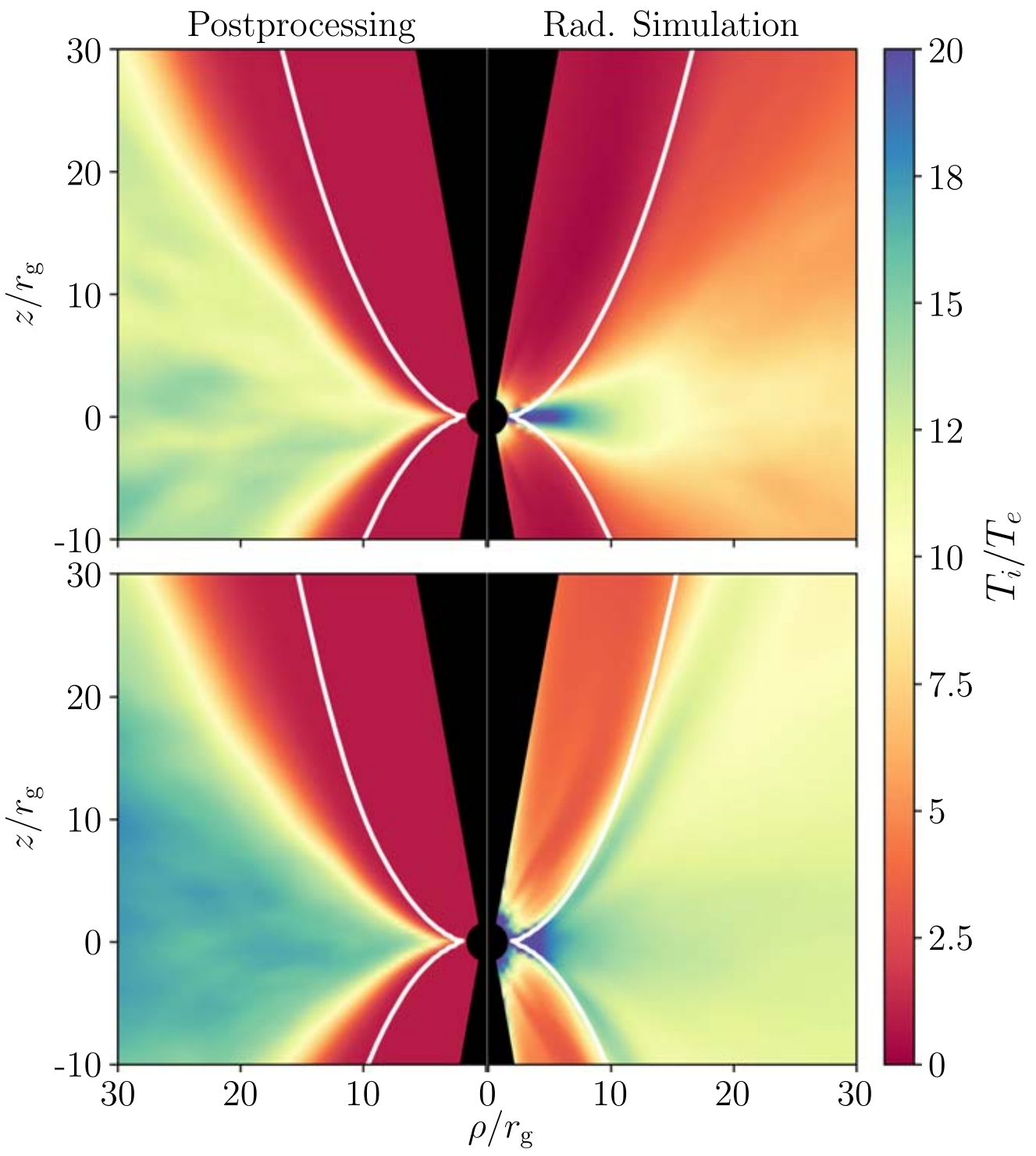

**Figure 17.** Comparison of the temperature ratio obtained directly from two radiative GRMHD simulations with the values obtained from the postprocessing model used in this work. The top and bottom rows show, respectively, time- and azimuth-averaged data from the two radiative MAD simulations of M87* presented in Chael et al. (2019). The top row shows data from simulation `H10`, where electrons are heated by the turbulent heating prescription of Howes (2010), and the bottom row shows the `R17` simulation, where electrons are heated by the magnetic reconnection prescription of Rowan et al. (2017). The left column shows the spatial distribution of the ion-to-electron temperature ratio obtained from these simulations by applying our $\beta$-dependent postprocessing equation to the total gas temperature (Equation (15) with $R_{\rm low} = 1$, $R_{\rm high} = 20$). The right column shows the temperature ratio obtained directly from the independently evolved electron and ion species in the radiative simulations. The white contour shows the $\sigma = 1$ surface.

While the disk electrons are cooler than the jet electrons in both radiative simulations shown in Figure 17, $T_i/T_e$ in the disk of model `H10` increases at small radii. In contrast, model `R17` better matches the postprocessing model prediction of a roughly constant disk temperature ratio. (Note that the 230 GHz emission is entirely produced at radii $r \lesssim 10r_g$.)

In addition to the MAD simulations from Chael et al. (2019) in Figure 17, we have also checked the temperature ratios in the radiative SANE simulations of Ryan et al. (2018). In these simulations, the average temperature ratio in the EHT 230 GHz emission region can also be roughly approximated by the Mościbrodzka et al. (2016) prescription, with $R_{\rm low} = 1$ and $R_{\rm high}$ in the range $R_{\rm high} \sim 10$–20.

Our preliminary results indicate that while some important features of the temperature-ratio distributions produced in radiative simulations can be described by the $R_{\rm low}$, $R_{\rm high}$ model (Equation (15)), the current postprocessing model cannot capture all of the behavior produced in radiative simulations. A more detailed comparison is left for future work.

## 7. Predictions

We have identified a subset of a large parameter space of GRMHD models that is consistent with constraints derived from current EHT total intensity and polarimetric observations of M87*. The models that pass our constraints on the polarimetric structure and jet power from M87* are all magnetically arrested (MAD) accretion flows. Here we make predictions for testing our interpretation with future observations.

### 7.1. Repeated Observations

Repeated EHT observations of M87* at 230 GHz (both in total intensity, e.g., Wielgus et al. 2020, and in linear polarization) will continue to constrain the model parameter space. Figure 18 shows the time evolution of $\beta_2$ amplitude and phase for 200 snapshots of three viable library models: MAD $a_* = -0.5$, $R_{\rm low} = 10$, $R_{\rm high} = 20$; MAD $a_* = +0.5$, $R_{\rm low} = 10$, $R_{\rm high} = 80$; and MAD $a_* = +0.94$, $R_{\rm low} = 10$, $R_{\rm high} = 80$. The observer inclination was 17 deg and 163 deg for the retrograde and prograde models, respectively.

Both quantities show variations on timescales from days to months. The phase and amplitude of $\beta_2$ should change over the course of a week of observations. In EHTC VII, we observe changes in the the $\beta_2$ amplitude and phase over the week of observations in 2017, and use the results from two epochs to define our acceptable parameter ranges. Figure 18 suggests that occasionally the observed changes in $\beta_2$ on ~week timescales can be much more dramatic than we observe in 2017, with variations in $\beta_2$ phase of 90 deg for some models on short timescales.

The scatter in both quantities on longer ~month timescales is much larger than the uncertainty range derived from the EHT 2017 measurements. If our passing GRMHD models accurately describe the 230 GHz emitting region in M87*, future EHT observations should detect variability in the polarization structure. According to current models, the time-averaged $\beta_2$ amplitude and $\langle |m| \rangle$ should remain similar to the current values for prograde spin models, and tend toward larger values for retrograde spin models. For high-prograde spin (or many SANE models), the $\beta_2$ phase should on average be closer to zero than we observe in 2017.

### 7.2. Future Observations at 260 and 345 GHz

In selecting models, we have focused on metrics corresponding to salient features of the data. We have not attempted to compare models in detail to specific features of the reconstructed polarimetric images, most notably the apparently depolarized bright patch in the eastern part of the image (Figure 1). We do note that such depolarized features occur in many of our library images, particularly in MAD models with $R_{\rm low} = 10$. If the eastern patch in the 2017 image is depolarized due to Faraday rotation, it may be possible to tell with future higher-frequency observations. Figure 19 shows images of two of the favored MAD models at the current observing frequency of 230 GHz and at two additional frequencies planned for the future EHT observing campaigns. In addition to internal

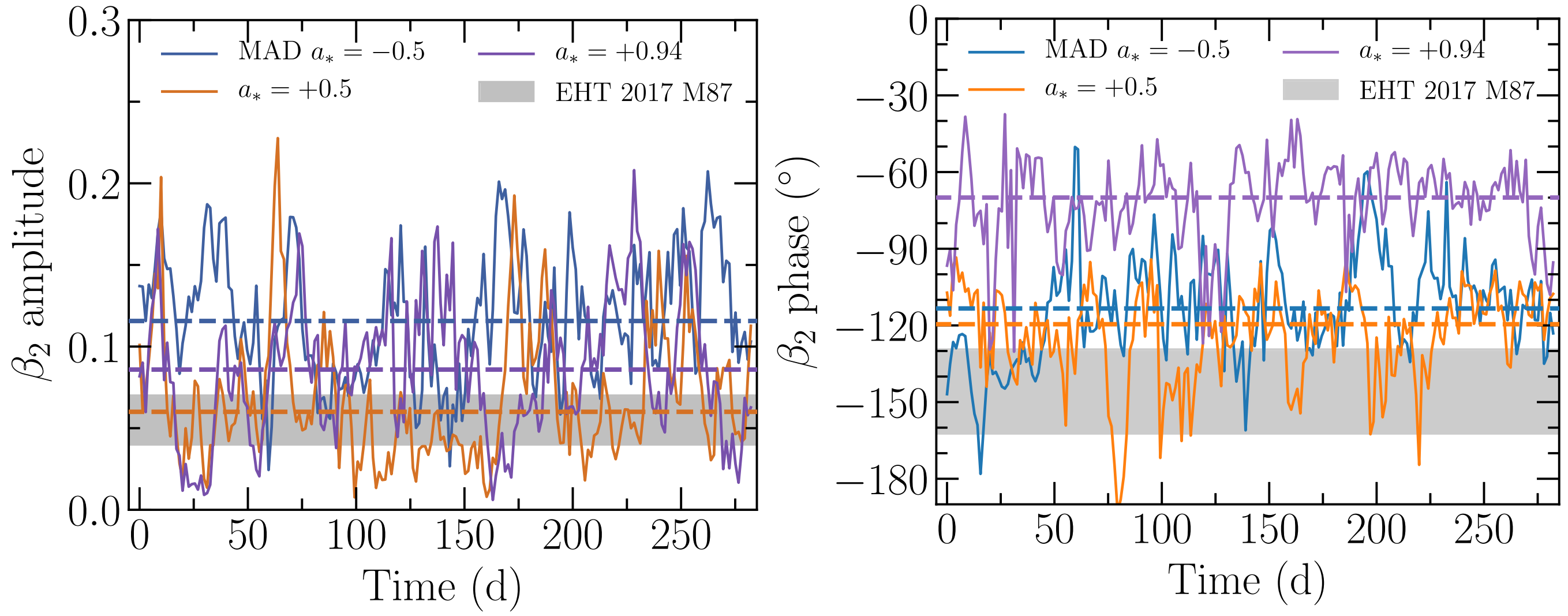

**Figure 18.** Amplitude (left panel) and phase (right panel) of $\beta_2$ as a function of time for three viable GRMHD library models identified here (points, all with $R_{\rm low} = 10$) compared to ranges measured from EHT 2017 M87* data (gray shaded region). The dashed lines show the median values for each model. The retrograde spin model predicts higher $\beta_2$ amplitude in future observations. In the high-prograde spin model, the median $\beta_2$ phase is closer to zero than the observed range in 2017. Changes in both quantities occur on timescales of weeks to months, and should be apparent in future EHT data sets.

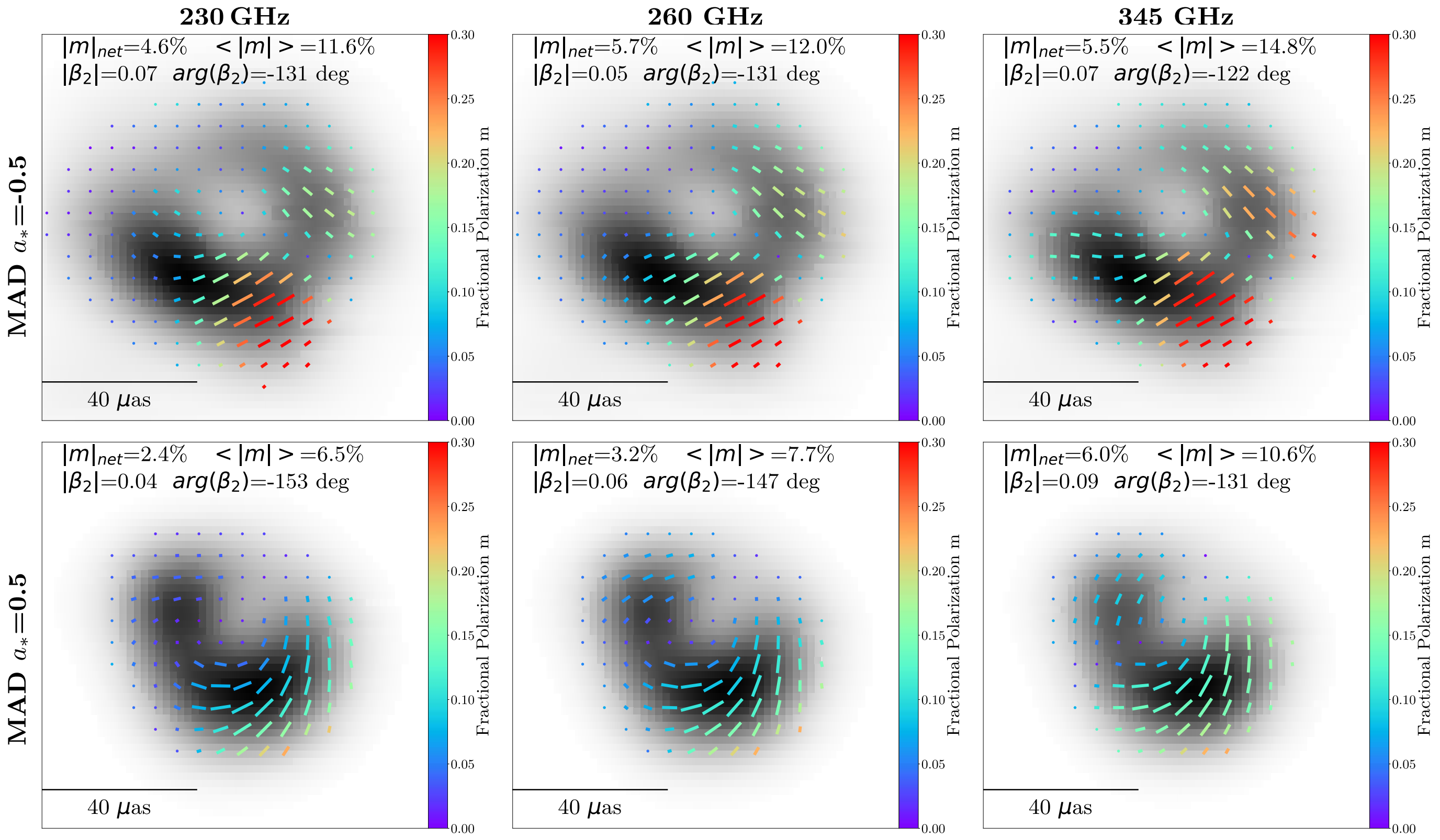

**Figure 19.** Random snapshot of MAD models with $a_* = -0.5$ (top row) and with $a_* = 0.5$ (bottom row) at the current EHT frequency of 230 GHz and two higher frequencies of 260 and 345 GHz, which are planned for the future EHT observations. All images are convolved with a 20 $\mu$as Gaussian. In all images the black hole spin vector is pointing to the left and away from the observer. In all cases, the ring fractional polarization increases slightly with frequency. The EVPA pattern, as measured by the $\beta_2$, is similar at all three frequencies.

Faraday rotation, the sense of the EVPA pattern may also be subject to a net, coherent rotation due to external Faraday rotation. At higher frequency, Faraday rotation is suppressed and EHT observations will see the intrinsic magnetic field pattern more clearly.

For a snapshot from the MAD $a_* = -0.5$ ($R_{\rm high} = 160$ and $R_{\rm low} = 1$) model, Figure 19 shows that the $|m|_{\rm net}$ and $\langle|m|\rangle$ values are predicted to increase with frequency. The 230, 260, and 345 GHz net EVPAs are −77, −70, and −82 deg, respectively, corresponding to an (apparent) rotation measure RM $\sim 1 \times 10^5$ rad m$^{-2}$ between 230 and 345 GHz. The net circular polarization $|v|_{\rm net}$ remains small and nearly constant with frequency; it is 0.42%, 0.35%, 0.32% for 230, 260, and 345 GHz, respectively. A similar trend is observed in a MAD

$a_* = 0.5$ ($R_{\rm high} = 80$ and $R_{\rm low} = 10$) model. The image $|m|_{\rm net}$ and average polarization $\langle|m|\rangle$ are again expected to increase with frequency. The corresponding net EVPAs are $-38$, $-40$, and $-30$ deg, corresponding to an apparent rotation measure of $\mathrm{RM} \sim -1 \times 10^5\,\mathrm{rad\,m^{-2}}$. The net circular polarization fraction $|v|_{\rm net}$ remains roughly constant and close to zero, 0.33, 0.06, and 0.2% from low to high frequency.

Both of these models display similar EVPA structure at all three frequencies, indicating that in this example the net EVPA pattern is due to magnetic field structure rather than coherent Faraday rotation. Future multi-frequency observations will be able to infer the core RM and intrinsic EVPA pattern set by the near-horizon magnetic fields.

## 8. Conclusions

The EHT has produced resolved polarized intensity maps in the near-horizon region around the supermassive black hole in M87. Taken together with image-integrated data from simultaneous observations with ALMA, these images constrain the space of accretion flow and jet models used to interpret the EHT total intensity image with broad implications for jet launching near a black hole event horizon. Here we summarize the main results of that analysis.

1. We interpret the depolarization seen in EHT images as the result of beam depolarization due to Faraday rotation internal to the emission region. In the context of one-zone models and combined with the size and brightness temperature of the total intensity image, we estimate an average emission region plasma density of $n_e \sim 10^{4-7}\,\mathrm{cm^{-3}}$, magnetic field strength of $B \sim (1\text{–}30)$ G, and $T_e = (1\text{–}12) \times 10^{10}$ K.
2. The net EVPA pattern of the M87* polarization maps is predominantly azimuthal. In the context of semi-analytic models with imposed, idealized magnetic field geometry, such a pattern can be reproduced using a significant component of poloidal (radial and/or vertical) magnetic field. The presence of such magnetic fields in a rotating fluid would imply that the magnetic fields are dynamically important. However, significant Faraday rotation may be present and it is not clear whether the observed EVPA pattern can be interpreted in terms of magnetic field structure alone.
3. To capture the effects of realistic magnetic field structure, plasma conditions, and Faraday rotation and conversion, we have compared salient observables to a large library of images from the GRMHD simulation library of EHTC V. The observables are the net circular polarization fraction constrained by ALMA ($|v|_{\rm net}$), the net and image-averaged linear polarization fraction measured by the EHT ($|m|_{\rm net}$ and $\langle|m|\rangle$), and the $m = 2$ coefficient of a Fourier expansion of the azimuthal EVPA pattern ($\beta_2$). Of these, $\beta_2$ is the most constraining metric.
4. The model-scoring procedures disfavor most models from the GRMHD image library from polarimetric observations alone. Many weakly magnetized (SANE) models are too depolarized, or show an EVPA pattern that is too radial. Many strongly magnetized (MAD) models are too coherently polarized. The polarization fraction is generally set by the Faraday rotation depth close to the emission region. MAD models more frequently produce azimuthal EVPA patterns, as expected for magnetic field structures that include a significant poloidal field component. Combined with a conservative lower limit on the jet power of M87, only strongly magnetized (MAD) models remain viable. We use those remaining models to estimate the mass accretion rate onto the central supermassive black hole as $\dot{M} = (3\text{–}20) \times 10^{-4}\,M_\odot\,\mathrm{yr^{-1}}$. The average plasma parameters found from GRMHD images are in good agreement with those inferred from one-zone models.
5. The model space considered in this Letter is incomplete, and systematic uncertainties remain a challenge. While the radiative efficiency that we find is relatively high, we consider only non-radiative GRMHD models. We do not consider GRMHD models with misalignment between the disk and the black hole angular momentum. We also only consider one parameterization for determining the electron distribution function from the simulation data. Of these three major areas of uncertainty, we have explored a small sample of alternative models for determining the electron distribution function, including both alternative prescriptions for electron heating in strongly magnetized regions and including a nonthermal component. The quantitative estimates of mass accretion rate and jet power found here depend on the assumed electron distribution function and are uncertain at the order unity level. The alternative electron distribution functions considered here do not change the main finding that MAD models with dynamically important near-horizon magnetic fields appear more viable for explaining the first polarimetric EHT observations of M87*.
6. Our favored models show time variability in the polarization metrics used here. The median values found at several epochs should be sufficiently well measured to distinguish between the current retrograde and prograde spin models. At higher frequencies of 260 and 345 GHz, weaker Faraday effects should result in an increased degree of polarization. Measurements of the EVPA pattern at higher frequencies can distinguish between Faraday rotation along the line of sight and the imprint of the underlying magnetic field structure. Continued imaging with the EHT and advances in radiative and nonthermal theoretical models will further constrain the electron distribution and magnetic field structure in the jet-launching region near the supermassive black hole event horizon in M87.

The authors of the present Letter thank the following organizations and programs: the Academy of Finland (projects 274477, 284495, 312496, 315721); Agencia Nacional de Investigación y Desarrollo (ANID), Chile via NCN19_058 (TITANs), and Fondecyt 3190878; the Alexander von Humboldt Stiftung; an Alfred P. Sloan Research Fellowship; Allegro, the European ALMA Regional Centre node in the Netherlands, the NL astronomy research network NOVA and the astronomy institutes of the University of Amsterdam, Leiden University and Radboud University; the black hole Initiative at Harvard University, through a grant (60477) from the John Templeton Foundation; the China Scholarship Council; Consejo Nacional de Ciencia y Tecnología (CONACYT, Mexico, projects U0004-246083, U0004-259839, F0003-272050, M0037-279006, F0003-281692, 104497, 275201, 263356, 57265507); the Delaney Family via the

Delaney Family John A. Wheeler Chair at Perimeter Institute; Dirección General de Asuntos del Personal Académico-Universidad Nacional Autónoma de México (DGAPA-UNAM, projects IN112417 and IN112820); the EACOA Fellowship of the East Asia Core Observatories Association; the European Research Council Synergy Grant "BlackHoleCam: Imaging the Event Horizon of Black Holes" (grant 610058); the Generalitat Valenciana postdoctoral grant APOSTD/2018/177 and GenT Program (project CIDEGENT/2018/021); MICINN Research Project PID2019-108995GB-C22; the Gordon and Betty Moore Foundation (grants GBMF- 3561, GBMF-5278); the Istituto Nazionale di Fisica Nucleare (INFN) sezione di Napoli, iniziative specifiche TEONGRAV; the International Max Planck Research School for Astronomy and Astrophysics at the Universities of Bonn and Cologne; Joint Princeton/Flatiron and Joint Columbia/Flatiron Postdoctoral Fellowships, research at the Flatiron Institute is supported by the Simons Foundation; the Japanese Government (Monbukagakusho: MEXT) Scholarship; the Japan Society for the Promotion of Science (JSPS) Grant-in-Aid for JSPS Research Fellowship (JP17J08829); the Key Research Program of Frontier Sciences, Chinese Academy of Sciences (CAS, grants QYZDJ-SSW-SLH057, QYZDJSSW- SYS008, ZDBS-LY-SLH011).

We further thank the Leverhulme Trust Early Career Research Fellowship; the Max-Planck-Gesellschaft (MPG); the Max Planck Partner Group of the MPG and the CAS; the MEXT/JSPS KAKENHI (grants 18KK0090, JP18K13594, JP18K03656, JP18H03721, 18K03709, 18H01245, JP19H01943, 25120007); the Malaysian Fundamental Research Grant Scheme (FRGS) FRGS/1/2019/STG02/UM/02/6; the MIT International Science and Technology Initiatives (MISTI) Funds; the Ministry of Science and Technology (MOST) of Taiwan (105-2112-M-001-025-MY3, 106-2112-M-001-011, 106-2119- M-001-027, 107-2119-M-001-017, 107-2119-M-001-020, 107-2119-M-110-005, 108-2112-M-001-048, and 109-2124-M-001-005); the National Aeronautics and Space Administration (NASA, Fermi Guest Investigator grant 80NSSC20K1567, NASA Astrophysics Theory Program grant 80NSSC20K0527, NASA grant NNX17AL82G, Hubble Fellowship grant HST-HF2-51431.001-A awarded by the Space Telescope Science Institute, which is operated by the Association of Universities for Research in Astronomy, Inc., for NASA, under contract NAS5-26555, and NASA NuSTAR award 80NSSC20K0645); the National Institute of Natural Sciences (NINS) of Japan; the National Key Research and Development Program of China (grant 2016YFA0400704, 2016YFA0400702); the National Science Foundation (NSF, grants AST-0096454, AST-0352953, AST-0521233, AST-0705062, AST-0905844, AST-0922984, AST-1126433, AST-1140030, DGE-1144085, AST-1207704, AST-1207730, AST-1207752, MRI-1228509, OPP-1248097, AST-1310896, AST-1337663, AST-1440254, AST-1555365, AST-1615796, AST-1715061, AST-1716327, AST-1716536, OISE-1743747, AST-1816420, AST-1903847, AST-1935980, AST-2034306); the Natural Science Foundation of China (grants 11573051, 11633006, 11650110427, 10625314, 11721303, 11725312, 11933007, 11991052, 11991053); a fellowship of China Postdoctoral Science Foundation (2020M671266); the Natural Sciences and Engineering Research Council of Canada (NSERC, including a Discovery Grant and the NSERC Alexander Graham Bell Canada Graduate Scholarships-Doctoral Program); the National Research Foundation of Korea (the Global PhD Fellowship Grant: grants 2014H1A2A1018695, NRF-2015H1A2A1033752, 2015-R1D1A1A01056807, the Korea Research Fellowship Program: NRF-2015H1D3A1066561, Basic Research Support Grant 2019R1F1A1059721); the Netherlands Organization for Scientific Research (NWO) VICI award (grant 639.043.513) and Spinoza Prize SPI 78-409; the New Scientific Frontiers with Precision Radio Interferometry Fellowship awarded by the South African Radio Astronomy Observatory (SARAO), which is a facility of the National Research Foundation (NRF), an agency of the Department of Science and Innovation (DSI) of South Africa; the South African Research Chairs Initiative of the Department of Science and Innovation and National Research Foundation; the Onsala Space Observatory (OSO) national infrastructure, for the provisioning of its facilities/observational support (OSO receives funding through the Swedish Research Council under grant 2017-00648) the Perimeter Institute for Theoretical Physics (research at Perimeter Institute is supported by the Government of Canada through the Department of Innovation, Science and Economic Development and by the Province of Ontario through the Ministry of Research, Innovation and Science); the Spanish Ministerio de Economía y Competitividad (grants PGC2018-098915-B-C21, AYA2016-80889-P, PID2019-108995GB-C21); the State Agency for Research of the Spanish MCIU through the "Center of Excellence Severo Ochoa" award for the Instituto de Astrofísica de Andalucía (SEV-2017-0709); the Toray Science Foundation; the Consejería de Economía, Conocimiento, Empresas y Universidad of the Junta de Andalucía (grant P18-FR-1769), the Consejo Superior de Investigaciones Científicas (grant 2019AEP112); the US Department of Energy (USDOE) through the Los Alamos National Laboratory (operated by Triad National Security, LLC, for the National Nuclear Security Administration of the USDOE (Contract 89233218CNA000001); the European Union's Horizon 2020 research and innovation program under grant agreement No 730562 RadioNet; ALMA North America Development Fund; the Academia Sinica; Chandra TM6-17006X and DD7-18089X; the GenT Program (Generalitat Valenciana) Project CIDEGENT/2018/021.

This work used the Extreme Science and Engineering Discovery Environment (XSEDE), supported by NSF grant ACI-1548562, and CyVerse, supported by NSF grants DBI-0735191, DBI-1265383, and DBI-1743442. XSEDE Stampede2 resource at TACC was allocated through TG-AST170024 and TG-AST080026N. XSEDE JetStream resource at PTI and TACC was allocated through AST170028. The simulations were performed in part on the SuperMUC cluster at the LRZ in Garching, on the LOEWE cluster in CSC in Frankfurt, and on the HazelHen cluster at the HLRS in Stuttgart. This research was enabled in part by support provided by Compute Ontario (http://computeontario.ca), Calcul Quebec (http://www.calculquebec.ca) and Compute Canada (http://www.computecanada.ca). We thank the staff at the participating observatories, correlation centers, and institutions for their enthusiastic support.

This Letter makes use of the following ALMA data: ADS/JAO.ALMA#2016.1.01154.V. ALMA is a partnership of the European Southern Observatory (ESO; Europe, representing its member states), NSF, and National Institutes of Natural Sciences of Japan, together with National Research Council (Canada), Ministry of Science and Technology (MOST; Taiwan), Academia Sinica Institute of Astronomy and Astrophysics (ASIAA; Taiwan), and Korea Astronomy and Space Science Institute (KASI; Republic of Korea), in cooperation with the Republic of Chile. The Joint

ALMA Observatory is operated by ESO, Associated Universities, Inc. (AUI)/NRAO, and the National Astronomical Observatory of Japan (NAOJ). The NRAO is a facility of the NSF operated under cooperative agreement by AUI. This paper has made use of the following APEX data: Project ID T-091.F-0006-2013. APEX is a collaboration between the Max-Planck-Institut für Radioastronomie (Germany), ESO, and the Onsala Space Observatory (Sweden). The SMA is a joint project between the SAO and ASIAA and is funded by the Smithsonian Institution and the Academia Sinica. The JCMT is operated by the East Asian Observatory on behalf of the NAOJ, ASIAA, and KASI, as well as the Ministry of Finance of China, Chinese Academy of Sciences, and the National Key R&D Program (No. 2017YFA0402700) of China. Additional funding support for the JCMT is provided by the Science and Technologies Facility Council (UK) and participating universities in the UK and Canada. The LMT is a project operated by the Instituto Nacional de Astrófisica, Óptica, y Electrónica (Mexico) and the University of Massachusetts at Amherst (USA), with financial support from the Consejo Nacional de Ciencia y Tecnología and the National Science Foundation. The IRAM 30-m telescope on Pico Veleta, Spain is operated by IRAM and supported by CNRS (Centre National de la Recherche Scientifique, France), MPG (Max-Planck- Gesellschaft, Germany) and IGN (Instituto Geográfico Nacional, Spain). The SMT is operated by the Arizona Radio Observatory, a part of the Steward Observatory of the University of Arizona, with financial support of operations from the State of Arizona and financial support for instrumentation development from the NSF. The SPT is supported by the National Science Foundation through grant PLR- 1248097. Partial support is also provided by the NSF Physics Frontier Center grant PHY-1125897 to the Kavli Institute of Cosmological Physics at the University of Chicago, the Kavli Foundation and the Gordon and Betty Moore Foundation grant GBMF 947. The SPT hydrogen maser was provided on loan from the GLT, courtesy of ASIAA. The EHTC has received generous donations of FPGA chips from Xilinx Inc., under the Xilinx University Program. The EHTC has benefited from technology shared under open-source license by the Collaboration for Astronomy Signal Processing and Electronics Research (CASPER). The EHT project is grateful to T4Science and Microsemi for their assistance with Hydrogen Masers. This research has made use of NASA's Astrophysics Data System. We gratefully acknowledge the support provided by the extended staff of the ALMA, both from the inception of the ALMA Phasing Project through the observational campaigns of 2017 and 2018. We would like to thank A. Deller and W. Brisken for EHT-specific support with the use of DiFX. We acknowledge the significance that Maunakea, where the SMA and JCMT EHT stations are located, has for the indigenous Hawaiian people.

## Appendix A
## Relationship between $\beta_2$ Coefficient and *E*- and *B*-modes

The $\beta_2$ coefficients of the azimuthal decomposition of the complex linear polarization $\mathcal{P} = \mathcal{Q} + i\mathcal{U}$ used in Section 5 are directly related to the decomposition of polarization fields into *E*- and *B*-modes familiar from cosmology. In this Appendix, we illustrate that relationship and demonstrate that the information in the image-space decomposition of GRMHD library images in Section 5 can also be accessed directly in calibrated visibility domain data sampled on EHT 2017 baselines, provided the data are accurately phase calibrated.

### A.1. Definitions

In defining flat sky *E*- and *B*-modes, we follow the conventions of Kamionkowski & Kovetz (2016), Section 4.1 (up to factors of $\sqrt{2}$). *E*- and *B*-modes are naturally defined in the visibility space sampled by an interferometer. For a baseline vector $\boldsymbol{u}$ with a magnitude $u$ and PA $\theta$, the $\tilde{E}$ and $\tilde{B}$ visibilities are related to the Stokes visibilities $\tilde{\mathcal{Q}}$ and $\tilde{\mathcal{U}}$ by a rotation of $2\theta$ in the Fourier plane

$$\begin{pmatrix} \tilde{E}(u,\theta) \\ \tilde{B}(u,\theta) \end{pmatrix} = \begin{bmatrix} \cos 2\theta & \sin 2\theta \\ -\sin 2\theta & \cos 2\theta \end{bmatrix} \begin{pmatrix} \tilde{\mathcal{Q}}(u,\theta) \\ \tilde{\mathcal{U}}(u,\theta) \end{pmatrix}. \tag{A1}$$

In real space, the *E*- and *B*-mode fields are analogous to the gradient and curl of the polarization tensor:

$$\nabla^2 E = \partial_a \partial_b \boldsymbol{P}_{ab}, \quad \nabla^2 B = \epsilon_{ac} \partial_b \partial_c \boldsymbol{P}_{ab}, \tag{A2}$$

where $\epsilon_{ac}$ is the 2D Levi-Cevita symbol and the polarimetric tensor is

$$\boldsymbol{P}_{ab} = \begin{bmatrix} \mathcal{Q} & \mathcal{U} \\ \mathcal{U} & -\mathcal{Q} \end{bmatrix}. \tag{A3}$$

$\boldsymbol{P}$ transforms as a trace-free tensor under rotations; for a rotation matrix $\boldsymbol{R}(\alpha)$ that rotates the coordinate axes by an angle $\alpha$, $\boldsymbol{P} \rightarrow \boldsymbol{R}(\alpha)\boldsymbol{P}\boldsymbol{R}^T(\alpha)$ (equivalently, the complex field $\mathcal{P} \rightarrow \mathcal{P}e^{2i\alpha}$). While the values of the $\mathcal{Q}$ and $\mathcal{U}$ images depend on the choices of coordinate axes, the real space *E*- and *B*-mode images are coordinate-independent scalars.

### A.2. Relationship between (E, B) and $\beta_2$ Coefficients

Consider a linearly polarized image $\mathcal{P} = \mathcal{Q} + i\mathcal{U}$ in 2D image-domain polar coordinates $(\rho, \phi)$. We can expand the image in a multipole series:

$$\mathcal{P}(\rho, \phi) = I_0 \sum_{m=-\infty}^{\infty} \beta_m f_m(\rho) e^{im\phi}. \tag{A4}$$

In the decomposition of Equation (A4), $I_0$ is the total flux density of the Stokes $\mathcal{I}$ image, the $\beta_m$ coefficients are complex, and the radial envelope function $f_m(\rho)$ is normalized so that

$$2\pi \int_{\rho_{\min}}^{\rho_{\max}} f_m(\rho) \rho \, d\rho = 1. \tag{A5}$$

The $\beta_m$ coefficients defined in this way then correspond to those defined in Equation (9):

$$\beta_m = \frac{1}{I_0} \int_{\rho_{\min}}^{\rho_{\max}} \int_0^{2\pi} \mathcal{P}(\rho, \phi) e^{-im\phi} \rho \, d\rho \, d\phi. \tag{A6}$$

In particular, $\beta_0$ is the image-integrated complex fractional polarization, and $\beta_2$ encodes the same information on the gradient and curl of the polarization field that is available in the *E*- and *B*-modes. Note that because $\mathcal{P}$ is a complex image, in general $\beta_m \neq \beta_{-m}^*$.

The Fourier transform of $P(\rho, \phi)$ is

$$\tilde{P}(u, \theta) = 2\pi I_0 \sum_{m=-\infty}^{\infty} i^{-m} \beta_m e^{im\theta} F_m(u), \tag{A7}$$

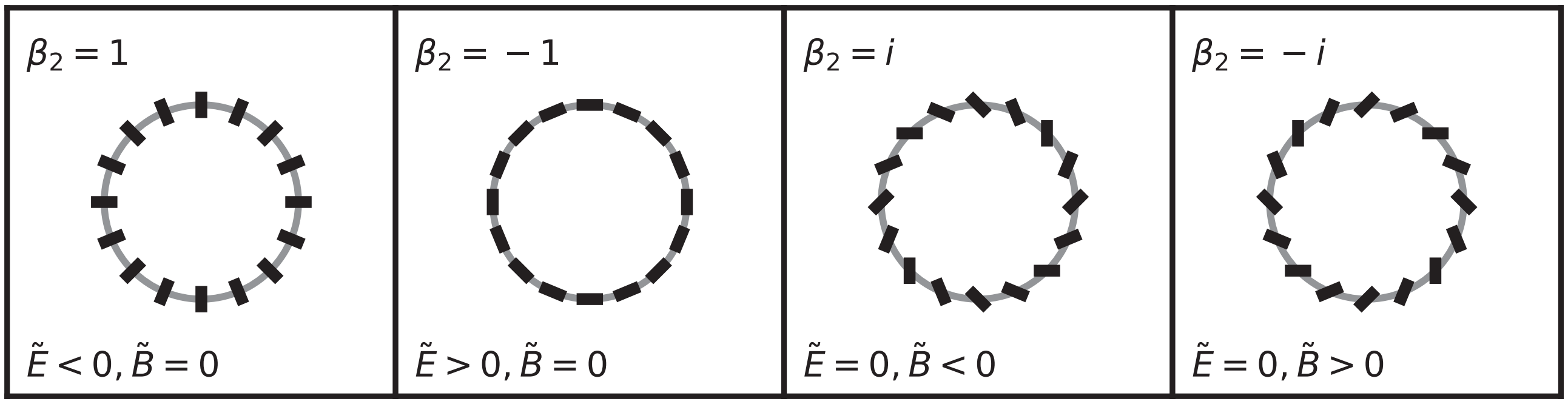

**Figure 20.** Schematic guide to EVPA patterns for ring images with power only in the $\beta_2$ mode in an azimuthal decomposition, and the corresponding signs of the $\tilde{E}$ and $\tilde{B}$ mode visibilities.

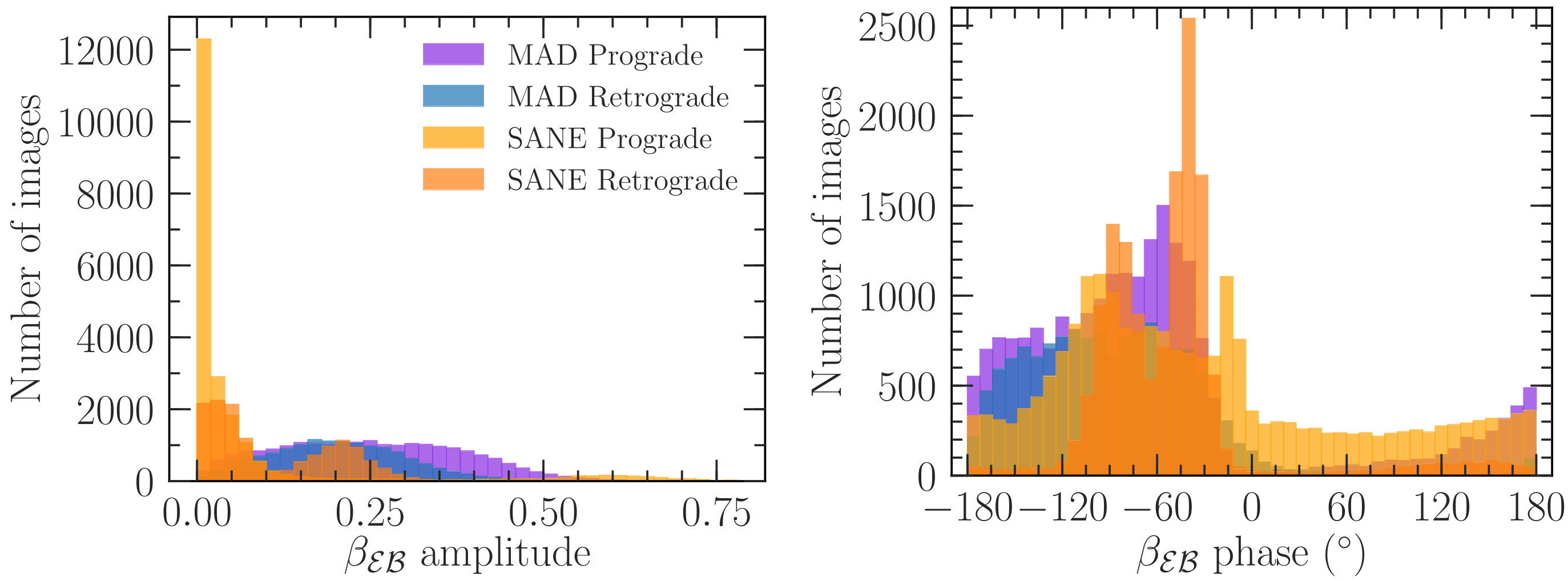

**Figure 21.** Distributions of the amplitude (left panel) and phase (right panel, in degrees) of the complex quantity $\beta_{\mathcal{EB}} = -\mathcal{E}_R - i\mathcal{B}_R$ computed from simulated data on EHT 2017 baselines for the full GRMHD image library considered in this work. The distributions are broadly consistent with the $\beta_2$ results measured in the image domain in Figure 9, illustrating the relationship between the $\beta_2$ metric and average $\tilde{E}$ and $\tilde{B}$ mode visibilities.

where $F_m(u)$ is the $m$th order Hankel transform of the radial function $f_m(\rho)$:

$$F_m(u) = \int_{\rho_{\min}}^{\rho_{\max}} f_m(\rho) J_m(2\pi\rho u)\rho \, d\rho. \tag{A8}$$

The Fourier transforms of $\mathcal{Q}$ and $\mathcal{U}$ (the linear Stokes visibilities) are then

$$\tilde{\mathcal{Q}}(u, \theta) = 2\pi I_0 \sum_{m=-\infty}^{\infty} i^{-m} \,\mathrm{Re}\,[\beta_m e^{im\theta} F_m(u)],$$
$$\tilde{\mathcal{U}}(u, \theta) = 2\pi I_0 \sum_{m=-\infty}^{\infty} i^{-m} \,\mathrm{Im}\,[\beta_m e^{im\theta} F_m(u)], \tag{A9}$$

and from the definition in Equation (A1) the $E$- and $B$-mode visibilites are

$$\tilde{E}(u, \theta) = 2\pi I_0 \sum_{m=-\infty}^{\infty} i^{-m} \,\mathrm{Re}\,[\beta_m e^{i(m-2)\theta} F_m(u)],$$
$$\tilde{B}(u, \theta) = 2\pi I_0 \sum_{m=-\infty}^{\infty} i^{-m} \,\mathrm{Im}\,[\beta_m e^{i(m-2)\theta} F_m(u)]. \tag{A10}$$

From Equation (A10), we can see immediately that an image with real $\beta_2$ and all other $\beta_{m\neq 2} = 0$ is a pure $E$-mode; for instance, if $\beta_2 = 1$ (radial polarization vectors), $\tilde{E} < 0$. If $\beta_2 = -1$ (toroidal polarization vectors), $\tilde{E} > 0$. Similarly, if an image has purely imaginary $\beta_2$ (and all other $\beta_{m\neq 2} = 0$) it is a pure $B$-mode; if e.g., $\beta_2 = i$ (right-handed helical polarization vectors), then $\tilde{B} < 0$, and if $\beta_2 = -i$ (left-handed helix), $\tilde{B} > 0$. Figure 20 illustrates the values of the $\beta_2$ and the corresponding signs of the $\tilde{E}$ and $\tilde{B}$ visibilities for these four azimuthally symmetric cases.

### A.3. $\tilde{E}$ and $\tilde{B}$ Distributions of GRMHD Library Images

Starting with Equation (A10), and integrating over the baseline angle $\theta$, only the $\beta_2$ mode survives.

$$\int_0^{2\pi} \tilde{E}(u, \theta) d\theta = -2\pi I_0 \,\mathrm{Re}\,[\beta_2 F_2(u)] \tag{A11}$$

$$\int_0^{2\pi} \tilde{B}(u, \theta) d\theta = -2\pi I_0 \,\mathrm{Im}\,[\beta_2 F_2(u)]. \tag{A12}$$

Thus, if we have calibrated visibility measurements of $\tilde{Q}$ and $\tilde{U}$ (and thus $\tilde{E}$ and $\tilde{B}$), we can measure the $\beta_2$ mode by averaging $\tilde{E}$ and $\tilde{B}$ in visibility space.

To illustrate this connection between $E$- and $B$-mode visibilities and the $\beta_2$ coefficient, we compute the following averages on images in the GRMHD library

$$\mathcal{E}_R = \frac{\langle \mathrm{Re}[\tilde{E}] \rangle_{(u,v)}}{\langle |\tilde{I}| \rangle_{(u,v)}}, \quad \mathcal{B}_R = \frac{\langle \mathrm{Re}[\tilde{B}] \rangle_{(u,v)}}{\langle |\tilde{I}| \rangle_{(u,v)}}, \tag{A13}$$

where we take the average only over ($u$, $v$) points sampled by the EHT in 2017, including conjugate baselines. As atmospheric phase errors and $D$-term miscalibration will affect the phase of the $\tilde{E}$ and $\tilde{B}$ visibility and thus the results for the averages, we perform this test on synthetic data from the image library using perfectly calibrated data with no noise.

Because we include conjugates in the average over all data points, the average of the imaginary part is zero. We perform the averaging only over a $u$, $v$ range $[1, 10]G\lambda$ in order to remove any effects from large-scale structure, which may have a different net sense of polarization than the resolved emission ring.

We finally combine $\mathcal{E}_{\rm R}$ and $\mathcal{B}_{\rm R}$ in a complex quantity:

$$\beta_{\mathcal{E}\mathcal{B}} = -\mathcal{E}_{\rm R} - i\mathcal{B}_{\rm R}, \tag{A14}$$

where the negative signs are chosen to match the angle convention for $\beta_2$ in Equation (9). In Figure 21 we show histograms of the magnitude and angle of the $\beta_{\mathcal{E}\mathcal{B}}$ for comparison with the $\beta_2$ histograms in Figure 9. The distributions broken down by model type (MAD/SANE, prograde/retrograde) reproduce the general behavior of $\beta_2$ amplitude and phase from the image-domain calculations in Figure 9, although the normalization of the $\beta_{\mathcal{E}\mathcal{B}}$ amplitude is different. Comparing Figure 9 with Figure 21, it is apparent that all of the essential information on the EVPA structure used in this Letter can, in principle, be extracted from EHT visibilities without image reconstruction. However, because phase and amplitude calibration of the EHT visibilities is necessary for extracting the $E$- and $B$-modes from the visibilities, modeling the source structure in the image domain would remain a necessary part of the analysis even if we were to use $\mathcal{E}_{\rm R}$ and $\mathcal{B}_{\rm R}$ instead of $\beta_2$.

## Appendix B
## Faraday Rotation in GRMHD Models of M87*

As linear polarization travels through magnetized plasma, Faraday rotation shifts its EVPA by $\tau_{\rho_V}/2$ radians. If $\tau_{\rho_V} \gg 1$, as is the case for most of our models (see Figure 7), Faraday rotation can, in principle, scramble otherwise observable polarimetric signals. In this section, we explore in more detail the sources of Faraday rotation in our models, and demonstrate that observable linear polarization signals can, in fact, exist in models with $\tau_{\rho_V} \gg 1$. This is because Faraday rotation occurs co-spatially with the emission and should not be conceived of as a purely external screen.

Ricarte et al. (2020) studied the resolved Faraday rotation properties of a subset of the same models used in this work. Figure 14 shows their inferred |RM| versus $|m|_{\rm net}$ for those images. For each model, 11 snapshots spaced between 7500 and 10000 $r_g/c$ are included. Each of these models was found to pass the constraints of EHTC V. Snapshots with positive RMs are plotted with filled symbols, while those with negative RMs are plotted with open symbols. In gray, we overplot the allowed range of $|m|_{\rm net}$ as well as the range of RM for the core region inferred from simultaneous ALMA-only observations.

Despite the large Faraday depths of these models, many of them are capable of producing RMs that are consistent with the observed data. RM and $|m|_{\rm net}$ are anti-correlated, as expected, because a greater amount of Faraday rotation should both increase the RM and cause a greater amount of scrambling of the polarized emission. Note that the RM varies by orders of magnitude and even flips sign over time in these models. This is due to the summation of time-variable regions with significantly different and even oppositely signed Faraday rotation depths, which also contributes to highly non-$\lambda^2$ evolution of the EVPA with wavelength.

One potential source of uncertainty is the limited volume of our GRMHD simulations. In our image library, the radiative transfer equation is solved within a radius of only 50 or 100 $r_g$ depending on the model, while in principle significant Faraday rotation may occur at much larger radius. Figure 22 demonstrates that for the gas distributions studied, most of the Faraday rotation occurs at radii much smaller than the outer domain. Five example models are visualized here: (a) MAD $a_* = +0.94$ $R_{\rm high} = 20$, (b) SANE $a_* = +0.5$ $R_{\rm high} = 1$, (c) MAD $a_* = -0.5$ $R_{\rm high} = 160$, (d) SANE $a_* = +0.5$ $R_{\rm high} = 160$, and (e) SANE $a_* = 0$ $R_{\rm high} = 80$. The brightness of each pixel scales with the total intensity (intentionally saturating 0.3% of the pixels), while the color of each pixel scales with the intensity-weighted Faraday depth $\tau_{\rho_V,I}$, as shown in the colorbar. $\tau_{\rho_V,I}$ is distinct from $\tau_{\rho_V}$ because it is intensity weighted along each ray, such that in each pixel

$$\tau_{\rho_V,I} \equiv \frac{1}{I}\int |\rho_V| I(s) ds. \tag{B1}$$

As also shown in Figure 7, these models span a wide range of Faraday depths. Typically, SANE models and models with a larger $R_{\rm high}$ have larger Faraday depths than MAD models and those with smaller $R_{\rm high}$. SANE models require a larger accretion rate to reproduce the total intensity of M87*, which increases the amount of Faraday rotating material. Meanwhile, increasing $R_{\rm high}$ lowers the temperature of the midplane by construction. This makes Faraday rotation more efficient, and also requires a larger accretion rate to compensate for the lower electron temperatures (see Mościbrodzka et al. 2017, for an extended discussion).

In Figure 23, we confirm that the linear polarization parameters used in this study are not strongly evolving at the outer simulation domain. Here, we provide polarization maps and the linear polarization parameters for these models blurred with a Gaussian beam with a FWHM of 20 $\mu$as. The rows display models with outer integration radii of 10, 20, 40, and 50 $r_g$. For the three left-most models, there is little difference between images constructed with $r_{\rm max} = 10\ r_g$ and those with $r_{\rm max} = 50\ r_g$, echoing our previous findings that there are small fractional differences in the Faraday depth between these scales. Faraday rotation thick models (d) and (e) show modest differences between images calculated with outer boundaries of 10 $r_g$ and 50 $r_g$. Those images also appear to be converged by 40 $r_g$.

Some of our models produce observable polarimetric signatures despite $\tau_{\rho_V}$ being large enough to potentially depolarize all of the emission. This apparent contradiction is resolved by the fact that not all emission is Faraday rotated by the same amount. Because Faraday rotation occurs co-spatially with the emission, instead of as an external screen, there can exist emission traveling on Faraday-thin paths to the camera even in models where $\tau_{\rho_V} \gg 1$ when integrated along the entire geodesic. In Figure 24, we illustrate this phenomenon by splitting these images into the emission originating in front of or behind the midplane. Here, models are plotted as in Figure 22 with $r_{\rm max} = 50\ r_g$. Emission originating from the back half of the simulation domain exhibits larger values of $\tau_{\rho_V,I}$, as it must travel through more Faraday rotating material. Notice the clearly Faraday-thin regions in panels (c) and (e), despite the enormous values of $\langle\tau_{\rho_V,I}\rangle$ for the image overall. Emission from in front of the midplane that is observed within the photon ring has a much larger $\tau_{\rho_V,I}$ than the rest of the image because those geodesics pass through the midplane and around the black hole through Faraday-thick material.

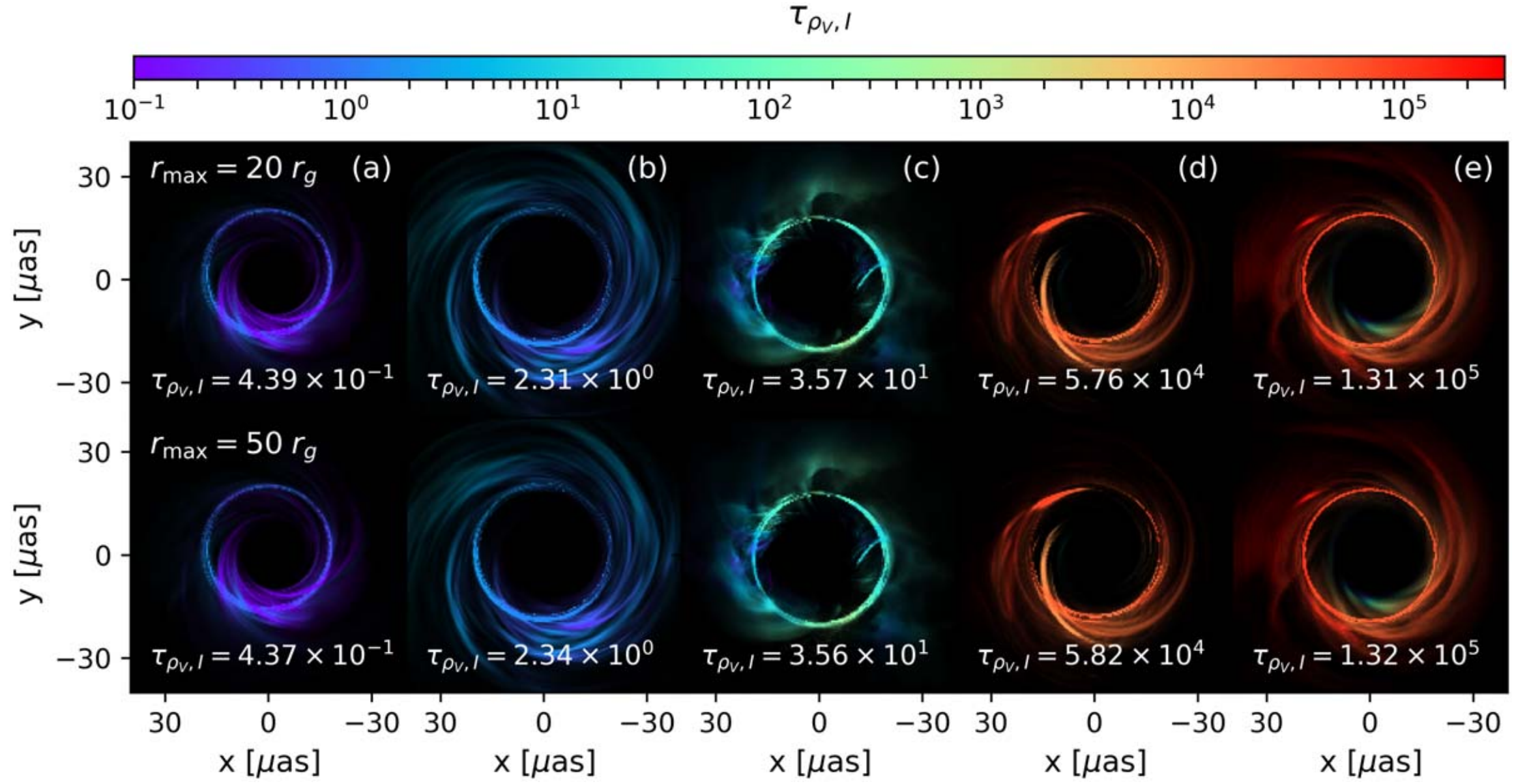

**Figure 22.** Intensity-weighted Faraday depths visualized with five example models: (a) MAD $a_* = +0.94$ $R_{\rm high} = 20$, (b) SANE $a_* = +0.5$ $R_{\rm high} = 1$, (c) MAD $a_* = -0.5$ $R_{\rm high} = 160$, (d) SANE $a_* = +0.5$ $R_{\rm high} = 160$, and (e) SANE $a_* = 0$ $R_{\rm high} = 80$. The brightness of each pixel scales with the total intensity (intentionally saturating 0.3% of the pixels), while the color indicates the intensity-weighted Faraday depth, $\tau_{\rho_V,I}$. In the top row, the maximum integration radius is set to 20 $r_g$, while in the bottom row, the maximum integration radius is set to 50 $r_g$. We find very little difference, confirming that our results should be insensitive to the outer radius of the simulation domain.

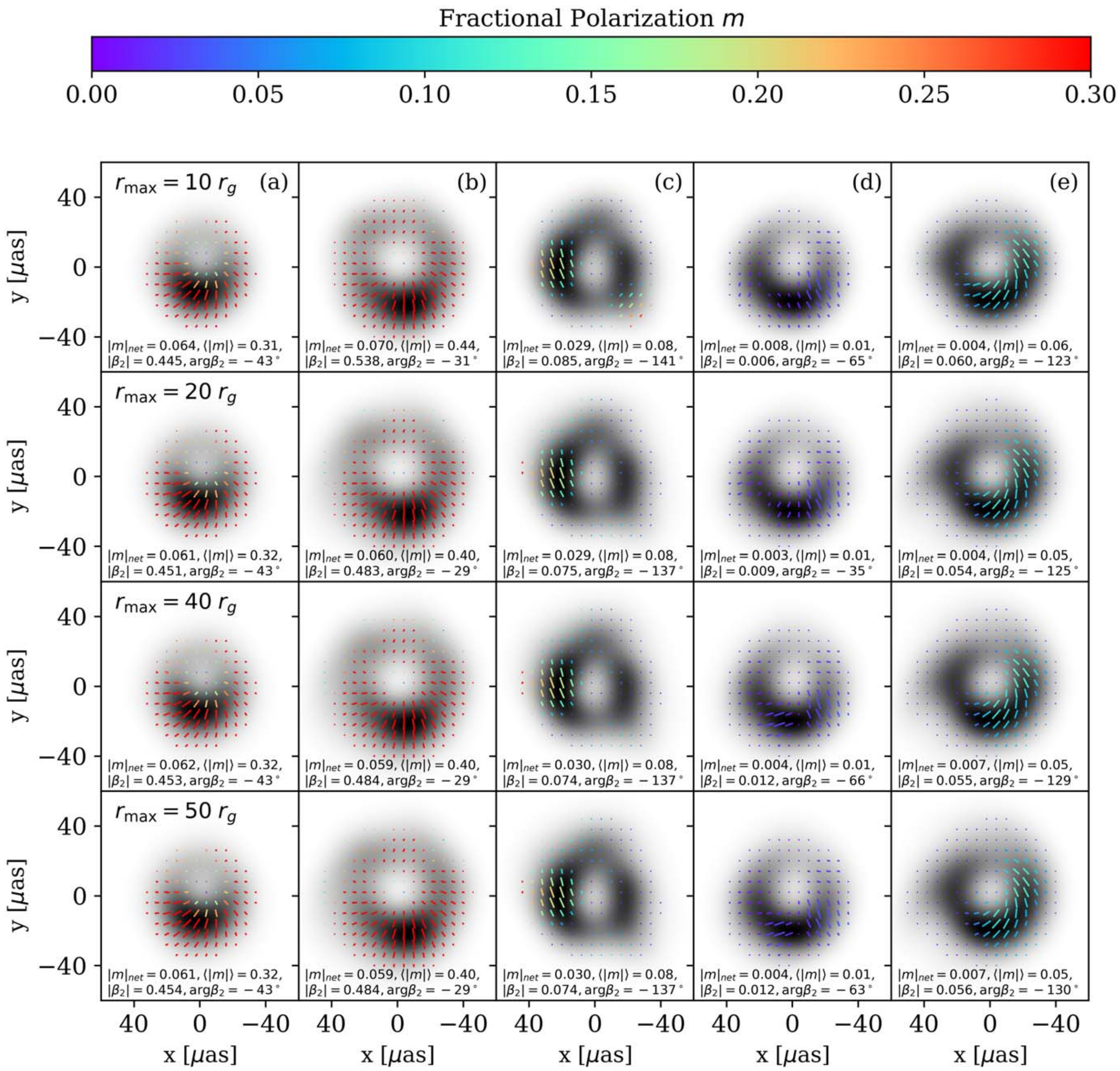

**Figure 23.** Polarimetric images of five example models: (a) MAD $a_* = +0.94$ $R_{\rm high} = 20$, (b) SANE $a_* = +0.5$ $R_{\rm high} = 1$, (c) MAD $a_* = -0.5$ $R_{\rm high} = 160$, (d) SANE $a_* = +0.5$ $R_{\rm high} = 160$, and (e) SANE $a_* = 0$ $R_{\rm high} = 80$. The maximum integration radius is set to 10, 20, 40, and 50 $r_g$ in each row from top to bottom. For models (a)–(c), there is little difference between images computed with $r_{\rm max} = 20\ r_g$ and those computed with $r_{\rm max} = 50\ r_g$. The significantly more Faraday-thick models (d) and (e) appear to be converged by a radius of 40 $r_g$.

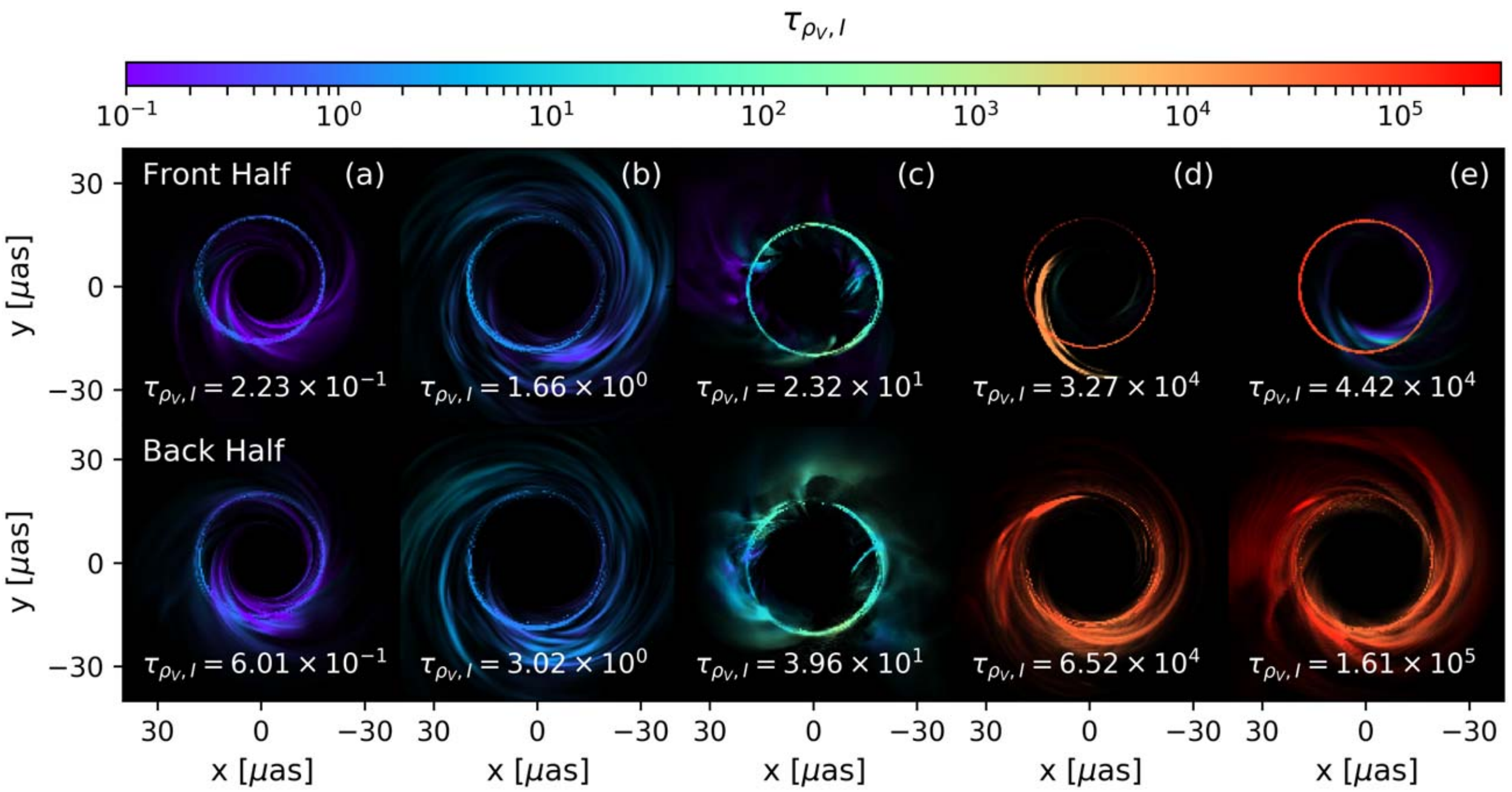

**Figure 24.** Faraday depth visualizations as in Figure 22, but with emission origin split into the front and back halves of the simulation domain. $r_{\max} = 50\, r_g$ for all of these images. Because the Faraday rotation occurs co-spatially with the emission, radiation originating from the front half of the simulation domain has smaller $\tau_{\rho_V,I}$ than emission originating from the back half. In panels (c) and (e), notice the Faraday-thin ($\tau_{\rho_V,I} < 1$) regions (purple) in the front half images even though $\tau_{\rho_V,I} \gg 1$ for the model overall.

## Appendix C
## Distributions of Theory Metrics for Each Model

Figures 25–29 show distributions of the metrics used to score models, split out for each model individually.

Prograde SANE models show rapidly decreasing $|m|_{\rm net}$ with increasing $R_{\rm high}$ (Figure 25) and are significantly depolarized when $R_{\rm high} > 10$. This behavior was previously demonstrated by Mościbrodzka et al. (2017). The accretion flow electron temperature decreases with increasing $R_{\rm high}$, increasing the strength of Faraday rotation while also concentrating the emission at high latitudes behind the black hole (see also EHTC V). The emission is then depolarized when traveling through the Faraday-thick midplane plasma.

Retrograde SANE models, however, show nearly the opposite behavior, with depolarization maximized for $R_{\rm high} = 1$. At larger values of $R_{\rm high}$, linearly polarized emission appears on the near side of the midplane, producing coherent linear polarization structure that is not Faraday depolarized.

MAD models at all spins show a mild degree of depolarization with increasing $R_{\rm high}$. The accretion flow electron temperature remains high even for large values of $R_{\rm high}$, as much of the plasma has $\beta \simeq 1$.

Similar qualitative behavior is seen in $\langle |m| \rangle$ (Figure 26) and the amplitude of $\beta_2$ (Figure 27). However, those quantities show less time variability (narrower distributions) than is seen in $|m|_{\rm net}$. As a result, observed ranges of those values are more constraining. In particular, the MAD models show consistent offsets where $\langle |m| \rangle$ and $|\beta_2|$ are lower for $R_{\rm low} = 10$ than $R_{\rm low} = 1$ models. Some spin dependence is also apparent, with high-prograde spin usually corresponding to the highest degrees of ordered polarization.

When the $\beta_2$ amplitude is not strongly suppressed (e.g., by Faraday rotation), the $\beta_2$ phase distributions are related to intrinsic magnetic field structure (e.g., Figure 3 and PWP). Prograde spin, $R_{\rm high} = 1$ SANE models and retrograde spin, large $R_{\rm high}$ SANE models both show radial EVPA patterns, resulting in $\beta_2$ phase distributions near zero. MAD models show spin-dependent $\beta_2$ phase distributions for low values of $R_{\rm high}$, ranging from spiral patterns ($\angle\beta_2 \simeq -90$ deg) for retrograde spin to more radial patterns at high-prograde spin. The patterns are relatively constant functions of $R_{\rm high}$ and $R_{\rm low}$, although with some shift of MAD prograde distributions to twistier EVPA patterns, particularly for $R_{\rm low} = 10$.

Most models show distributions of $v_{\rm net}$ centered on zero, near the observed range (Figure 29). MAD models generally show low circular polarization fractions, while heavily depolarized SANE models (retrograde low $R_{\rm high}$, prograde large $R_{\rm high}$) tend to show larger $|v_{\rm net}|$ than observed, which can be explained by stronger Faraday conversion in the emission region.

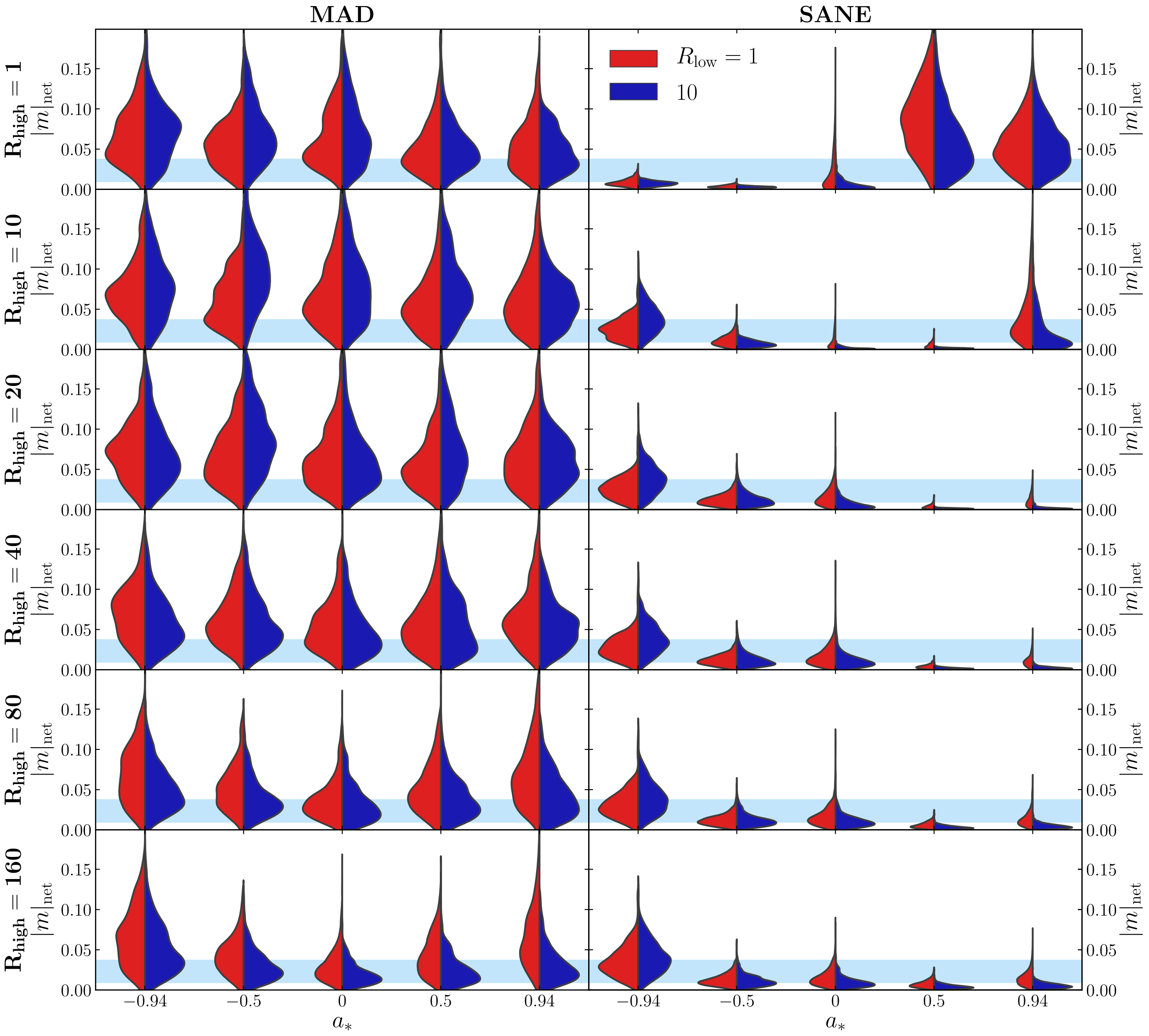

**Figure 25.** Distributions of net polarization fraction $|m|_{\rm net}$ for all models. MAD and SANE simulations are shown in the left and right panels. Black hole spin $a_*$ varies along the $x$ axis, $R_{\rm high}$ varies in each row, and the distributions at $R_{\rm low} = 1$ and 10 are shown in red and blue in each case.

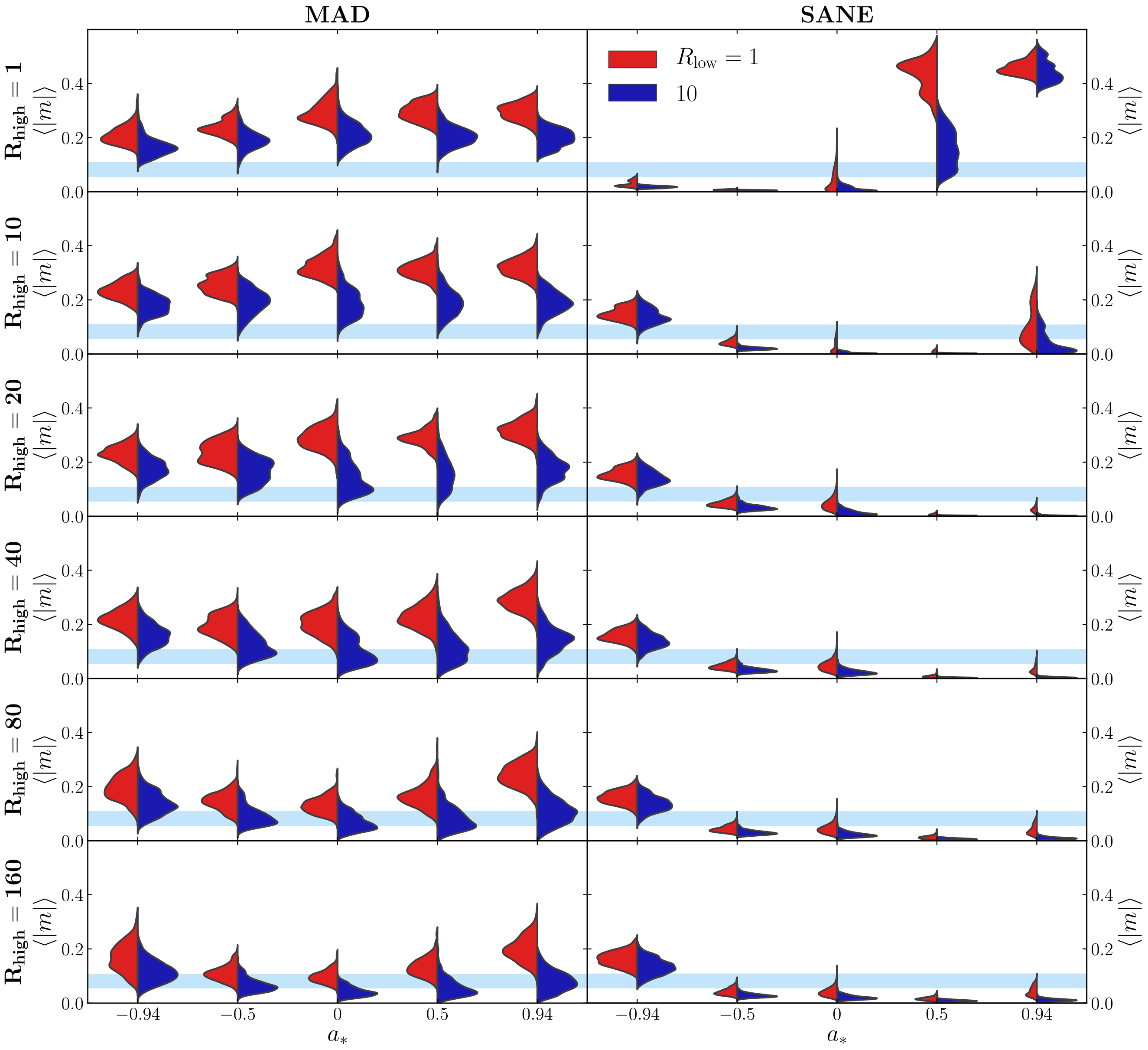

**Figure 26.** Same as in Figure 25, but for the image-averaged polarization fraction $\langle|m|\rangle$.

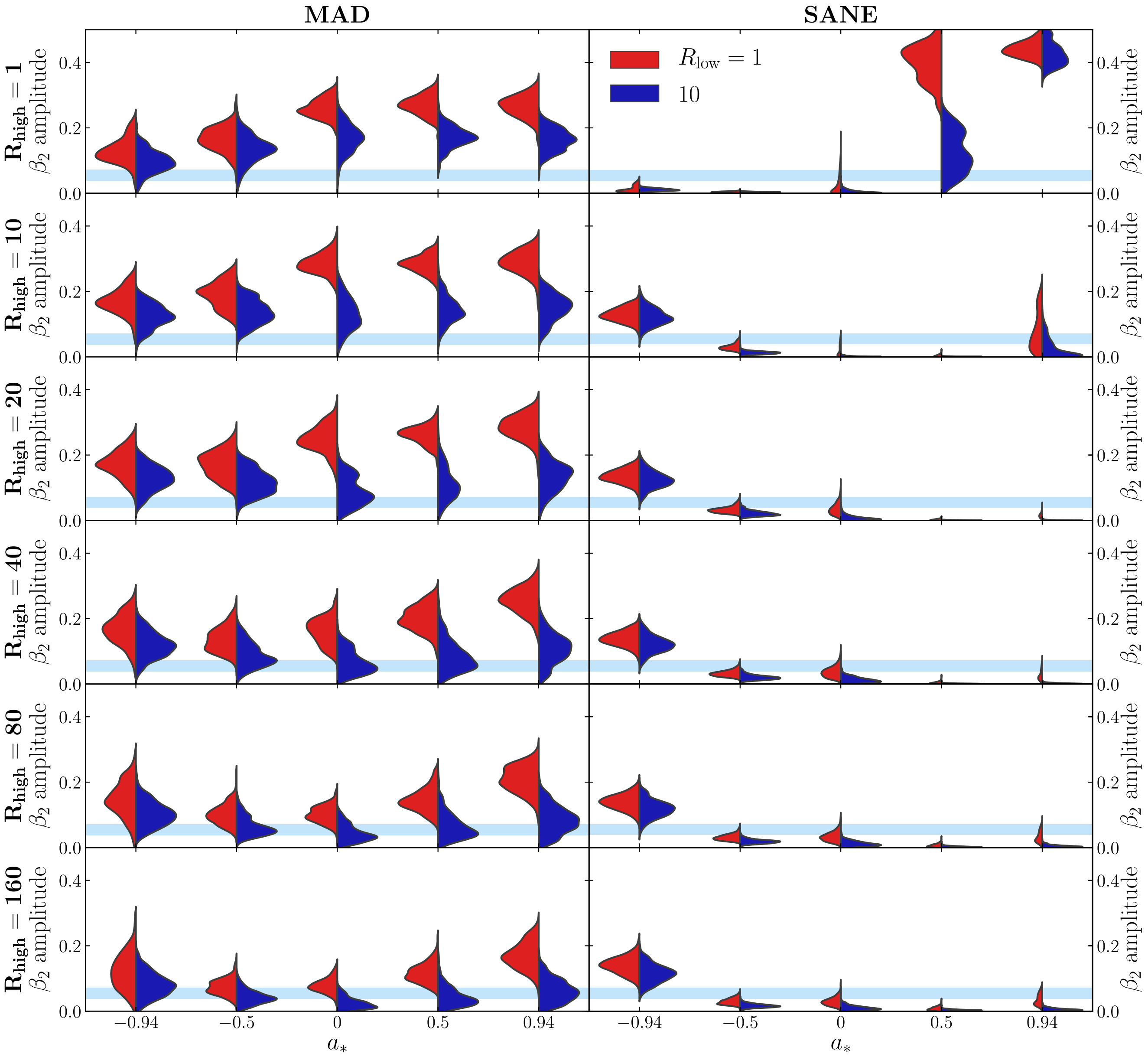

**Figure 27.** Same as in Figure 25, but for the amplitude of $\beta_2$, $|\beta_2|$.

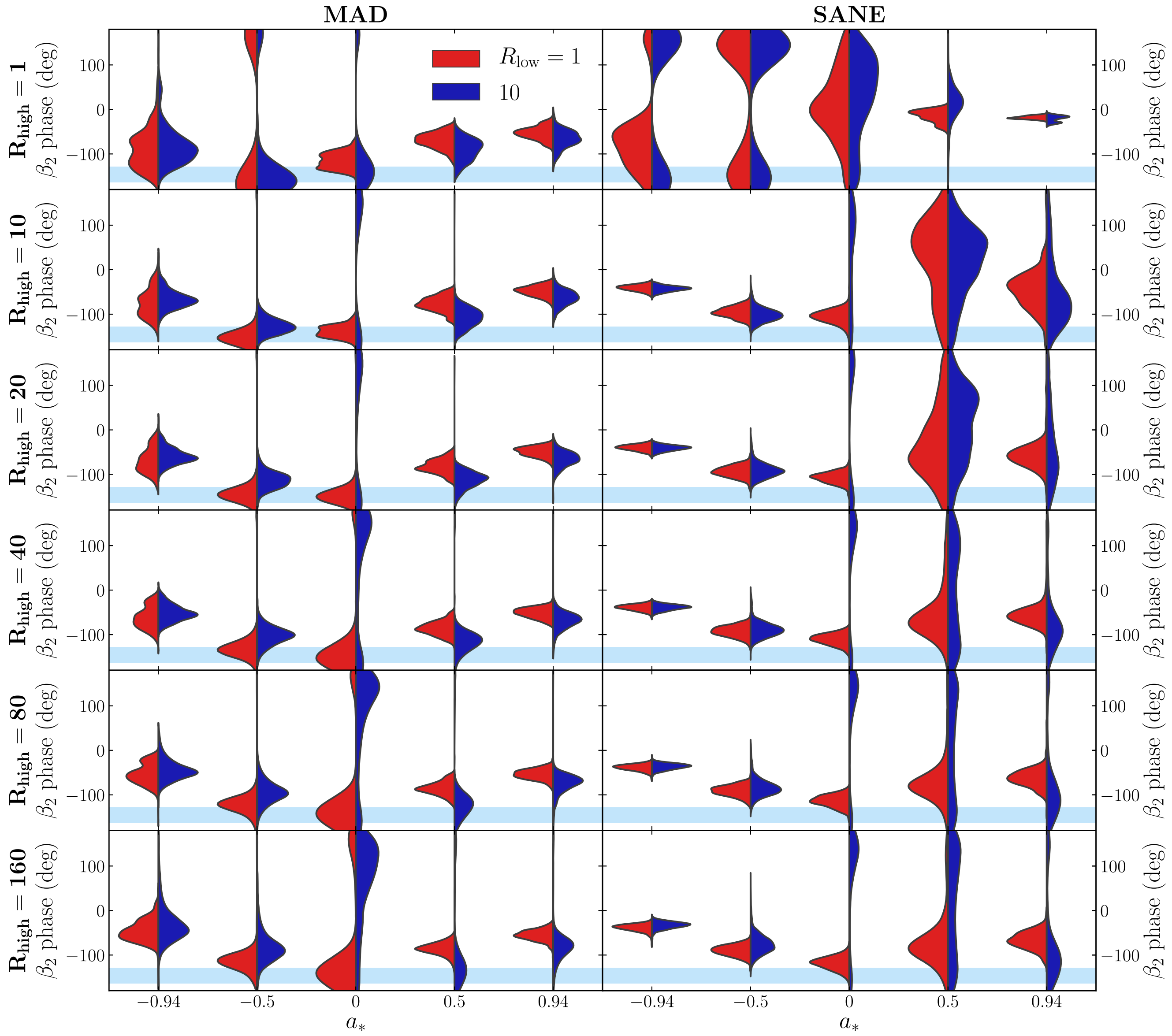

**Figure 28.** Same as in Figure 25, but for the phase of $\beta_2$.

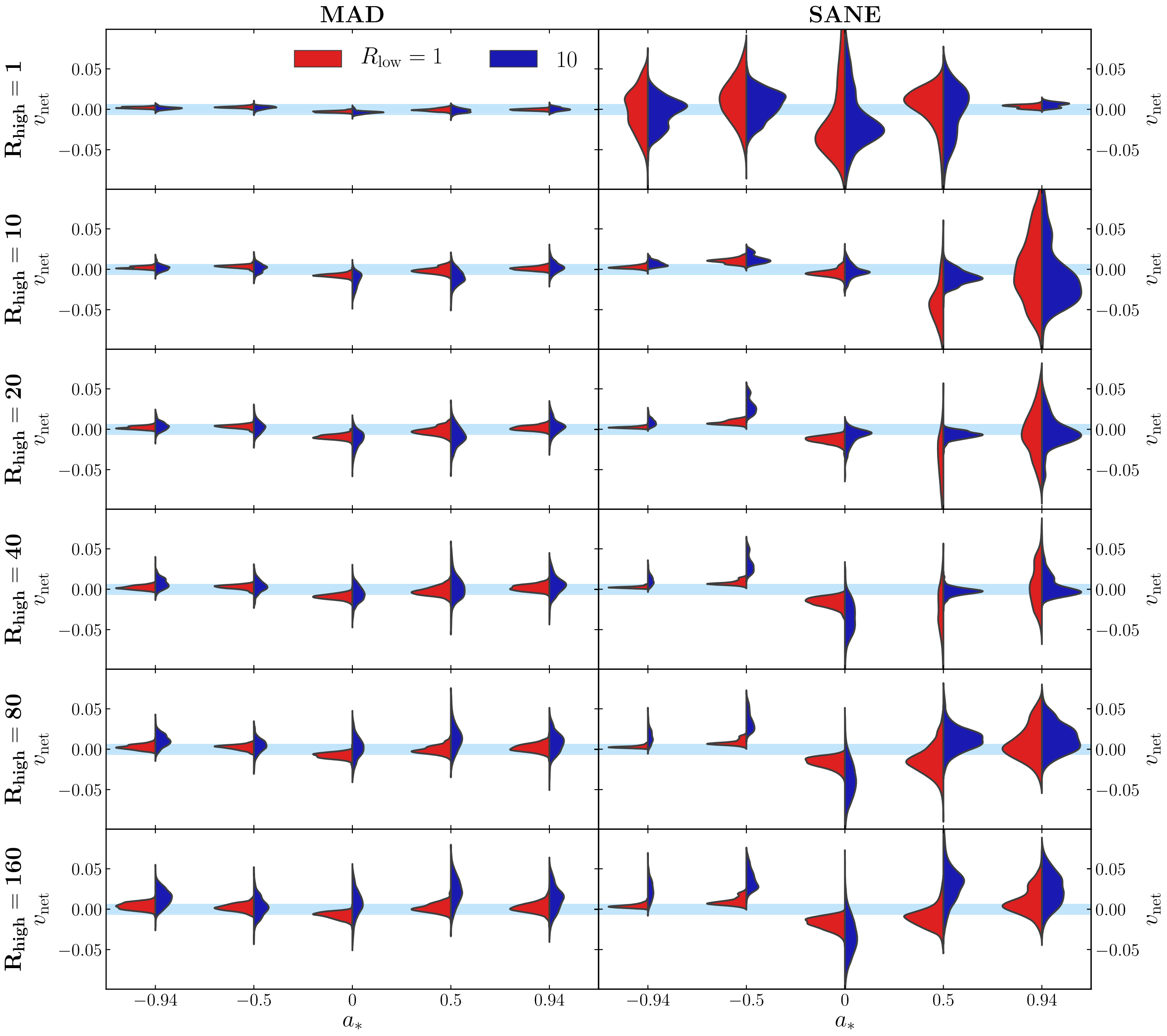

**Figure 29.** Same as in Figure 25, but for the net circular polarization fraction $v_{\rm net}$.

## Appendix D
## Detailed Model Scoring Results

In Table 3 we provide a summary of the number of images for each model that fall within the observed range of each individual theory metric (used in the joint scoring procedure) and within the observed ranges of all metrics simultaneously (used in the simultaneous scoring). Boldfaced type is used for models that are deemed viable by one of the scoring systems. For simultaneous scoring, a viable model contains at least one image that simultaneously satisfies all constraints. For joint scoring, a viable model has a joint likelihood >1% that of the maximum found across all models. We also provide a summary score—"pass" indicates a model that satisfies the polarimetric constraints according to either scoring procedure, as well as the jet-power cut of $P_{\rm jet} > 10^{42}$ erg s$^{-1}$ (EHTC V).

**Table 3**
Scoring Results for the Models Used Here

| Flux | $a_*$ | $R_{\rm low}$ | $R_{\rm high}$ | $\dot{M}_{-3}$ | $P_{\rm jet,42}$ | $N_{m_{\rm net}}$ | $N_{v_{\rm net}}$ | $N_{\langle\|m\|\rangle}$ | $N_{\|\beta_2\|}$ | $N_{\arg\beta_2}$ | $N_{\rm sim}$ | Summary |
|---|---|---|---|---|---|---|---|---|---|---|---|---|
| SANE | −0.94 | 1 | 1 | 3.79 | 1.19 | 7 | 578 | 0 | 0 | 126 | 0 | fail |
| SANE | −0.94 | 1 | 10 | 16.30 | 5.11 | 504 | 95 | 0 | 0 | 0 | 0 | fail |
| SANE | −0.94 | 1 | 20 | 19.04 | 6.00 | 482 | 24 | 0 | 0 | 0 | 0 | fail |
| SANE | −0.94 | 1 | 40 | 22.88 | 7.26 | 460 | 27 | 0 | 0 | 0 | 0 | fail |
| SANE | −0.94 | 1 | 80 | 29.18 | 9.19 | 375 | 31 | 0 | 0 | 0 | 0 | fail |
| SANE | −0.94 | 1 | 160 | 39.32 | 12.40 | 301 | 41 | 0 | 0 | 0 | 0 | fail |
| SANE | −0.94 | 10 | 1 | 6.24 | 1.96 | 0 | 506 | 0 | 0 | 185 | 0 | fail |
| SANE | −0.94 | 10 | 10 | 48.18 | 15.16 | 263 | 26 | 28 | 12 | 0 | 0 | fail |
| SANE | −0.94 | 10 | 20 | 55.54 | 17.47 | 256 | 32 | 27 | 6 | 0 | 0 | fail |
| SANE | −0.94 | 10 | 40 | 66.19 | 20.82 | 256 | 36 | 34 | 7 | 0 | 0 | fail |
| SANE | −0.94 | 10 | 80 | 83.97 | 26.42 | 251 | 30 | 39 | 11 | 0 | 0 | fail |
| SANE | −0.94 | 10 | 160 | 115.94 | 36.47 | 232 | 27 | 49 | 15 | 0 | 0 | fail |
| SANE | −0.5 | 1 | 1 | 2.48 | 0.04 | 0 | 407 | 0 | 0 | 143 | 0 | fail |
| SANE | −0.5 | 1 | 10 | 18.36 | 0.30 | 51 | 0 | 10 | 15 | 0 | 0 | fail |
| SANE | −0.5 | 1 | 20 | 21.37 | 0.35 | 78 | 0 | 12 | 28 | 0 | 0 | fail |
| SANE | −0.5 | 1 | 40 | 24.80 | 0.41 | 61 | 2 | 10 | 16 | 0 | 0 | fail |
| SANE | −0.5 | 1 | 80 | 30.28 | 0.50 | 63 | 13 | 5 | 12 | 0 | 0 | fail |
| SANE | −0.5 | 1 | 160 | 39.73 | 0.66 | 80 | 37 | 5 | 7 | 0 | 0 | fail |
| SANE | −0.5 | 10 | 1 | 3.75 | 0.06 | 0 | 491 | 0 | 0 | 134 | 0 | fail |
| SANE | −0.5 | 10 | 10 | 64.76 | 1.06 | 9 | 3 | 0 | 0 | 13 | 0 | fail |
| SANE | −0.5 | 10 | 20 | 84.43 | 1.39 | 107 | 0 | 0 | 0 | 10 | 0 | fail |
| SANE | −0.5 | 10 | 40 | 92.84 | 1.53 | 95 | 0 | 0 | 0 | 13 | 0 | fail |
| SANE | −0.5 | 10 | 80 | 107.39 | 1.77 | 70 | 0 | 0 | 0 | 9 | 0 | fail |
| SANE | −0.5 | 10 | 160 | 134.03 | 2.20 | 48 | 0 | 0 | 0 | 11 | 0 | fail |
| SANE | 0.0 | 1 | 1 | 0.89 | 0.00 | **555** | **513** | **90** | **96** | **38** | 0 | fail |
| SANE | 0.0 | 1 | 10 | 17.40 | 0.00 | 78 | 174 | 12 | 29 | 11 | 0 | fail |
| SANE | 0.0 | 1 | 20 | 31.92 | 0.00 | 334 | 7 | 45 | 164 | 3 | 0 | fail |
| SANE | 0.0 | 1 | 40 | 36.44 | 0.00 | 356 | 7 | 46 | 159 | 1 | 0 | fail |
| SANE | 0.0 | 1 | 80 | 41.10 | 0.00 | 312 | 8 | 31 | 77 | 5 | **1** | fail |
| SANE | 0.0 | 1 | 160 | 48.91 | 0.00 | 238 | 7 | 13 | 21 | 39 | **1** | fail |
| SANE | 0.0 | 10 | 1 | 1.26 | 0.00 | 28 | 419 | 0 | 0 | 42 | 0 | fail |
| SANE | 0.0 | 10 | 10 | 30.20 | 0.00 | 0 | 386 | 0 | 0 | 53 | 0 | fail |
| SANE | 0.0 | 10 | 20 | 134.87 | 0.00 | 28 | 153 | 0 | 0 | 262 | 0 | fail |
| SANE | 0.0 | 10 | 40 | 211.01 | 0.00 | 73 | 46 | 3 | 3 | 122 | 0 | fail |
| SANE | 0.0 | 10 | 80 | 223.92 | 0.00 | 60 | 56 | 4 | 3 | 98 | 0 | fail |
| SANE | 0.0 | 10 | 160 | 249.08 | 0.00 | 45 | 99 | 3 | 2 | 87 | 0 | fail |
| SANE | 0.5 | 1 | 1 | 0.28 | 0.01 | 74 | 584 | 0 | 0 | 0 | 0 | fail |
| SANE | 0.5 | 1 | 10 | 2.05 | 0.04 | 1 | 69 | 0 | 0 | 23 | 0 | fail |
| SANE | 0.5 | 1 | 20 | 4.38 | 0.08 | 0 | 114 | 0 | 0 | 96 | 0 | fail |
| SANE | 0.5 | 1 | 40 | 8.60 | 0.16 | 0 | 211 | 0 | 0 | 67 | 0 | fail |
| SANE | 0.5 | 1 | 80 | 13.63 | 0.25 | 0 | 245 | 0 | 0 | 53 | 0 | fail |
| SANE | 0.5 | 1 | 160 | 18.22 | 0.33 | 2 | 215 | 0 | 0 | 76 | 0 | fail |
| SANE | 0.5 | 10 | 1 | 0.45 | 0.01 | 203 | 557 | 198 | 189 | 3 | 0 | fail |
| SANE | 0.5 | 10 | 10 | 5.53 | 0.10 | 0 | 185 | 0 | 0 | 22 | 0 | fail |
| SANE | 0.5 | 10 | 20 | 17.96 | 0.33 | 0 | 58 | 0 | 0 | 39 | 0 | fail |
| SANE | 0.5 | 10 | 40 | 56.18 | 1.03 | 0 | 278 | 0 | 0 | 137 | 0 | fail |
| SANE | 0.5 | 10 | 80 | 110.23 | 2.02 | 2 | 204 | 0 | 0 | 125 | 0 | fail |
| SANE | 0.5 | 10 | 160 | 140.88 | 2.58 | 2 | 107 | 0 | 0 | 93 | 0 | fail |
| SANE | 0.94 | 1 | 1 | 0.05 | 0.02 | 165 | 59 | 0 | 0 | 0 | 0 | fail |
| SANE | 0.94 | 1 | 10 | 0.32 | 0.13 | **336** | **594** | **481** | **439** | **21** | 0 | fail |
| SANE | 0.94 | 1 | 20 | 0.73 | 0.30 | 61 | 521 | 0 | 0 | 12 | 0 | fail |
| SANE | 0.94 | 1 | 40 | 1.35 | 0.56 | 45 | 525 | 6 | 35 | 3 | 0 | fail |
| SANE | 0.94 | 1 | 80 | 2.00 | 0.83 | 122 | 420 | 14 | 121 | 1 | 0 | fail |
| SANE | 0.94 | 1 | 160 | 2.78 | 1.20 | 182 | 321 | 13 | 163 | 0 | 0 | fail |
| SANE | 0.94 | 10 | 1 | 0.07 | 0.03 | 181 | 33 | 0 | 0 | 0 | 0 | fail |
| SANE | 0.94 | 10 | 10 | 0.49 | 0.20 | **582** | **469** | **113** | **107** | **116** | 0 | fail |
| SANE | 0.94 | 10 | 20 | 1.57 | 0.65 | 0 | 427 | 0 | 0 | 137 | 0 | fail |
| SANE | 0.94 | 10 | 40 | 5.86 | 2.44 | 0 | 493 | 0 | 0 | 122 | 0 | fail |
| SANE | 0.94 | 10 | 80 | 12.76 | 5.31 | 0 | 272 | 0 | 0 | 178 | 0 | fail |
| SANE | 0.94 | 10 | 160 | 17.71 | 7.37 | 1 | 144 | 0 | 0 | 205 | 0 | fail |
| MAD | −0.94 | 1 | 1 | 0.13 | 1.74 | 118 | 23 | 2 | 41 | 109 | 0 | fail |
| MAD | −0.94 | 1 | 10 | 0.20 | 2.60 | 112 | 116 | 0 | 3 | 8 | 0 | fail |
| MAD | −0.94 | 1 | 20 | 0.24 | 3.15 | 94 | 199 | 0 | 0 | 0 | 0 | fail |
| MAD | −0.94 | 1 | 40 | 0.31 | 3.98 | 92 | 256 | 0 | 7 | 0 | 0 | fail |
| MAD | −0.94 | 1 | 80 | 0.41 | 5.31 | 134 | 261 | 4 | 51 | 1 | 0 | fail |

**Table 3**
(Continued)

| Flux | $a_*$ | $R_{low}$ | $R_{high}$ | $\dot{M}_{-3}$ | $P_{jet,42}$ | $N_{m_{net}}$ | $N_{v_{net}}$ | $N_{\langle\|m\|\rangle}$ | $N_{\|\beta_2\|}$ | $N_{\arg\beta_2}$ | $N_{sim}$ | Summary |
|---|---|---|---|---|---|---|---|---|---|---|---|---|
| MAD | −0.94 | 1 | 160 | 0.58 | 7.57 | 160 | 253 | 59 | 134 | 1 | 0 | fail |
| MAD | −0.94 | 10 | 1 | 0.29 | 3.75 | 117 | 239 | 15 | 134 | 48 | 0 | fail |
| MAD | −0.94 | 10 | 10 | 0.51 | 6.66 | 113 | 392 | 2 | 27 | 4 | 0 | fail |
| MAD | −0.94 | 10 | 20 | 0.68 | 8.78 | 131 | 310 | 10 | 33 | 0 | 0 | fail |
| MAD | −0.94 | 10 | 40 | 0.93 | 12.12 | 157 | 166 | 43 | 68 | 0 | 0 | fail |
| MAD | −0.94 | 10 | 80 | 1.37 | 17.76 | 180 | 105 | 105 | 134 | 0 | 0 | fail |
| MAD | −0.94 | 10 | 160 | 2.12 | 27.45 | **226** | **118** | **247** | **304** | **20** | 0 | **pass** |
| MAD | −0.5 | 1 | 1 | 0.12 | 0.53 | 148 | 27 | 0 | 4 | 585 | 0 | fail |
| MAD | −0.5 | 1 | 10 | 0.19 | 0.82 | 147 | 53 | 0 | 0 | 587 | 0 | fail |
| MAD | −0.5 | 1 | 20 | 0.23 | 1.00 | 121 | 87 | 0 | 0 | 374 | 0 | fail |
| MAD | −0.5 | 1 | 40 | 0.29 | 1.25 | 126 | 157 | 1 | 11 | 139 | 0 | fail |
| MAD | −0.5 | 1 | 80 | 0.38 | 1.63 | 168 | 256 | 26 | 85 | 89 | **9** | **pass** |
| MAD | −0.5 | 1 | 160 | 0.52 | 2.26 | **229** | **377** | **185** | **260** | **71** | **19** | **pass** |
| MAD | −0.5 | 10 | 1 | 0.27 | 1.15 | 128 | 213 | 0 | 40 | 529 | **1** | **pass** |
| MAD | −0.5 | 10 | 10 | 0.52 | 2.25 | 62 | 471 | 1 | 10 | 141 | **5** | **pass** |
| MAD | −0.5 | 10 | 20 | 0.71 | 3.08 | 71 | 447 | 21 | 64 | 53 | **1** | **pass** |
| MAD | −0.5 | 10 | 40 | 0.99 | 4.28 | **157** | **366** | **215** | **224** | **31** | **1** | **pass** |
| MAD | −0.5 | 10 | 80 | 1.40 | 6.06 | **270** | **368** | **475** | **454** | **35** | 0 | **pass** |
| MAD | −0.5 | 10 | 160 | 2.05 | 8.87 | **438** | **462** | **324** | **352** | **48** | 0 | **pass** |
| MAD | 0.0 | 1 | 1 | 0.10 | 0.00 | 140 | 7 | 0 | 0 | 20 | 0 | fail |
| MAD | 0.0 | 1 | 10 | 0.16 | 0.00 | 146 | 3 | 0 | 0 | 405 | 0 | fail |
| MAD | 0.0 | 1 | 20 | 0.20 | 0.00 | 149 | 2 | 0 | 0 | 596 | 0 | fail |
| MAD | 0.0 | 1 | 40 | 0.25 | 0.00 | 173 | 5 | 0 | 1 | 406 | 0 | fail |
| MAD | 0.0 | 1 | 80 | 0.32 | 0.00 | 297 | 11 | 58 | 35 | 310 | **5** | fail |
| MAD | 0.0 | 1 | 160 | 0.43 | 0.00 | **432** | **34** | **300** | **168** | **288** | **19** | fail |
| MAD | 0.0 | 10 | 1 | 0.19 | 0.00 | 147 | 46 | 3 | 7 | 303 | 0 | fail |
| MAD | 0.0 | 10 | 10 | 0.41 | 0.00 | 116 | 77 | 19 | 85 | 218 | 0 | fail |
| MAD | 0.0 | 10 | 20 | 0.59 | 0.00 | **145** | **135** | **169** | **299** | **276** | 0 | fail |
| MAD | 0.0 | 10 | 40 | 0.85 | 0.01 | **245** | **323** | **445** | **461** | **135** | 0 | fail |
| MAD | 0.0 | 10 | 80 | 1.23 | 0.01 | **410** | **542** | **456** | **422** | **64** | **1** | fail |
| MAD | 0.0 | 10 | 160 | 1.80 | 0.01 | 490 | 371 | 18 | 50 | 29 | 0 | fail |
| MAD | 0.5 | 1 | 1 | 0.07 | 0.70 | 198 | 367 | 0 | 0 | 0 | 0 | fail |
| MAD | 0.5 | 1 | 10 | 0.12 | 1.22 | 190 | 407 | 0 | 0 | 0 | 0 | fail |
| MAD | 0.5 | 1 | 20 | 0.15 | 1.55 | 163 | 371 | 0 | 0 | 0 | 0 | fail |
| MAD | 0.5 | 1 | 40 | 0.20 | 2.00 | 176 | 400 | 0 | 0 | 0 | 0 | fail |
| MAD | 0.5 | 1 | 80 | 0.26 | 2.63 | 197 | 548 | 33 | 10 | 0 | 0 | fail |
| MAD | 0.5 | 1 | 160 | 0.36 | 3.60 | 290 | 445 | 72 | 20 | 0 | 0 | fail |
| MAD | 0.5 | 10 | 1 | 0.13 | 1.33 | 162 | 320 | 0 | 0 | 10 | 0 | fail |
| MAD | 0.5 | 10 | 10 | 0.30 | 3.02 | 132 | 182 | 9 | 19 | 39 | 0 | fail |
| MAD | 0.5 | 10 | 20 | 0.45 | 4.49 | **152** | **287** | **138** | **148** | **48** | 0 | **pass** |
| MAD | 0.5 | 10 | 40 | 0.67 | 6.74 | **229** | **533** | **347** | **318** | **60** | **4** | **pass** |
| MAD | 0.5 | 10 | 80 | 1.02 | 10.20 | **303** | **221** | **556** | **479** | **125** | **4** | **pass** |
| MAD | 0.5 | 10 | 160 | 1.57 | 15.72 | **528** | **120** | **315** | **462** | **230** | **1** | **pass** |
| MAD | 0.94 | 1 | 1 | 0.04 | 1.96 | 199 | 401 | 0 | 0 | 0 | 0 | fail |
| MAD | 0.94 | 1 | 10 | 0.06 | 3.02 | 176 | 408 | 0 | 0 | 0 | 0 | fail |
| MAD | 0.94 | 1 | 20 | 0.08 | 3.72 | 157 | 375 | 0 | 0 | 0 | 0 | fail |
| MAD | 0.94 | 1 | 40 | 0.10 | 4.72 | 143 | 413 | 0 | 0 | 0 | 0 | fail |
| MAD | 0.94 | 1 | 80 | 0.13 | 6.24 | 142 | 432 | 0 | 0 | 0 | 0 | fail |
| MAD | 0.94 | 1 | 160 | 0.19 | 8.72 | 144 | 433 | 4 | 0 | 0 | 0 | fail |
| MAD | 0.94 | 10 | 1 | 0.08 | 3.75 | 235 | 575 | 0 | 0 | 0 | 0 | fail |
| MAD | 0.94 | 10 | 10 | 0.15 | 7.06 | 145 | 392 | 1 | 7 | 0 | 0 | fail |
| MAD | 0.94 | 10 | 20 | 0.21 | 9.74 | 132 | 381 | 33 | 57 | 1 | 0 | fail |
| MAD | 0.94 | 10 | 40 | 0.30 | 14.12 | 168 | 362 | 129 | 162 | 1 | 0 | fail |
| MAD | 0.94 | 10 | 80 | 0.45 | 21.18 | **298** | **325** | **344** | **283** | **8** | **1** | **pass** |
| MAD | 0.94 | 10 | 160 | 0.72 | 33.45 | **431** | **261** | **598** | **488** | **35** | 0 | **pass** |

**Note.** The number of images passing each polarimetric constraint are given along with the number $N_{sim}$ simultaneously passing all of them. The accretion rate $\dot{M}_{-3}$ is in units of $10^{-3}\ M_\odot\ \mathrm{yr}^{-1}$, and $P_{jet,42}$ is the jet power in units of $10^{42}\ \mathrm{erg\ s}^{-1}$. Models that pass according to either the simultaneous or joint scoring method (boldface) and have $P_{jet,42} \geqslant 1$ are given a summary score of pass.

## ORCID iDs

Kazunori Akiyama https://orcid.org/0000-0002-9475-4254
Juan Carlos Algaba https://orcid.org/0000-0001-6993-1696
Antxon Alberdi https://orcid.org/0000-0002-9371-1033
Richard Anantua https://orcid.org/0000-0003-3457-7660

Rebecca Azulay https://orcid.org/0000-0002-2200-5393
Anne-Kathrin Baczko https://orcid.org/0000-0003-3090-3975
Mislav Baloković https://orcid.org/0000-0003-0476-6647
John Barrett https://orcid.org/0000-0002-9290-0764
Bradford A. Benson https://orcid.org/0000-0002-5108-6823
Lindy Blackburn https://orcid.org/0000-0002-9030-642X
Raymond Blundell https://orcid.org/0000-0002-5929-5857
Katherine L. Bouman https://orcid.org/0000-0003-0077-4367
Geoffrey C. Bower https://orcid.org/0000-0003-4056-9982
Hope Boyce https://orcid.org/0000-0002-6530-5783
Christiaan D. Brinkerink https://orcid.org/0000-0002-2322-0749
Roger Brissenden https://orcid.org/0000-0002-2556-0894
Silke Britzen https://orcid.org/0000-0001-9240-6734
Avery E. Broderick https://orcid.org/0000-0002-3351-760X
Do-Young Byun https://orcid.org/0000-0003-1157-4109
Andrew Chael https://orcid.org/0000-0003-2966-6220
Chi-kwan Chan https://orcid.org/0000-0001-6337-6126
Shami Chatterjee https://orcid.org/0000-0002-2878-1502
Koushik Chatterjee https://orcid.org/0000-0002-2825-3590
Paul M. Chesler https://orcid.org/0000-0001-6327-8462
Ilje Cho https://orcid.org/0000-0001-6083-7521
Pierre Christian https://orcid.org/0000-0001-6820-9941
John E. Conway https://orcid.org/0000-0003-2448-9181
James M. Cordes https://orcid.org/0000-0002-4049-1882
Thomas M. Crawford https://orcid.org/0000-0001-9000-5013
Geoffrey B. Crew https://orcid.org/0000-0002-2079-3189
Alejandro Cruz-Osorio https://orcid.org/0000-0002-3945-6342
Yuzhu Cui https://orcid.org/0000-0001-6311-4345
Jordy Davelaar https://orcid.org/0000-0002-2685-2434
Mariafelicia De Laurentis https://orcid.org/0000-0002-9945-682X
Roger Deane https://orcid.org/0000-0003-1027-5043
Jessica Dempsey https://orcid.org/0000-0003-1269-9667
Gregory Desvignes https://orcid.org/0000-0003-3922-4055
Jason Dexter https://orcid.org/0000-0003-3903-0373
Sheperd S. Doeleman https://orcid.org/0000-0002-9031-0904
Ralph P. Eatough https://orcid.org/0000-0001-6196-4135
Heino Falcke https://orcid.org/0000-0002-2526-6724
Joseph Farah https://orcid.org/0000-0003-4914-5625
Vincent L. Fish https://orcid.org/0000-0002-7128-9345
Ed Fomalont https://orcid.org/0000-0002-9036-2747
H. Alyson Ford https://orcid.org/0000-0002-9797-0972
Raquel Fraga-Encinas https://orcid.org/0000-0002-5222-1361
Per Friberg https://orcid.org/0000-0002-8010-8454
Antonio Fuentes https://orcid.org/0000-0002-8773-4933
Peter Galison https://orcid.org/0000-0002-6429-3872
Charles F. Gammie https://orcid.org/0000-0001-7451-8935
Roberto García https://orcid.org/0000-0002-6584-7443
Zachary Gelles https://orcid.org/0000-0001-8053-4392
Boris Georgiev https://orcid.org/0000-0002-3586-6424
Ciriaco Goddi https://orcid.org/0000-0002-2542-7743
Roman Gold https://orcid.org/0000-0003-2492-1966
José L. Gómez https://orcid.org/0000-0003-4190-7613
Arturo I. Gómez-Ruiz https://orcid.org/0000-0001-9395-1670
Minfeng Gu (顾敏峰) https://orcid.org/0000-0002-4455-6946
Mark Gurwell https://orcid.org/0000-0003-0685-3621
Kazuhiro Hada https://orcid.org/0000-0001-6906-772X
Daryl Haggard https://orcid.org/0000-0001-6803-2138
Ronald Hesper https://orcid.org/0000-0003-1918-6098
Luis C. Ho (何子山) https://orcid.org/0000-0001-6947-5846
Mareki Honma https://orcid.org/0000-0003-4058-9000
Chih-Wei L. Huang https://orcid.org/0000-0001-5641-3953
Lei Huang (黄磊) https://orcid.org/0000-0002-1923-227X
Shiro Ikeda https://orcid.org/0000-0002-2462-1448
Sara Issaoun https://orcid.org/0000-0002-5297-921X
David J. James https://orcid.org/0000-0001-5160-4486
Michael Janssen https://orcid.org/0000-0001-8685-6544
Britton Jeter https://orcid.org/0000-0003-2847-1712
Wu Jiang (江悟) https://orcid.org/0000-0001-7369-3539
Alejandra Jimenez-Rosales https://orcid.org/0000-0002-2662-3754
Michael D. Johnson https://orcid.org/0000-0002-4120-3029
Svetlana Jorstad https://orcid.org/0000-0001-6158-1708
Taehyun Jung https://orcid.org/0000-0001-7003-8643
Mansour Karami https://orcid.org/0000-0001-7387-9333
Ramesh Karuppusamy https://orcid.org/0000-0002-5307-2919
Tomohisa Kawashima https://orcid.org/0000-0001-8527-0496
Garrett K. Keating https://orcid.org/0000-0002-3490-146X
Mark Kettenis https://orcid.org/0000-0002-6156-5617
Dong-Jin Kim https://orcid.org/0000-0002-7038-2118
Jae-Young Kim https://orcid.org/0000-0001-8229-7183
Jongsoo Kim https://orcid.org/0000-0002-1229-0426
Junhan Kim https://orcid.org/0000-0002-4274-9373
Motoki Kino https://orcid.org/0000-0002-2709-7338
Jun Yi Koay https://orcid.org/0000-0002-7029-6658
Patrick M. Koch https://orcid.org/0000-0003-2777-5861
Shoko Koyama https://orcid.org/0000-0002-3723-3372
Michael Kramer https://orcid.org/0000-0002-4175-2271
Carsten Kramer https://orcid.org/0000-0002-4908-4925
Thomas P. Krichbaum https://orcid.org/0000-0002-4892-9586
Cheng-Yu Kuo https://orcid.org/0000-0001-6211-5581
Tod R. Lauer https://orcid.org/0000-0003-3234-7247
Sang-Sung Lee https://orcid.org/0000-0002-6269-594X
Aviad Levis https://orcid.org/0000-0001-7307-632X
Yan-Rong Li (李彦荣) https://orcid.org/0000-0001-5841-9179
Zhiyuan Li (李志远) https://orcid.org/0000-0003-0355-6437
Michael Lindqvist https://orcid.org/0000-0002-3669-0715
Rocco Lico https://orcid.org/0000-0001-7361-2460
Greg Lindahl https://orcid.org/0000-0002-6100-4772
Jun Liu (刘俊) https://orcid.org/0000-0002-7615-7499
Kuo Liu https://orcid.org/0000-0002-2953-7376
Elisabetta Liuzzo https://orcid.org/0000-0003-0995-5201
Laurent Loinard https://orcid.org/0000-0002-5635-3345
Ru-Sen Lu (路如森) https://orcid.org/0000-0002-7692-7967
Nicholas R. MacDonald https://orcid.org/0000-0002-6684-8691
Jirong Mao (毛基荣) https://orcid.org/0000-0002-7077-7195
Nicola Marchili https://orcid.org/0000-0002-5523-7588
Sera Markoff https://orcid.org/0000-0001-9564-0876
Daniel P. Marrone https://orcid.org/0000-0002-2367-1080
Alan P. Marscher https://orcid.org/0000-0001-7396-3332

Iván Martí-Vidal https://orcid.org/0000-0003-3708-9611
Satoki Matsushita https://orcid.org/0000-0002-2127-7880
Lynn D. Matthews https://orcid.org/0000-0002-3728-8082
Lia Medeiros https://orcid.org/0000-0003-2342-6728
Karl M. Menten https://orcid.org/0000-0001-6459-0669
Izumi Mizuno https://orcid.org/0000-0002-7210-6264
Yosuke Mizuno https://orcid.org/0000-0002-8131-6730
James M. Moran https://orcid.org/0000-0002-3882-4414
Kotaro Moriyama https://orcid.org/0000-0003-1364-3761
Monika Moscibrodzka https://orcid.org/0000-0002-4661-6332
Cornelia Müller https://orcid.org/0000-0002-2739-2994
Gibwa Musoke https://orcid.org/0000-0003-1984-189X
Alejandro Mus Mejías https://orcid.org/0000-0003-0329-6874
Daniel Michalik https://orcid.org/0000-0002-7618-6556
Andrew Nadolski https://orcid.org/0000-0001-9479-9957
Hiroshi Nagai https://orcid.org/0000-0003-0292-3645
Neil M. Nagar https://orcid.org/0000-0001-6920-662X
Masanori Nakamura https://orcid.org/0000-0001-6081-2420
Ramesh Narayan https://orcid.org/0000-0002-1919-2730
Iniyan Natarajan https://orcid.org/0000-0001-8242-4373
Joey Neilsen https://orcid.org/0000-0002-8247-786X
Roberto Neri https://orcid.org/0000-0002-7176-4046
Chunchong Ni https://orcid.org/0000-0003-1361-5699
Aristeidis Noutsos https://orcid.org/0000-0002-4151-3860
Michael A. Nowak https://orcid.org/0000-0001-6923-1315
Héctor Olivares https://orcid.org/0000-0001-6833-7580
Gisela N. Ortiz-León https://orcid.org/0000-0002-2863-676X
Daniel C. M. Palumbo https://orcid.org/0000-0002-7179-3816
Jongho Park https://orcid.org/0000-0001-6558-9053
Ue-Li Pen https://orcid.org/0000-0003-2155-9578
Dominic W. Pesce https://orcid.org/0000-0002-5278-9221
Richard Plambeck https://orcid.org/0000-0001-6765-9609
Oliver Porth https://orcid.org/0000-0002-4584-2557
Felix M. Pötzl https://orcid.org/0000-0002-6579-8311
Ben Prather https://orcid.org/0000-0002-0393-7734
Jorge A. Preciado-López https://orcid.org/0000-0002-4146-0113
Dimitrios Psaltis https://orcid.org/0000-0003-4058-2837
Hung-Yi Pu https://orcid.org/0000-0001-9270-8812
Venkatessh Ramakrishnan https://orcid.org/0000-0002-9248-086X
Ramprasad Rao https://orcid.org/0000-0002-1407-7944
Mark G. Rawlings https://orcid.org/0000-0002-6529-202X
Alexander W. Raymond https://orcid.org/0000-0002-5779-4767
Luciano Rezzolla https://orcid.org/0000-0002-1330-7103
Angelo Ricarte https://orcid.org/0000-0001-5287-0452
Bart Ripperda https://orcid.org/0000-0002-7301-3908
Freek Roelofs https://orcid.org/0000-0001-5461-3687
Eduardo Ros https://orcid.org/0000-0001-9503-4892
Mel Rose https://orcid.org/0000-0002-2016-8746
Alan L. Roy https://orcid.org/0000-0002-1931-0135
Chet Ruszczyk https://orcid.org/0000-0001-7278-9707
Kazi L. J. Rygl https://orcid.org/0000-0003-4146-9043
David Sánchez-Arguelles https://orcid.org/0000-0002-7344-9920
Mahito Sasada https://orcid.org/0000-0001-5946-9960
Tuomas Savolainen https://orcid.org/0000-0001-6214-1085
Lijing Shao https://orcid.org/0000-0002-1334-8853
Zhiqiang Shen (沈志强) https://orcid.org/0000-0003-3540-8746
Des Small https://orcid.org/0000-0003-3723-5404
Bong Won Sohn https://orcid.org/0000-0002-4148-8378
Jason SooHoo https://orcid.org/0000-0003-1938-0720
He Sun (孙赫) https://orcid.org/0000-0003-1526-6787
Fumie Tazaki https://orcid.org/0000-0003-0236-0600
Alexandra J. Tetarenko https://orcid.org/0000-0003-3906-4354
Paul Tiede https://orcid.org/0000-0003-3826-5648
Remo P. J. Tilanus https://orcid.org/0000-0002-6514-553X
Michael Titus https://orcid.org/0000-0002-3423-4505
Kenji Toma https://orcid.org/0000-0002-7114-6010
Pablo Torne https://orcid.org/0000-0001-8700-6058
Efthalia Traianou https://orcid.org/0000-0002-1209-6500
Sascha Trippe https://orcid.org/0000-0003-0465-1559
Ilse van Bemmel https://orcid.org/0000-0001-5473-2950
Huib Jan van Langevelde https://orcid.org/0000-0002-0230-5946
Daniel R. van Rossum https://orcid.org/0000-0001-7772-6131
Jan Wagner https://orcid.org/0000-0003-1105-6109
Derek Ward-Thompson https://orcid.org/0000-0003-1140-2761
John Wardle https://orcid.org/0000-0002-8960-2942
Jonathan Weintroub https://orcid.org/0000-0002-4603-5204
Robert Wharton https://orcid.org/0000-0002-7416-5209
Maciek Wielgus https://orcid.org/0000-0002-8635-4242
George N. Wong https://orcid.org/0000-0001-6952-2147
Qingwen Wu (吴庆文) https://orcid.org/0000-0003-4773-4987
Doosoo Yoon https://orcid.org/0000-0001-8694-8166
André Young https://orcid.org/0000-0003-0000-2682
Ken Young https://orcid.org/0000-0002-3666-4920
Ziri Younsi https://orcid.org/0000-0001-9283-1191
Feng Yuan (袁峰) https://orcid.org/0000-0003-3564-6437
J. Anton Zensus https://orcid.org/0000-0001-7470-3321
Guang-Yao Zhao https://orcid.org/0000-0002-4417-1659
Shan-Shan Zhao https://orcid.org/0000-0002-9774-3606

---

Kazunori Akiyama[1,2,3], Juan Carlos Algaba[4], Antxon Alberdi[5], Walter Alef[6], Richard Anantua[3,7,8], Keiichi Asada[9], Rebecca Azulay[6,10,11], Anne-Kathrin Baczko[6], David Ball[12], Mislav Baloković[13,14], John Barrett[1], Bradford A. Benson[15,16], Dan Bintley[17], Lindy Blackburn[3,7], Raymond Blundell[7], Wilfred Boland[18], Katherine L. Bouman[3,7,19], Geoffrey C. Bower[20], Hope Boyce[21,22], Michael Bremer[23], Christiaan D. Brinkerink[24], Roger Brissenden[3,7], Silke Britzen[6], Avery E. Broderick[25,26,27], Dominique Broguiere[23], Thomas Bronzwaer[24], Do-Young Byun[28,29], John E. Carlstrom[16,30,31,32], Andrew Chael[33,126], Chi-kwan Chan[12,34], Shami Chatterjee[35], Koushik Chatterjee[36], Ming-Tang Chen[20], Yongjun Chen (陈永军)[37,38], Paul M. Chesler[3], Ilje Cho[28,29], Pierre Christian[39], John E. Conway[40], James M. Cordes[35], Thomas M. Crawford[16,30], Geoffrey B. Crew[1], Alejandro Cruz-Osorio[41], Yuzhu Cui[42,43], Jordy Davelaar[8,24,44], Mariafelicia De Laurentis[41,45,46], Roger Deane[47,48,49], Jessica Dempsey[17], Gregory Desvignes[50], Jason Dexter[51], Sheperd S. Doeleman[3,7], Ralph P. Eatough[6,52], Heino Falcke[24], Joseph Farah[3,7,53], Vincent L. Fish[1], Ed Fomalont[54], H. Alyson Ford[12], Raquel Fraga-Encinas[24], Per Friberg[17], Christian M. Fromm[3,7,41], Antonio Fuentes[5], Peter Galison[3,55,56], Charles F. Gammie[57,58], Roberto García[23], Zachary Gelles[3,7], Olivier Gentaz[23], Boris Georgiev[26,27], Ciriaco Goddi[24,59], Roman Gold[25,60], José L. Gómez[5], Arturo I. Gómez-Ruiz[61,62], Minfeng Gu (顾敏峰)[37,63], Mark Gurwell[7], Kazuhiro Hada[42,43],

Daryl Haggard[21,22], Michael H. Hecht[1], Ronald Hesper[64], Elizabeth Himwich[3,65], Luis C. Ho (何子山)[66,67], Paul Ho[9], Mareki Honma[42,43,68], Chih-Wei L. Huang[9], Lei Huang (黄磊)[37,63], David H. Hughes[61], Shiro Ikeda[2,69,70,71], Makoto Inoue[9], Sara Issaoun[24], David J. James[72], Buell T. Jannuzi[12], Michael Janssen[6], Britton Jeter[26,27], Wu Jiang (江悟)[38], Alejandra Jimenez-Rosales[24], Michael D. Johnson[3,7], Svetlana Jorstad[73,74], Taehyun Jung[28,29], Mansour Karami[25,26], Ramesh Karuppusamy[6], Tomohisa Kawashima[75], Garrett K. Keating[7], Mark Kettenis[76], Dong-Jin Kim[6], Jae-Young Kim[6,28], Jongsoo Kim[28], Junhan Kim[12,19], Motoki Kino[2,77], Jun Yi Koay[9], Yutaro Kofuji[42,68], Patrick M. Koch[9], Shoko Koyama[9], Michael Kramer[6], Carsten Kramer[23], Thomas P. Krichbaum[6], Cheng-Yu Kuo[78,9], Tod R. Lauer[79], Sang-Sung Lee[28], Aviad Levis[19], Yan-Rong Li (李彦荣)[80], Zhiyuan Li (李志远)[81,82], Michael Lindqvist[40], Rocco Lico[5,6], Greg Lindahl[7], Jun Liu (刘俊)[6], Kuo Liu[6], Elisabetta Liuzzo[83], Wen-Ping Lo[9,84], Andrei P. Lobanov[6], Laurent Loinard[85,86], Colin Lonsdale[1], Ru-Sen Lu (路如森)[37,38,6], Nicholas R. MacDonald[6], Jirong Mao (毛基荣)[87,88,89], Nicola Marchili[6,83], Sera Markoff[36,90], Daniel P. Marrone[12], Alan P. Marscher[73], Iván Martí-Vidal[10,11], Satoki Matsushita[9], Lynn D. Matthews[1], Lia Medeiros[12,91], Karl M. Menten[6], Izumi Mizuno[17], Yosuke Mizuno[41,92], James M. Moran[3,7], Kotaro Moriyama[1,42], Monika Moscibrodzka[24], Cornelia Müller[6,24], Gibwa Musoke[24,36], Alejandro Mus Mejías[10,11], Daniel Michalik[93,94], Andrew Nadolski[58], Hiroshi Nagai[2,43], Neil M. Nagar[95], Masanori Nakamura[9,96], Ramesh Narayan[3,7], Gopal Narayanan[97], Iniyan Natarajan[47,49,98], Antonios Nathanail[41,99], Joey Neilsen[100], Roberto Neri[23], Chunchong Ni[26,27], Aristeidis Noutsos[6], Michael A. Nowak[101], Hiroki Okino[42,68], Héctor Olivares[24], Gisela N. Ortiz-León[6], Tomoaki Oyama[42], Feryal Özel[12], Daniel C. M. Palumbo[3,7], Jongho Park[9,127], Nimesh Patel[7], Ue-Li Pen[25,102,103,104], Dominic W. Pesce[3,7], Vincent Piétu[23], Richard Plambeck[105], Aleksandar PopStefanija[97], Oliver Porth[36,41], Felix M. Pötzl[6], Ben Prather[57], Jorge A. Preciado-López[25], Dimitrios Psaltis[12], Hung-Yi Pu[9,25,106], Venkatessh Ramakrishnan[95], Ramprasad Rao[20], Mark G. Rawlings[17], Alexander W. Raymond[3,7], Luciano Rezzolla[41,107,108], Angelo Ricarte[3,7], Bart Ripperda[8,109], Freek Roelofs[24], Alan Rogers[1], Eduardo Ros[6], Mel Rose[12], Arash Roshanineshat[12], Helge Rottmann[6], Alan L. Roy[6], Chet Ruszczyk[1], Kazi L. J. Rygl[83], Salvador Sánchez[110], David Sánchez-Arguelles[61,62], Mahito Sasada[42,111], Tuomas Savolainen[6,112,113], F. Peter Schloerb[97], Karl-Friedrich Schuster[23], Lijing Shao[6,67], Zhiqiang Shen (沈志强)[37,38], Des Small[76], Bong Won Sohn[28,29,114], Jason SooHoo[1], He Sun (孙赫)[19], Fumie Tazaki[42], Alexandra J. Tetarenko[17], Paul Tiede[26,27], Remo P. J. Tilanus[12,24,59,115], Michael Titus[1], Kenji Toma[116,117], Pablo Torne[6,110], Tyler Trent[12], Efthalia Traianou[6], Sascha Trippe[118], Ilse van Bemmel[76], Huib Jan van Langevelde[76,119], Daniel R. van Rossum[24], Jan Wagner[6], Derek Ward-Thompson[120], John Wardle[121], Jonathan Weintroub[3,7], Norbert Wex[6], Robert Wharton[6], Maciek Wielgus[3,7], George N. Wong[57], Qingwen Wu (吴庆文)[122], Doosoo Yoon[36], André Young[24], Ken Young[7], Ziri Younsi[41,123,128], Feng Yuan (袁峰)[37,63,124], Ye-Fei Yuan (袁业飞)[125], J. Anton Zensus[6], Guang-Yao Zhao[5], and Shan-Shan Zhao[37]

The Event Horizon Telescope Collaboration

[1] Massachusetts Institute of Technology Haystack Observatory, 99 Millstone Road, Westford, MA 01886, USA
[2] National Astronomical Observatory of Japan, 2-21-1 Osawa, Mitaka, Tokyo 181-8588, Japan
[3] Black Hole Initiative at Harvard University, 20 Garden Street, Cambridge, MA 02138, USA
[4] Department of Physics, Faculty of Science, University of Malaya, 50603 Kuala Lumpur, Malaysia
[5] Instituto de Astrofísica de Andalucía-CSIC, Glorieta de la Astronomía s/n, E-18008 Granada, Spain
[6] Max-Planck-Institut für Radioastronomie, Auf dem Hügel 69, D-53121 Bonn, Germany
[7] Center for Astrophysics|Harvard & Smithsonian, 60 Garden Street, Cambridge, MA 02138, USA
[8] Center for Computational Astrophysics, Flatiron Institute, 162 Fifth Avenue, New York, NY 10010, USA
[9] Institute of Astronomy and Astrophysics, Academia Sinica, 11F of Astronomy-Mathematics Building, AS/NTU No. 1, Section 4, Roosevelt Rd, Taipei 10617, Taiwan, R.O.C.
[10] Departament d'Astronomia i Astrofísica, Universitat de València, C. Dr. Moliner 50, E-46100 Burjassot, València, Spain
[11] Observatori Astronòmic, Universitat de València, C. Catedrático José Beltrán 2, E-46980 Paterna, València, Spain
[12] Steward Observatory and Department of Astronomy, University of Arizona, 933 North Cherry Avenue, Tucson, AZ 85721, USA
[13] Yale Center for Astronomy & Astrophysics, 52 Hillhouse Avenue, New Haven, CT 06511, USA
[14] Department of Physics, Yale University, P.O. Box 2018120, New Haven, CT 06520, USA
[15] Fermi National Accelerator Laboratory, MS209, P.O. Box 500, Batavia, IL, 60510, USA
[16] Department of Astronomy and Astrophysics, University of Chicago, 5640 South Ellis Avenue, Chicago, IL 60637, USA
[17] East Asian Observatory, 660 North A'ohoku Place, Hilo, HI 96720, USA
[18] Nederlandse Onderzoekschool voor Astronomie (NOVA), PO Box 9513, 2300 RA Leiden, The Netherlands
[19] California Institute of Technology, 1200 East California Boulevard, Pasadena, CA 91125, USA
[20] Institute of Astronomy and Astrophysics, Academia Sinica, 645 North A'ohoku Place, Hilo, HI 96720, USA
[21] Department of Physics, McGill University, 3600 rue University, Montréal, QC H3A 2T8, Canada
[22] McGill Space Institute, McGill University, 3550 rue University, Montréal, QC H3A 2A7, Canada
[23] Institut de Radioastronomie Millimétrique, 300 rue de la Piscine, F-38406 Saint Martin d'Hères, France
[24] Department of Astrophysics, Institute for Mathematics, Astrophysics and Particle Physics (IMAPP), Radboud University, P.O. Box 9010, 6500 GL Nijmegen, The Netherlands
[25] Perimeter Institute for Theoretical Physics, 31 Caroline Street North, Waterloo, ON, N2L 2Y5, Canada
[26] Department of Physics and Astronomy, University of Waterloo, 200 University Avenue West, Waterloo, ON, N2L 3G1, Canada
[27] Waterloo Centre for Astrophysics, University of Waterloo, Waterloo, ON, N2L 3G1, Canada
[28] Korea Astronomy and Space Science Institute, Daedeok-daero 776, Yuseong-gu, Daejeon 34055, Republic of Korea

[29] University of Science and Technology, Gajeong-ro 217, Yuseong-gu, Daejeon 34113, Republic of Korea
[30] Kavli Institute for Cosmological Physics, University of Chicago, 5640 South Ellis Avenue, Chicago, IL 60637, USA
[31] Department of Physics, University of Chicago, 5720 South Ellis Avenue, Chicago, IL 60637, USA
[32] Enrico Fermi Institute, University of Chicago, 5640 South Ellis Avenue, Chicago, IL 60637, USA
[33] Princeton Center for Theoretical Science, Jadwin Hall, Princeton University, Princeton, NJ 08544, USA
[34] Data Science Institute, University of Arizona, 1230 North Cherry Avenue, Tucson, AZ 85721, USA
[35] Cornell Center for Astrophysics and Planetary Science, Cornell University, Ithaca, NY 14853, USA
[36] Anton Pannekoek Institute for Astronomy, University of Amsterdam, Science Park 904, 1098 XH, Amsterdam, The Netherlands
[37] Shanghai Astronomical Observatory, Chinese Academy of Sciences, 80 Nandan Road, Shanghai 200030, People's Republic of China
[38] Key Laboratory of Radio Astronomy, Chinese Academy of Sciences, Nanjing 210008, People's Republic of China
[39] Physics Department, Fairfield University, 1073 North Benson Road, Fairfield, CT 06824, USA
[40] Department of Space, Earth and Environment, Chalmers University of Technology, Onsala Space Observatory, SE-43992 Onsala, Sweden
[41] Institut für Theoretische Physik, Goethe-Universität Frankfurt, Max-von-Laue-Straße 1, D-60438 Frankfurt am Main, Germany
[42] Mizusawa VLBI Observatory, National Astronomical Observatory of Japan, 2-12 Hoshigaoka, Mizusawa, Oshu, Iwate 023-0861, Japan
[43] Department of Astronomical Science, The Graduate University for Advanced Studies (SOKENDAI), 2-21-1 Osawa, Mitaka, Tokyo 181-8588, Japan
[44] Department of Astronomy and Columbia Astrophysics Laboratory, Columbia University, 550 W 120th Street, New York, NY 10027, USA
[45] Dipartimento di Fisica "E. Pancini," Universitá di Napoli "Federico II", Compl. Univ. di Monte S. Angelo, Edificio G, Via Cinthia, I-80126, Napoli, Italy
[46] INFN Sez. di Napoli, Compl. Univ. di Monte S. Angelo, Edificio G, Via Cinthia, I-80126, Napoli, Italy
[47] Wits Centre for Astrophysics, University of the Witwatersrand, 1 Jan Smuts Avenue, Braamfontein, Johannesburg 2050, South Africa
[48] Department of Physics, University of Pretoria, Hatfield, Pretoria 0028, South Africa
[49] Centre for Radio Astronomy Techniques and Technologies, Department of Physics and Electronics, Rhodes University, Makhanda 6140, South Africa
[50] LESIA, Observatoire de Paris, Université PSL, CNRS, Sorbonne Université, Université de Paris, 5 place Jules Janssen, F-92195 Meudon, France
[51] JILA and Department of Astrophysical and Planetary Sciences, University of Colorado, Boulder, CO 80309, USA
[52] National Astronomical Observatories, Chinese Academy of Sciences, 20A Datun Road, Chaoyang District, Beijing 100101, People's Republic of China
[53] University of Massachusetts Boston, 100 William T. Morrissey Boulevard, Boston, MA 02125, USA
[54] National Radio Astronomy Observatory, 520 Edgemont Road, Charlottesville, VA 22903, USA
[55] Department of History of Science, Harvard University, Cambridge, MA 02138, USA
[56] Department of Physics, Harvard University, Cambridge, MA 02138, USA
[57] Department of Physics, University of Illinois, 1110 West Green Street, Urbana, IL 61801, USA
[58] Department of Astronomy, University of Illinois at Urbana-Champaign, 1002 West Green Street, Urbana, IL 61801, USA
[59] Leiden Observatory—Allegro, Leiden University, P.O. Box 9513, 2300 RA Leiden, The Netherlands
[60] CP3-Origins, University of Southern Denmark, Campusvej 55, DK-5230 Odense M, Denmark
[61] Instituto Nacional de Astrofísica, Óptica y Electrónica. Apartado Postal 51 y 216, 72000. Puebla Pue., México
[62] Consejo Nacional de Ciencia y Tecnología, Av. Insurgentes Sur 1582, 03940, Ciudad de México, México
[63] Key Laboratory for Research in Galaxies and Cosmology, Chinese Academy of Sciences, Shanghai 200030, People's Republic of China
[64] NOVA Sub-mm Instrumentation Group, Kapteyn Astronomical Institute, University of Groningen, Landleven 12, 9747 AD Groningen, The Netherlands
[65] Center for the Fundamental Laws of Nature, Harvard University, Cambridge, MA 02138, USA
[66] Department of Astronomy, School of Physics, Peking University, Beijing 100871, People's Republic of China
[67] Kavli Institute for Astronomy and Astrophysics, Peking University, Beijing 100871, People's Republic of China
[68] Department of Astronomy, Graduate School of Science, The University of Tokyo, 7-3-1 Hongo, Bunkyo-ku, Tokyo 113-0033, Japan
[69] The Institute of Statistical Mathematics, 10-3 Midori-cho, Tachikawa, Tokyo, 190-8562, Japan
[70] Department of Statistical Science, The Graduate University for Advanced Studies (SOKENDAI), 10-3 Midori-cho, Tachikawa, Tokyo 190-8562, Japan
[71] Kavli Institute for the Physics and Mathematics of the Universe, The University of Tokyo, 5-1-5 Kashiwanoha, Kashiwa, 277-8583, Japan
[72] ASTRAVEO, LLC, PO Box 1668, Gloucester, MA 01931, USA
[73] Institute for Astrophysical Research, Boston University, 725 Commonwealth Avenue, Boston, MA 02215, USA
[74] Astronomical Institute, St. Petersburg University, Universitetskij pr., 28, Petrodvorets,198504 St.Petersburg, Russia
[75] Institute for Cosmic Ray Research, The University of Tokyo, 5-1-5 Kashiwanoha, Kashiwa, Chiba 277-8582, Japan
[76] Joint Institute for VLBI ERIC (JIVE), Oude Hoogeveensedijk 4, 7991 PD Dwingeloo, The Netherlands
[77] Kogakuin University of Technology & Engineering, Academic Support Center, 2665-1 Nakano, Hachioji, Tokyo 192-0015, Japan
[78] Physics Department, National Sun Yat-Sen University, No. 70, Lien-Hai Rd, Kaosiung City 80424, Taiwan, R.O.C.
[79] National Optical Astronomy Observatory, 950 North Cherry Avenue, Tucson, AZ 85719, USA
[80] Key Laboratory for Particle Astrophysics, Institute of High Energy Physics, Chinese Academy of Sciences, 19B Yuquan Road, Shijingshan District, Beijing, People's Republic of China
[81] School of Astronomy and Space Science, Nanjing University, Nanjing 210023, People's Republic of China
[82] Key Laboratory of Modern Astronomy and Astrophysics, Nanjing University, Nanjing 210023, People's Republic of China
[83] Italian ALMA Regional Centre, INAF-Istituto di Radioastronomia, Via P. Gobetti 101, I-40129 Bologna, Italy
[84] Department of Physics, National Taiwan University, No.1, Sect. 4, Roosevelt Road, Taipei 10617, Taiwan, R.O.C.
[85] Instituto de Radioastronomía y Astrofísica, Universidad Nacional Autónoma de México, Morelia 58089, México
[86] Instituto de Astronomía, Universidad Nacional Autónoma de México, CdMx 04510, México
[87] Yunnan Observatories, Chinese Academy of Sciences, 650011 Kunming, Yunnan Province, People's Republic of China
[88] Center for Astronomical Mega-Science, Chinese Academy of Sciences, 20A Datun Road, Chaoyang District, Beijing, 100012, People's Republic of China
[89] Key Laboratory for the Structure and Evolution of Celestial Objects, Chinese Academy of Sciences, 650011 Kunming, People's Republic of China
[90] Gravitation Astroparticle Physics Amsterdam (GRAPPA) Institute, University of Amsterdam, Science Park 904, 1098 XH Amsterdam, The Netherlands
[91] School of Natural Sciences, Institute for Advanced Study, 1 Einstein Drive, Princeton, NJ 08540, USA
[92] Tsung-Dao Lee Institute and School of Physics and Astronomy, Shanghai Jiao Tong University, 800 Dongchuan Road, Shanghai, 200240, People's Republic of China
[93] Science Support Office, Directorate of Science, European Space Research and Technology Centre (ESA/ESTEC), Keplerlaan 1, 2201 AZ Noordwijk, The Netherlands
[94] University of Chicago, 5640 South Ellis Avenue, Chicago, IL 60637, USA
[95] Astronomy Department, Universidad de Concepción, Casilla 160-C, Concepción, Chile
[96] National Institute of Technology, Hachinohe College, 16-1 Uwanotai, Tamonoki, Hachinohe City, Aomori 039-1192, Japan
[97] Department of Astronomy, University of Massachusetts, Amherst, MA 01003, USA
[98] South African Radio Astronomy Observatory, Observatory 7925, Cape Town, South Africa
[99] Department of Physics, National and Kapodistrian University of Athens, Panepistimiopolis, GR 15783 Zografos, Greece
[100] Villanova University, Mendel Science Center Rm. 263B, 800 East Lancaster Avenue, Villanova PA 19085, USA

[101] Physics Department, Washington University CB 1105, St Louis, MO 63130, USA
[102] Canadian Institute for Theoretical Astrophysics, University of Toronto, 60 St. George Street, Toronto, ON M5S 3H8, Canada
[103] Dunlap Institute for Astronomy and Astrophysics, University of Toronto, 50 St. George Street, Toronto, ON M5S 3H4, Canada
[104] Canadian Institute for Advanced Research, 180 Dundas Street West, Toronto, ON M5G 1Z8, Canada
[105] Radio Astronomy Laboratory, University of California, Berkeley, CA 94720, USA
[106] Department of Physics, National Taiwan Normal University, No. 88, Section 4, Tingzhou Road, Taipei 116, Taiwan, R.O.C.
[107] Frankfurt Institute for Advanced Studies, Ruth-Moufang-Strasse 1, 60438 Frankfurt, Germany
[108] School of Mathematics, Trinity College, Dublin 2, Ireland
[109] Department of Astrophysical Sciences, Peyton Hall, Princeton University, Princeton, NJ 08544, USA
[110] Instituto de Radioastronomía Milimétrica, IRAM, Avenida Divina Pastora 7, Local 20, E-18012, Granada, Spain
[111] Hiroshima Astrophysical Science Center, Hiroshima University, 1-3-1 Kagamiyama, Higashi-Hiroshima, Hiroshima 739-8526, Japan
[112] Aalto University Department of Electronics and Nanoengineering, PL 15500, FI-00076 Aalto, Finland
[113] Aalto University Metsähovi Radio Observatory, Metsähovintie 114, FI-02540 Kylmälä, Finland
[114] Department of Astronomy, Yonsei University, Yonsei-ro 50, Seodaemun-gu, 03722 Seoul, Republic of Korea
[115] Netherlands Organisation for Scientific Research (NWO), Postbus 93138, 2509 AC Den Haag, The Netherlands
[116] Frontier Research Institute for Interdisciplinary Sciences, Tohoku University, Sendai 980-8578, Japan
[117] Astronomical Institute, Tohoku University, Sendai 980-8578, Japan
[118] Department of Physics and Astronomy, Seoul National University, Gwanak-gu, Seoul 08826, Republic of Korea
[119] Leiden Observatory, Leiden University, Postbus 2300, 9513 RA Leiden, The Netherlands
[120] Jeremiah Horrocks Institute, University of Central Lancashire, Preston PR1 2HE, UK
[121] Physics Department, Brandeis University, 415 South Street, Waltham, MA 02453, USA
[122] School of Physics, Huazhong University of Science and Technology, Wuhan, Hubei, 430074, People's Republic of China
[123] Mullard Space Science Laboratory, University College London, Holmbury St. Mary, Dorking, Surrey, RH5 6NT, UK
[124] School of Astronomy and Space Sciences, University of Chinese Academy of Sciences, No. 19A Yuquan Road, Beijing 100049, People's Republic of China
[125] Astronomy Department, University of Science and Technology of China, Hefei 230026, People's Republic of China